# Self-replicating fuels via autocatalytic molecular bond fission

Peter Agbo[1,2,3]*

[1]Chemical Sciences Division
Lawrence Berkeley National Laboratory
Berkeley, CA 94720 USA

[2]Molecular Biophysics & Integrated Bio-imaging Division
Lawrence Berkeley National Laboratory
Berkeley, CA 94720 USA

[3]Carbon Negative Initiative
Lawrence Berkeley National Laboratory
Berkeley, CA 94720, United States

*Corresponding author

**Abstract**

This computational study introduces a conceptual framework for practical, electrochemical fuel generation schemes that display exponential product yields as functions of time. Exponential reaction scaling for formate replication is simulated through an autocatalytic cycle that emulates the process of DNA replication facilitated by the polymerase chain reaction (PCR). Here, an initial buildup of formate into a two-carbon chain through $CO_2$ carboxylation forms oxalate. A subsequent, two-electron reduction yields glyoxylate, with base-mediated hydrolysis driving C-C bond breakage of glyoxylate into two formate equivalents. These products are then recycled to serve as reactants. This recursive process chemistry drives formate evolution that scales as $2^n$, where $n$ is the cycle number. Each step of the proposed fuel cycle is analogized to the steps of DNA annealing, nucleotide polymerization and hybridized strand fission that are responsible for the exponential product yields observed in PCR-mediated DNA synthesis. As a consequence of this replication behavior, rapid rates of fuel production become accessible, even when the individual rate constants for the cycle's constituent reactions are slow. Practical barriers to realizing this system are discussed, particularly the difficulty of formate carboxylation and the energy demands of chemical amplification.

## Introduction

Binary cell division (mitosis) displays exponential growth, with population sizes evolving according to a $2^n$ expansion rule, where $n$ is the doubling time[1]. Similarly, the process of DNA amplification using the polymerase chain reaction (PCR) involves chemical reactions where temporal product yields have an exponential time dependence, in contrast to the linear time dependencies found in most chemical systems[2]. In physics, such nonlinearities are also common. For example, both lasing[3] and nuclear fission[4,5] exhibit exponential increases in photon yield (as a function of input power) and $^{235}U/^{240}Pu$ fission rates (as a function of incident neutron flux), respectively. However, despite the clear prevalence of nonlinear processes in biology and physics, mainstream application of nonlinear reactions towards chemical energy storage remains a comparatively marginal subject area[6–8]. Instead, research in this domain generally involve fuel-forming reactions that exhibit steady-state behavior and product yields that scale linearly with time. In chemistry, autocatalysis has generally been applied towards origin of life chemistry[9,10], understanding the behavior of atmospheric aerosols[11], molecular synthesis/self-assembly[12–14], and nanoparticulate growth[15]. In close relation to this study, work in the field of capacitive energy storage has demonstrated the possibility of constructing networks of capacitors where rates of energy storage scale nonlinearly as a function of cycle number[16]. A key distinction between forms of autocatalysis such as nuclear fission and the chain reactions comprising chemical explosions, versus phenomena like DNA and cell replication, is that the former represent processes where autocatalytic behavior drives rapid energy release, whereas the latter are thermodynamically-driven processes demanding the uptake of energy from the environment. As a result, these driven processes effectively concentrate the diffuse energy of a thermal bath into energy-dense units such as the biosynthetic components of a cell or the phosphoanhydride bonds comprising replicating nucleic acids. This latter class of driven autocatalytic processes are effectively fuel-forming reactions, as they result in products where energy has been stored in chemical bonds; they inspire the type of self-replicating process discussed here. This understanding establishes autocatalytic fuel production as the effective conjugate of a process such as nuclear fission, where now, instead of rapid expansions in the rate of energy release over time, energy from some power source is rapidly incorporated as fuel at increasing rates with time.

This is distinct from conventional fuel-forming processes. The kinetics of energy conversion to fuel are typically constant with time, so fuel concentrations only scale linearly as the product of time and reaction rate. This serves as a key advantage of autocatalysis: because the reactant is also the product, this leads to higher-order scaling in reaction rates as functions of time and reactant concentration. This lies in stark contrast to linear processes where reaction orders are commonly 1$^{st}$ or 2$^{nd}$ order with respect to reactant concentration and invariant with time. As a consequence of this, an autocatalytic fuel cycle can achieve very fast rates of catalysis, even for systems comprised of individual steps with very slow rate constants and small initial reactant concentrations. In the context of electrochemical reactions, this behavior carries the additional possibility of constructing devices where mass-transport impedances actually decrease with time, as rates of substrate flux to electrode active sites rapidly increase with the rapidly growing substrate concentration. Furthermore, because these reactions can be efficiently run as batch processes (with respect to the reactant/product), the rapid product growth means they effectively incorporate the task of fuel production and fuel concentration into a single step.

This computational study explores a non-linear amplification mechanism for the generation of formate, a prospective hydrogen carrier and renewable fuel[17]. More broadly, this work aims to establish the general approach for rationally designing autocatalytic processes for renewable, self-replicating chemical fuel production. The reported concept adopts the autocatalytic mechanism for the *in-vitro* replication of DNA via PCR as a generic reaction template, upon which a specific process for the autocatalytic generation of formate is superimposed. Starting from the PCR template scheme, retrosynthetic analysis was applied to determine the intermediate compounds needed to generate two formate molecules from an initial input of one formate molecule and one molecule of carbon dioxide (Figure 1a). The recursive chemical mechanism resulting from this reverse-engineered approach displays exponential generation of formate with respect to time, in direct relation to the exponential yields of DNA realized through PCR. The proposed autocatalytic cycle for formate evolution is decomposed into three reaction steps. Each of these are explored at successively-increasing levels of mechanistic detail, following realistic strategies for driving them using existing chemistries. Through this exercise, we demonstrate that each of three basic reactions in the proposed cycle are possible, if challenging. Complexities that arise with increasing mechanistic granularity

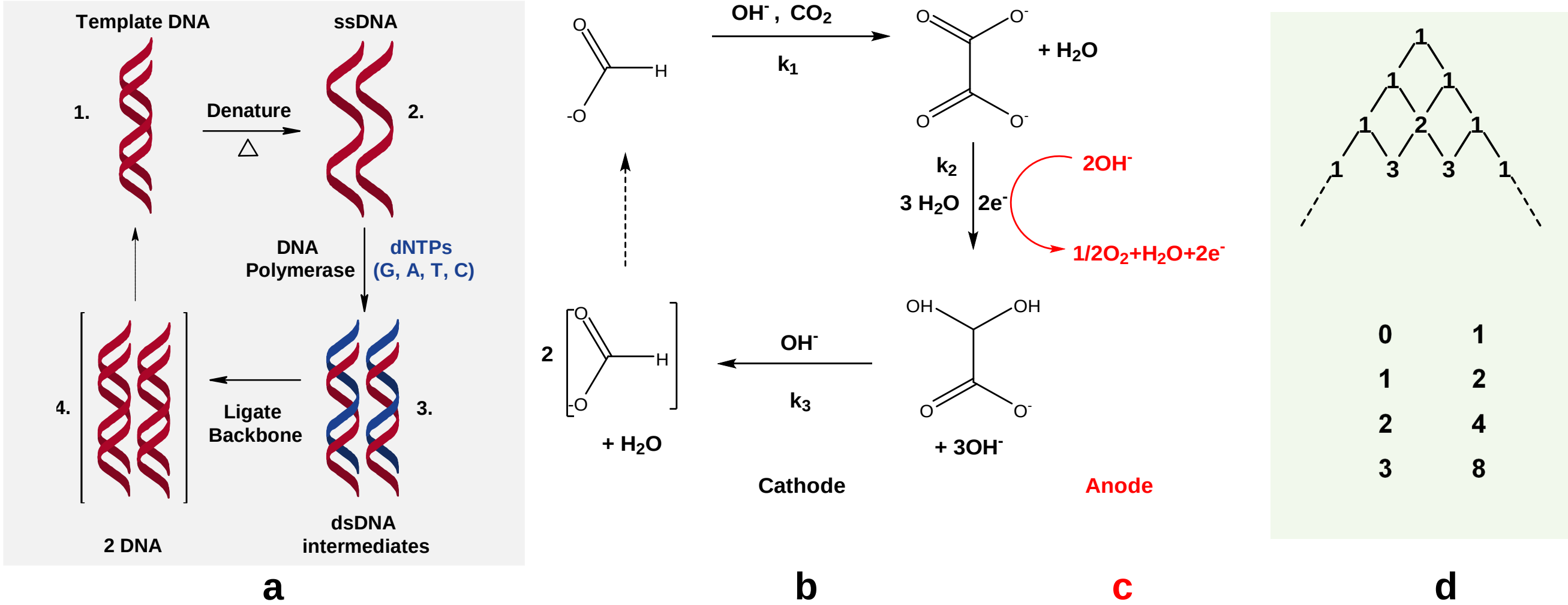

**Figure 1**

**a)** The chemical mechanism of the polymerase chain reaction. Replication of a single DNA molecule (and strand) exhibits exponential growth kinetics (amplification) in regimes where the reaction order is pseudo-first order in [DNA], with the deoxynucleotide triphosphates (dNTPs) serving as the monomer building blocks present in excess. In the ideal limit, reaction kinetics follow a $2^n$ expansion rule, where n is the cycle number. At high cycle numbers (longer times), dNTP consumption results in DNA replication kinetics that are limited by [dNTP], arresting polymerization at long times. **b)** A PCR-analogous scheme for formate generation. Exponential formate yields are initially seeded by a low concentration of the compound, with $CO_2$ serving as a building block analogous to the role played by dNTPs in PCR. The C-C coupling of $CO_2$ and formate in the first reaction step, and the hydrolytic C-C bond fission of glyoxylate to yield two formate molecules in step 3, can be viewed as equivalent to the respective steps of annealing and denaturation found in PCR cycles. **c)** The step of oxalate reduction (step 2) points to the possibility of driving this formate cycle by coupling this redox process to electrochemical or photoelectrochemical water-splitting. **d)** In the limit of perfect cycle efficiencies, product yields in these schemes follow Pascal's triangle.

are addressed, including observed deviations from ideal exponential behavior that inevitably occur as a result of substrate depletion in autocatalytic synthesis. We discuss the requirements that each elementary step of the proposed mechanism must satisfy for exponential growth kinetics to be observed, along with the thermodynamic constraints of a self-replicating fuel cycle. The limits of satisfying the exponentially-growing energy demands for sustaining such a cycle are discussed with respect to practical implementation. Practical implications, and a generalized set of rules governing the design of autocatalytic reactions dependent on molecular fission events are considered, with an emphasis on their applicability to a wider range of chemical transformations.

## Results

*Concept Framework*

As the analogy with PCR would suggest, mathematical descriptions of the proposed system for exponential formate production parallel those used to describe PCR-mediated DNA amplification[18]. The net reaction scheme (Figure 1b, c) is reducible to the simplest case of autocatalytic sets, of the form:

$$\mathrm{A} + \mathrm{B} \rightarrow 2\ \mathrm{B} \qquad \text{eq. 1a.}$$

The above reaction describes the most fundamental type of autocatalytic "core," according to taxonomic classifications for autocatalytic processes[19,20]. In this study, substrate A ($CO_2$) is supplied at a steady-state concentration required for sustaining the amplification of species B (formate). Here, formate generation per cycle follows a power rule (Figure 1d):

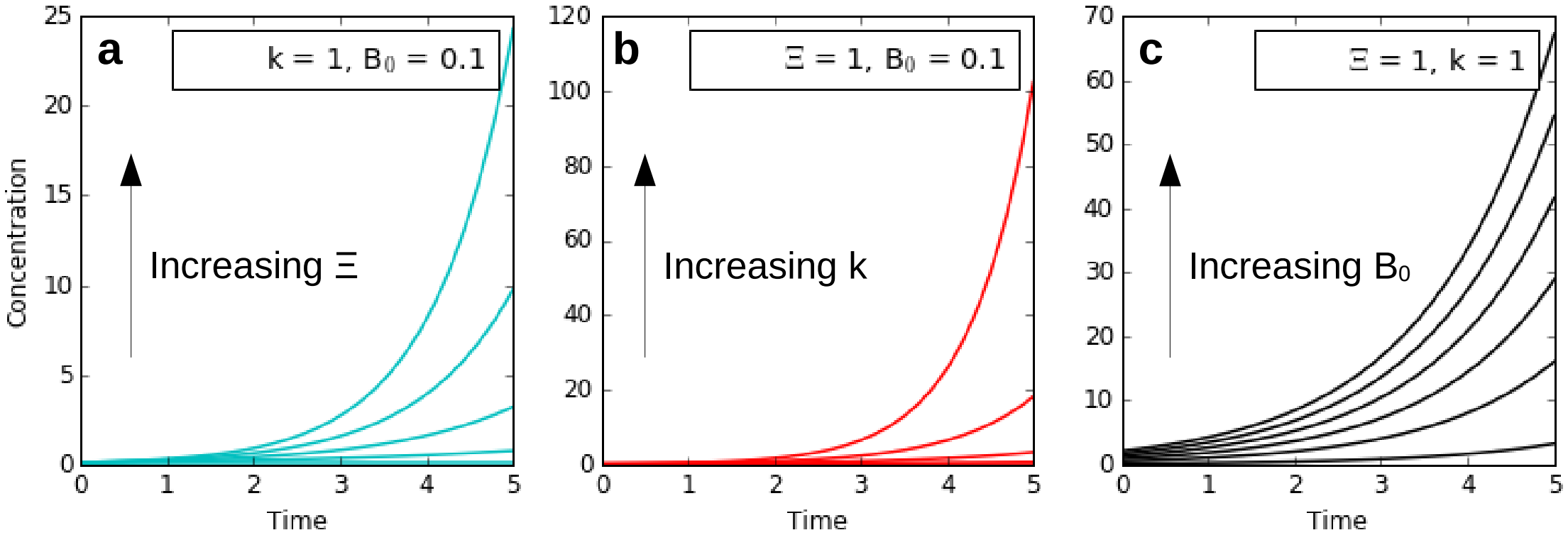

**Figure 2**

Dependencies of unbounded autocatalytic growth functions (exponential phase only), describing some arbitrary, chemical replicator, on parameters **(a)** Ξ, the reaction selectivity for a product, **(b)** the cycle rate constant, k = 1/τ and **(c)** the seed concentration of the replicator, $B_0$. While these functions are only idealized representations displaying no limits on growth, these relationships usefully describe the key dependencies of real autocatalytic replication during the exponential growth phase, for reaction types that include PCR, formate amplification in this study and reactions with $2^t$ scaling generally.

$$B(n)=B_0[1+\Xi_n]^n \qquad \text{eq. 1b,}$$

where $n$ refers to the $n^{th}$ reaction cycle, $B_0$ is the initial formate concentration, and $\Xi_n$ refers to the conversion selectivity towards formate of cycle $n$. In the ideal limit of no side reactions, $\Xi = 1$. Discrete cycles may be expressed as continuous functions of time ($t$) through the relation $n = t/\tau$, where $\tau$ is the characteristic time constant for each reaction cycle. The characteristic rate constant of replication is given by $k = 1/\tau$. These relations permit the description of concentration as a continuous function of time:

$$B(t) = B_0[1+\Xi(t)]^{\frac{t}{\tau}} = B_0[1+\Xi(t)]^{kt} \qquad \text{eq. 1c.}$$

Differentiation of eq. 1c with respect to time yields the rate equation for formate evolution:

$$r(t) = B_0(1+\Xi(t))^{kt}\left[k\ln(1+\Xi(t)) + \frac{kt}{1+\Xi(t)}\frac{d\Xi}{dt}\right] \qquad \text{eq. 1d.}$$

The general form of eqs. 1b-d speak to the exponential nature of formate production in the scheme described, with dependencies on the parameters τ, Ξ, and $B_0$ as depicted in Figure 2. The parameter $\tau = 1/k$ (where k is a unimolecular rate constant) describes the characteristic time required for the $n^{th}$ reaction cycle to complete (the replication constant). In the limit of a reaction cycle featuring no side reactions, $\Xi(t)$ converges to 1 and the bracketed quantity collapses to $2^n$ ($2^{t/\tau}$), resulting in a growth function displaying exponential behavior. While this represents a theoretical limit on yields per cycle, it should be noted that exponential growth may be observed for all $\Xi(t) > 0$, $k > 1$. In the limit $\Xi(t) \rightarrow 0$, reaction kinetics will converge towards arrested growth. It should be noted that equations 1b-d, while descriptive of the system explored here, are not generalized expressions for autocatalysis. Those laid out by Hanopolskyi et al.[21] or expressed by logistic growth equations[22], are rigorously derived and incorporate the effects of exponential plateau in the limit of reactant depletion. However, the use of equations 1 and 2 in this study was preferred as a result their expressing nonlinear expansion intuitively, while enabling a sharp parallel to be cast between the power functions commonly used to describe the PCR[18], cell replication[1] and nuclear fission processes that inspired this approach. In this current study, simulations are performed over a model system that is a semi-batch process, with $CO_2$ treated as a reactant supplied at a steady-state concentration. It is therefore not subject to depletion kinetics; all other species are treated as batch quantities, provided either at the start of the simulation or evolved through the course of autocatalytic cycling. It must be noted that steady-state $CO_2$ concentrations carry the implicit understanding that for a real system, this means that rates of $CO_2$ mass transport from the gas phase into the solution bulk must occur at rates fast enough to accommodate the rate of $CO_2$ consumption by autocatalysis, which will be increasing with time to sustain formate replication. Finally, rates of energy delivery into this thermodynamically-driven cycle serve as an ultimate limit on the system's replication kinetics. In this study, the endergonic conversions of formate to oxalate and oxalate reduction to glyoxylate are the points of external energy consumption (inputs) by the replicating system.

While some PCR models assume a static product selectivity, a generalized, time-dependent description of process selectivity will more accurately capture the system's time-evolution[18]. The overall, time-dependent selectivity of the entire fuel cycle is then given by the product of the individual selectivity (Ξ) of reactions 1-3 (S.5):

$$\Xi(t)=\prod_{j=1}^{3} \Xi_j(t) \qquad \text{eq. 2a.}$$

In this formulation, time dependence of individual reaction and overall cycle selectivity is captured by the rate constants defining product formation kinetics for each step (S.5).

As a time-dependent selectivity suggests, real replication processes will eventually deviate from perfect exponential growth. Such deviations occur for all real replicators bounded by a physical limit, such as limiting reactant or energy availability from the replicator's surroundings. As a result, determining key aspects of a replicator, such as the maximum value of its replication rate constant, becomes difficult and makes rigorous application of growth models such as those encapsulated by eqs. 1b-1d challenging when exponential growth phases deviate from ideality. In response, this paper introduces the following analytical approach, which exploits the fact that for an ideal replicator (or a non-ideal replicator at early times), the logarithm of its growth with respect to time is linear. Deviations from linearity indicate times where replication diverges from purely exponential growth as a result of limiting processes, setting the bound for curve-fits needed to extract replicator rate constants. Taking the logarithm of eq. 1c gives:

$$\ln(B(t)) = kt\ln[1+\Xi(t)] + \ln(B_0) \qquad \text{eq. 2b.}$$

For times where selectivity is constant, this yields a slope of $k\ln(B_0[1 + \Xi])$ and equals $k\ln(2B_0)$ for an ideal $2^t$ scaling replicator. Furthermore, taking the derivative of equation 2b yields:

$$\frac{d\ln(B(t))}{dt} = k\ln[1+\Xi(t)] + \frac{kt}{[1+\Xi(t)]}\frac{d\Xi}{dt} \qquad \text{eq. 2c.}$$

The time-invariant selectivity of an ideal replicator forces the differential term $d\Xi/dt$ to zero, causing the time dependent term of eq. 2c to vanish. In this limit, eq. 2c collapses to a time-independent term of constant Ξ, the case for an ideal replicator following purely exponential growth kinetics:

$$\frac{d\ln(B(t))}{dt} = k\ln[1+\Xi] \qquad \text{eq. 2d.}$$

As a result, using either eqs. 2b or 2d, we can readily determine the points where a non-ideal replicator begins to deviate from ideality and set the appropriate times for extracting values for $k$ (or $\Xi$) (S.1). The result expands the analytical utility of the idealized exponential replicator model for parameter extraction, even in the realistic cases of bounded, non-deal autocatalysts that deviate from exponential growth (S.1).

**Table 1 – Simulated 'Basic' System**

| **Reaction** | **Forward** | **Units** | **Reverse** | **Units** | **Ref** |
|---|---|---|---|---|---|
| Autocatalytic System | | | | | |
| (1) $HCO_2^- + CO_2 + H_2O \rightarrow C_2O_4^{2-} + H_3O^+$ | $k_1 = 10$ | $M^{-1}\ s^{-1}$ | -- | – | -- |
| (2) $C_2O_4^{2-} + 2\ H_2O \rightarrow C_2H_3O_4^- + OH^- + 0.5\ O_2$ | $k_2 = 0.01$ | $s^{-1}$ | -- | – | -- |
| (3) $C_2H_3O_4^- + OH^- \rightarrow 2\ HCO_2^- + H_2O$ | $k_3 = 1000$ | $M^{-1}\ s^{-1}$ | -- | – | -- |
| | | | | | |
| Buffering Equilibria ($pK_{a1} = 8$, $pK_{a2} = 13$) | | | | | |
| $H_3O^+ + B: \leftrightarrow BH + H_2O$ | $k_a = 1e8$ | $M^{-1}\ s^{-1}$ | $k_a' = 1$ | $s^{-1}$ | -- |
| $OH- + BH \leftrightarrow B: + H_2O$ | $k_b = 1e6$ | $M^{-1}\ s^{-1}$ | $k_a' = 1$ | $s^{-1}$ | -- |
| $BH + H_3O^+ \leftrightarrow BH_2 + H_2O$ | $k_a = 1e13$ | $M^{-1}\ s^{-1}$ | $k_a' = 1$ | $s^{-1}$ | -- |
| $OH- + BH_2 \leftrightarrow BH + H_2O$ | $k_b = 1e1$ | $M^{-1}\ s^{-1}$ | $k_a' = 1$ | $s^{-1}$ | -- |
| | | | | | |
| Solvent Autoionization | | | | | |
| $H_3O^+ + OH \leftrightarrow 2\ H_2O$ | $k_w = 1.12e11$ | $M^{-1}\ s^{-1}$ | $k_w' = 1.12e-3$ | $s^{-1}$ | [23] |

*Fuel replication kinetics – 'Basic System,' ideal replicator*

The fuel cycle under consideration is represented as reactions 1-3 in Table 1. Reactions (1) and (3) are difficult to facilitate. In the case of (3), no work to our knowledge exists detailing the kinetics of this reaction, although its thermodynamics have been reported[24]. As a result of the unavailability of rate constants for these reactions, for demonstration purposes, a range rate constants spanning orders of magnitude were used in simulations (S.4). In an effort to keep the simulation assumptions conservative as possible, the range of rate constants tested for reactions (1) and (3) span relatively low values, with mechanisms assumed to follow bimolecular kinetics. A base case assumes rate constants of 10 $M^{-1}\ s^{-1}$ for reaction (1) and 1000 $M^{-1}\ s^{-1}$ for reaction (3). Buffering reaction rates were determined using values for rate constants of proton/base recombination in autoionization equilibria[23], in conjunction with equilibrium constants for respective buffer $pK_a$ values and water dissociation processes (Table S1).

Simulations of the autocatalytic fuel cycle described here were carried out using Kinetiscope. As a stochastic kinetics simulator, Kinetiscope discretizes the problem of microkinetics simulation using the formalism pioneered by Gillespie[25], explicitly tracking the time evolution of all molecules in a reaction volume. Using the reaction scheme in Table 1, stochastic calculations of this system yield an amplifying formate concentration, in a manner analogous to the amplification of DNA produced via PCR. Rather than time-dependent linear increases in product concentration, formate production at steady-state $CO_2$ concentrations (100 μM) is characterized by a fast exponential rise, with the concentration of formate exceeding 1.5 M within the first 12000 s of the reaction (Figure 3a), for a reaction seeded with $B_0$ = 100 μM formate. It is worth observing that this autocatalytic cycle may, in principle, be seeded with any of the cycle intermediates – formate, oxalate or DHA (Figures 3b, 3c, 3d). In cases where bulk $CO_2$ concentrations are not held constant, $CO_2$ availability rapidly becomes limiting. This results in transient reactions whose rates quickly decay, with only negligible amounts of formate being generated (S.6). Modifying buffering equilibria in this toy model system to include actual values for phosphate and inorganic carbon (DIC) speciation has virtually no effect on the system behavior over the examined timescale (S.1, Figure S3).

The conversion of oxalate to DHA is a redox process; coupling this two-electron reduction to the oxygen evolution reaction yields the overall redox equation of reaction 2. We can therefore consider how an electrochemical cell integrating this type of autocatalysis will respond in this ideal limit. Comparing autocatalytic formate yields to the performance of an equivalent $CO_2$-to-formate electrolyzer illustrates the dramatic differences between the linear vs autocatalytic fuel synthesis. The autocatalytic electrolyzer current is calculated as the first-derivative of oxygen concentration over time:

$$I = n_e F \, \nu_{rxn} \frac{d[O_2]}{dt}$$

eq. 3a,

where $F$ is Faraday's constant, $n_e$ (= 4) is the number of electrons transferred per $O_2$ evolved, and $\nu_{rxn}$ = 1 $cm^3$ is the simulation reaction volume. This is in stark contrast to the current calculated for a comparable linear electrolyzer with a similar turnover rate constant, which yields a steady-state current according to:

$$I = n_e \nu_{rxn} F k_{cat} [CO_2]$$ eq. 3b.

Equation 3b assumes an ideal case for the standard formate electrolyzer, where current is only limited by $CO_2$ concentration (100 µM. Here $n_e = 2$, as $k_{cat}$ is for the $2e^-$ reduction of $CO_2$ to formate. The catalytic rate constant, $k_{cat} = 0.0013\ s^{-1}$. In this ideal limit, the steady-state current for such an electrolyzer is merely 25 µA $cm^{-3}$. This contrasts with the autocatalytic case, where current arising from oxalate reduction is 0 A at $t = 0$ but rapidly grows in proportion to formate and molecular oxygen growth, reaching 300 mA $cm^{-3}$ after 12000 s (Figure 3). Note that here, the 0-D nature of the simulation gives a current density in units of A $cm^{-3}$. We address this in subsequent simulations where system dimensionality is made explicit.

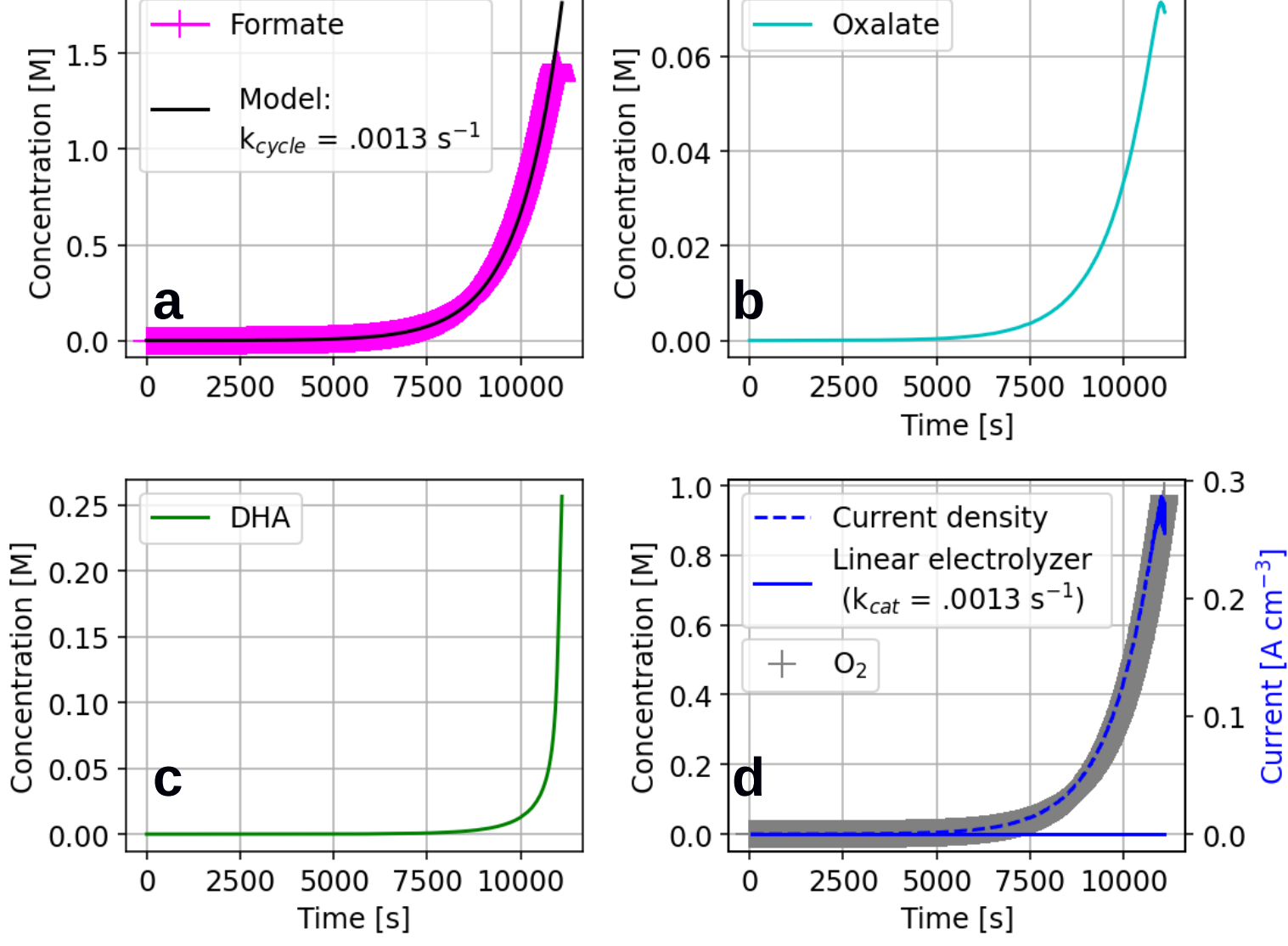

**Figure 3**

**a)** A stochastic kinetics simulation of ideal, unbounded formate autocatalysis according to a PCR-analogous mechanism described in Table 1. $B_0$ = 100 µM; $[CO_2]$ is held constant at 100 µM. As an ideal autocatalytic system, growth can be modeled with eqs. 1c and 1d, for $k_{cycle} = 1/\tau = 0.0013\ s^{-1}$, $\Xi$=1. Nonlinearity in the fuel cycle also results in exponential scaling in the growth of oxalate **(b)** and glyoxylate **(c)** intermediates as functions of time, with relative growth rates set by the relative rate constants feeding and consuming each species. **(d)** Coupling the reductive chemistry of the autocatalytic cycle to electrochemical oxygen evolution (gray) suggests the possbility of driving such processes electrochemically, provided that energetic sources are sufficiently power-dense to drive the process. Calculation of the first derivative of $O_2$ evolution (blue, dashed) is used to determine total current consumed in the generation of all reductive intermediates in the cycle for all times t. Comparison to an idealized, traditional electrolyzer system (blue line) operating within a regime only limited by substrate concentration (100 µM $CO_2$). Linear electrolyzer current is calculated using turnover rate constant equal to $k_{cycle}$, yielding a steady-state device current density of 25 µA $cm^{-3}$.

This carries direct implications for the expected current response for any electrolyzer employing this autocatalytic cycle. First, the exponential increase in the rate of formate production means the rate of charge flow must scale similarly, with electrical current increasing with time. In contrast to a typical electrolyzer, where catalytic turnover depletes available substrate, autocatalytic feedback between electrolyzer current and oxalate concentrations leads to increasing oxalate concentration. Now, the high current densities required by the system at long times are facilitated by a substrate availability that increases with time, despite the slow replication rate constant. In general, even slow catalysts should enable product synthesis that becomes more efficient as catalysis proceeds and substrate concentration and rate of production increases with successive turnovers, until some depletion condition is reached. Implementing such nonlinear schemes would offer practical utility for electrocatalysts that may be highly selective but have catalytic rate constants too low to be useful in linear electrolyzers.

A challenge with implementing this type of nonlinear scheme is highlighted by the requirement of supporting exponential current growth, which mandates that rates of $CO_2$ dissolution from the gas phase must keep pace with consumption. Additionally, the rates of side products evolved may become non-negligible, as they also may also display exponential amplification. However, any interference from amplified side products will depend on the particulars of the side reactions, especially their associated rate constants. Despite the focus on formate production here, each of the intermediates involved in the proposed autocatalytic cycle are subject to exponential growth. The relative growth rates of formate, oxalate, and DHA are functions of the rate constants for the reactions driving both their formation and depletion. This makes it possible to control the selectivity of the cycle for one intermediate over another, allowing for significant shifts in the product accumulation profile. While the base case tested ($k_1 = 10\ M^{-1}\ s^{-1}$, $k_2 = 0.01\ s^{-1}$, $k_3 = 1000\ M^{-1}\ s^{-1}$) causes formate growth to occur at rates significantly greater than that of oxalate or DHA, modifying the simulation by slowing down the depletion of DHA ($k_3 = 1\ s^{-1}$) results in a significantly faster rate of DHA growth. Similarly, increasing $k_1$ to 10 $s^{-1}$ or decreasing $k_2$ to 0.1 $s^{-1}$ increases the rate of oxalate accumulation (S.4).

*Fuel replication kinetics – 'Partial Mechanistic System,' non-ideal replicator*

Simulating this system using the stoichiometric rate laws for the net reactions requires the invocation of direct formate carboxylation in step (1). This involves the C-C coupling of two low-energy carbon compounds that are famously difficult to activate. In addition, reaction (2) represents a redox process, which must be expressed as an explicit sum of paired oxidation/reduction reactions in order to be meaningfully simulated. As a result, we now gradually build system complexity into the toy model described in Table 1, by decomposing these three reactions into the serial mechanistic processes that yield the net reactions. We start by first expressing reactions (2) and (3) as discrete, mechanistic steps while leaving (1) as a stoichiometric rate law for now (Table 2):

**Table 2 – Simulated 'Partial Mechanistic' System**

| Reaction | Forward | Units | Reverse | Units | Ref or SI |
|---|---|---|---|---|---|
| Autocatalytic System | | | | | |
| (1) $\mathbf{HCO_2^-}$ **+** $\mathbf{CO_2}$ + $H_2O$ → $\mathbf{C_2O_4^{2-}}$ + $H_3O^+$ | 10 | $M^{-1}\ s^{-1}$ | -- | -- | -- |
| (2) $C_2O_4^{2-}$ + 2 $H_2O$ → $C_2H_3O_4^-$ + $OH^-$ + 0.5 $O_2$ | | | | | |
| Cathode: $C_2O_4^{2-}$ + 3 $H_2O$ + 2e⁻ → $C_2H_3O_4^-$ + 3 $OH^-$ | | | | | |
| 2.1 $\mathbf{C_2O_4^{2-}}_{\mathbf{(aq)}}$ + A ↔ A-$C_2O_4^{2-}{}_{(ads)}$ | 2.4 | $s^{-1}$ | 0.8 | $s^{-1}$ | S.2 |
| 2.2 A-$C_2O_4^{2-}{}_{(ads)}$ + 2 $H_2O$ + 2$e^-$ ↔ A-$C_2HO_3^-{}_{(ads)}$ + 3 $OH^-$ | 215 | $s^{-1}$ | 0.002 | $s^{-1}$ | S.3 |
| 2.3 A-$C_2HO_3^-{}_{(ads)}$ ↔ $C_2HO_3^-{}_{(aq)}$ + A | 0.8 | $s^{-1}$ | 2.4 | $s^{-1}$ | S.2 |
| 2.4 $C_2HO_3^-{}_{(aq)}$ + $H_2O$ ↔ $\mathbf{C_2H_3O_4^-}_{\mathbf{(aq)}}$ | 70 | $s^{-1}$ | 1.14 | $s^{-1}$ | [26] |
| Anode: 2 $OH^-$ ↔ $H_2O$ + 0.5 $O_2$ + 2e- | 215 | $s^{-1}$ | 0.002 | $s^{-1}$ | S.3 |
| (3) $C_2H_3O_4^-$ + $OH^-$ → 2 $HCO_2^-$ + $H_2O$ | | | | | |
| 3.1 $\mathbf{C_2H_3O_4^-}$ + Cat ↔ Cat-$C_2H_3O_4^-$ | 100 | $M^{-1}\ s^{-1}$ | 1 | $s^{-1}$ | S.7 |
| 3.2 Cat-$C_2H_3O_4^-$ + $OH^-$ ↔ Cat-DHA* | 1000 | $M^{-1}\ s^{-1}$ | 1 | $s^{-1}$ | S.7 |
| 3.3 Cat-DHA* → **2** $\mathbf{HCO_2^-}$ + Cat + $H_2O$ | 10 | $s^{-1}$ | -- | -- | S.7 |
| | | | | | |
| Buffering Equilibria | | | | | |
| Phosphate – Values for $H_3PO_4^-$ (**$BH_2$**) ↔ $H_2PO_4^{2-}$ (**BH**) ↔ $HPO_4^{3-}$ (**B:**) equilibrium ($pK_{a1}$ = 7.20, $pK_{a2}$ = 12.32) | | | | | |
| $H_3O^+$ + B: ↔ BH + $H_2O$ | 1.58e7 | $M^{-1}\ s^{-1}$ | 1 | $s^{-1}$ | [27] |
| $OH^-$ + BH ↔ B: + $H_2O$ | 6.34e6 | $M^{-1}\ s^{-1}$ | 1 | $s^{-1}$ | [27] |
| $BH + H_3O^+ \leftrightarrow BH_2 + H_2O$ | 2.38e12 | $M^{-1}\,s^{-1}$ | 1 | $s^{-1}$ | [27] |
| $OH^- + BH_2 \leftrightarrow BH + H_2O$ | 4.21e1 | $M^{-1}\,s^{-1}$ | 1 | $s^{-1}$ | [27] |
| Carbonate | | | | | |
| $CO_2 + H_2O \leftrightarrow H_2CO_3$ | 0.04 | $s^{-1}$ | 12 | $s^{-1}$ | [28–31] |
| $CO_2 + OH^- \leftrightarrow HCO_3^-$ | 1.21e4 | $M^{-1}\,s^{-1}$ | 4e-4 | $s^{-1}$ | [31] |
| $H_2CO_3 \leftrightarrow H^+ + HCO_3^-$ | 1e7 | $s^{-1}$ | 5e10 | $M^{-1}\,s^{-1}$ | [30] |
| $HCO_3^- \leftrightarrow H^+ + CO_3^{2-}$ | 3 | $s^{-1}$ | 5e10 | $M^{-1}\,s^{-1}$ | [30] |
| | | | | | |
| Solvent Autoionization | | | | | |
| $H_3O^+ + OH \leftrightarrow 2\,H_2O$ | $k_w$ = 1.1e11 | $M^{-1}\,s^{-1}$ | $k_w'$ = 1.1e-3 | $s^{-1}$ | [23] |

The second step of the cycle, oxalate reduction to glyoxylate, is known electrochemistry. Previous work has demonstrated the possibility of reducing oxalate formed *in situ* (via $CO_2$ reduction) to glyoxylate under alkaline conditions[32–35], and in the case of a gallium tin oxide ($GaSnO_x$) catalyst, at faradaic yields over 97%[36]. In the mechanism for step (2), $A$ represents a free electrode active site, available for oxalate binding and subsequent reduction to glyoxylate. At time zero, $[A]_0$ equals the total active site concentration, $[A]_T$, with an active site population that evolves over the course of formate replication according to:

$$[A](t) = [A]_0 - [A\text{-oxa}_{(ads)}](t) - [A\text{-Gly}_{(ads)}](t). \qquad \text{eq. 4a.}$$

A value of $[A]_0 = 10^{17}$ $cm^{-2}$ is used for initializing these simulations. This represents a number within range of a typical, close-packed metal catalyst layer ($10^{14} - 10^{15} cm^{-2}$) deposited on a high-aspect ratio surface (100-1000x). Electrode adsorption of oxalate is treated as diffusion controlled, with oxalate adsorption ($A\text{-oxa}_{(ads)}$) at electrode active sites approximated as a pseudo-first order dependence of oxalate concentration at the electrode surface (S.2). The rate constant for oxalate diffusion is calculated according to a modified, Fickian description of diffusive mean squared displacement:

$$\frac{1}{k_{diff}} \approx \frac{\langle x^2 \rangle}{2 n_d \gamma_{mix} D} \qquad \text{eq. 4b,}$$

where $x$ is the diffusion length, $D$ is the diffusion constant, and $n_d$ gives the degrees of freedom in the system (taken as 3 for this system). The term $\gamma_{mix}$ is a purely empirical factor, reflecting the increased effective diffusion constant of a species under mixing conditions when $\gamma_{mix} > 1$; $\gamma_{mix} = 1$ reflects the case of pure diffusion in quiescent solution. Treatment of the reaction volume as a cubic, 1 $cm^3$ reactor featuring a 1 $cm^2$ electrode at one of the cube faces. The electrode surface region is defined as a volume element projected within one bond length (~0.2 nm) from the electrode surface. This yields an average distance of $x = 0.5$ cm for bulk to surface-region diffusion. From these parameters, a diffusion rate constant $k_{diff} = 2 \times 10^{-4}$ $s^{-1}$ was calculated for $D_{oxalate} = 1.03 \times 10^{-5}$ $cm^2$ $s^{-1}$[37]. Explicit measurements of the glyoxylate diffusion coefficient in water were approximated as the same order of magnitude of oxalate, at $10^{-5}$ $cm^2$ $s^{-1}$. Previous work suggests that glyoxylate, rather than its hydrate, dihydroxyacetate, as the species interacting with the electrode; an equilibrium process controls their interconversion with a $K = 0.016$[26]. Finally, we expand the buffering equilibria to now include the equilibrium speciation of dissolved phosphate and inorganic carbon (DIC) species, $H_2CO_3$, $HCO_3^-$ and $CO_3^{2-}$ into the simulation, with $[HCO_3^-] = 10$ mM at time zero.

Incorporation of the mass-transport of oxalate flux to the cathode surface results in a simulated growth curve for formate that retains significant exponential character at early times. This is evidenced by the formate subplot in Figure 4, which displays growth-limiting behavior at later times as a result of the depletion of unbound catalyst (Cat) in the glyoxylate fission reaction (Figure 4h, Cat subplot). Mass-transport constraints operate uniquely here, compared to a typical electrochemical cell where the electrode substrate is not under amplification. Here, the electrode substrate, oxalate, is also shown to be under nonlinear amplification (Figure 4, oxalate subplot). As a result, current draw from the system increases in proportion to this amplifying substrate concentration. This is captured by the anodic $O_2$ evolution reaction (Figure 4, $O_2$ subplot) and numerically differentiating oxygen concentration to give the electrolyzer current that peaks at ca. 250 mA $cm^{-2}$ (Figure 4o blue trace).

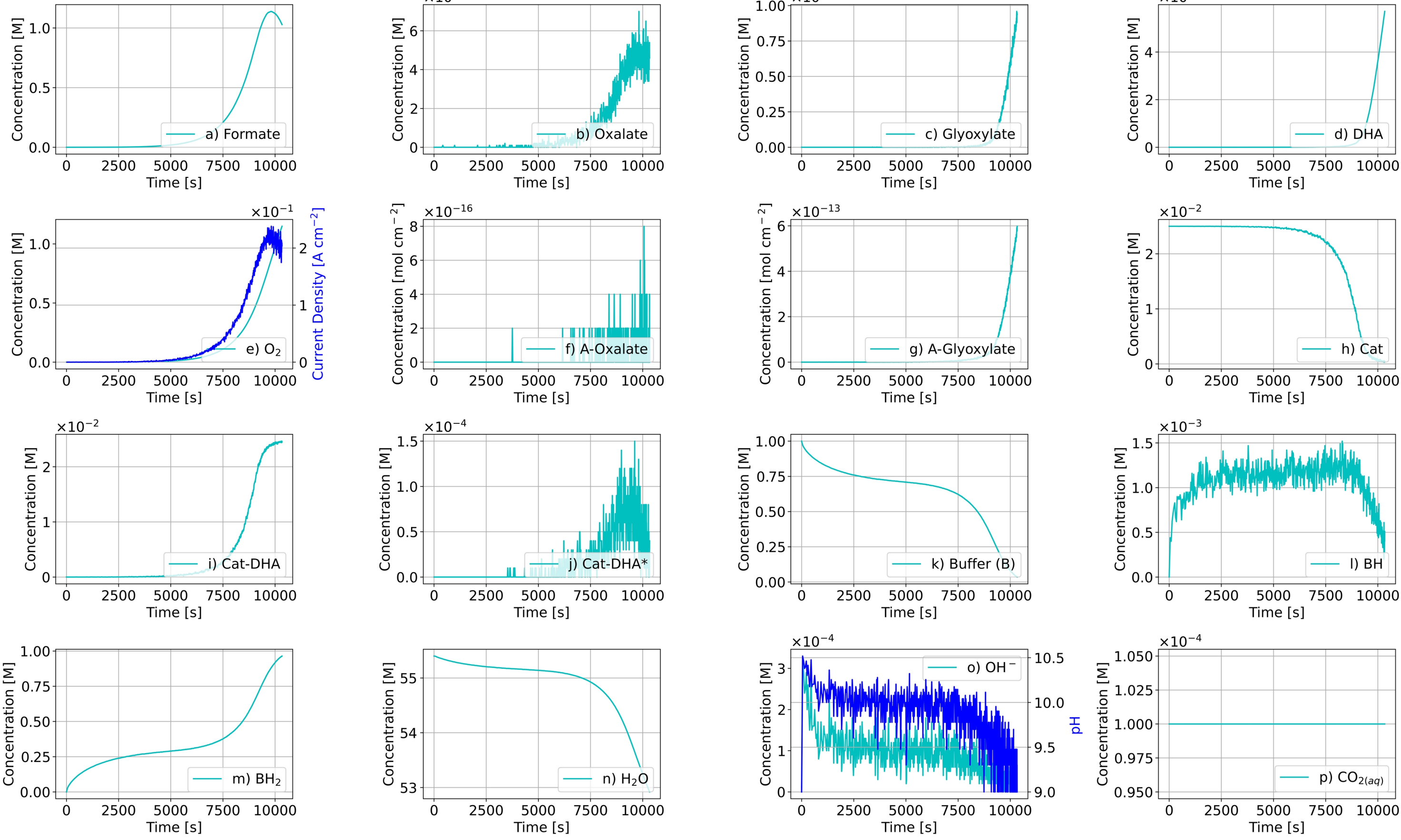

**Figure 4**
Autocatalytic formate replication ($[B]_0 = 100$ μM) coupled to anodic oxygen evolution reaction, shown for the exponential growth time domain. Bulk $[CO_2]$ is held constant at 100 μM (last panel). Incorporation of mechanistic steps for electrochemical oxalate reduction to glyoxylate and catalytic hydrolysis of glyoxylate/dihydroxyacetate (DHA) result in a slower catalytic cycle relative to the simplified base model.

In this work, oxalate reduction is treated as an electron transfer process controlled by a pseudo-first order dependence on oxalate adsorbate surface concentration, as suggested by Eggins[32,38]:

$$r_2(t) = k_{ET}[\text{A-oxa}_{(ads)}](t) - k'_{ET}[\text{Gly-ads}](t) \qquad \text{eq. 4c.}$$

Constants $k_{ET}$ and $k'_{ET}$ denote forward and reverse rate constants for the cathodic conversion of oxalate to glyoxylate using the traditional Butler-Volmer (BV) description for the electrode kinetics (S.3). Values of $k_0 = 0.624$ $s^{-1}$ ($j_0 = 10^{-5}$ A $cm^{-2}$ x 1000x aspect ratio), $\alpha = 0.5$, and $T = 298$ K are used as BV equation parameters. The overpotential (full cell), $\eta$, is taken to be -0.3 V. This gives respective magnitudes of $k_{ET}$ and $k'_{ET}$ of 215 and 0.002 $s^{-1}$ for the cathode's forward and back reactions.

In reaction (3), hydrolysis of dihydroxyacetate to two formate equivalents is decomposed into three mechanistic steps. The reaction is taken to proceed through an equilibrium between bound and unbound substrate at a catalyst (Cat), formation of an activated complex (Cat-DHA*), followed by decay of the activated transition state to products, in accordance with Transition State Theory[39]:

DHA + Cat ↔ [Cat-DHA]    eq. 4d,

[Cat-DHA] + $OH^-$ ↔ [Cat-DHA*]    eq. 4e,

[Cat-DHA*] → Cat + 2 $HCO_2^-$    eq. 4f.

Bracketed terms represent the proposed transition state in the process of DHA/glyoxylate hydrolysis, with the species DHA* representing the symmetric, pre-fission transition state that is proposed to form prior to C-C bond break and decay to two formate equivalents.

DHA*, $C_2$-Symmetric Transition State

**Figure 5**
A putative reaction mechanism for step 3, the base-mediated C-C bond break of glyoxylate (as its hydrate dihydroxyacetate), to form two formate equivalents. Evolution of two molecules of formate arises from C-C bond breakage following deprotonation of the alcohol functionality, yielding a $C_2$-symmetric transition state (DHA*) that may decay into two molecules of formate.

Hydrolysis of DHA to give two molecules of formate represents a less-studied reaction but is again anticipated to be feasible (Figure 5). While this process is expected to be slow under ambient conditions, its possibility is evidenced by the presence of residual formate that can be eventually detected in a solution of 0.1 M glyoxylate in pH 9.0, 1 M bicarbonate incubated at room temperature (S.11). Furthermore, preliminary experimental work has demonstrated the possibility of decomposing glyoyxlate/DHA solutions into formate (3) using the Mn-metallated N-macrocycle, 1,4,8,11-tetraazacyclotetradecane (Mn-cyclam) as a catalyst (S.11). This putative catalyst was inspired by the related, Mn-N(histidine) glyoxylate chemistry of glycolate oxidases[40]. In this study, glyoxylate decay to two formate molecules is proposed as a process first-order with respect to a pre-fission transition state, DHA* (Figure 5). Symmetric C-C bond breakage across the transition state's horizontal mirror plane, $\sigma_h$ (and a coincident $C_2$ axis), bisects the molecule, evolving two formate species according to the rate law:

$$r_3(t) = k_3[DHA^*](t) \quad \text{eq. 5.}$$

This rate law is informed by the likely mechanism for glyxoylate/DHA fission (Figure 5), which decomposes step 3 into a pre-equilibrium between glyoxylate and the DHA* and the first-order decay of DHA* to two formate equivalents. The bar bisecting the proposed glyoxylate fission transition state denotes the plane of mirror symmetry accounting for the 2:1 formate:glyoxylate stoichiometry of this reaction. This mechanistic sequence has not been characterized, making selection of rate constants for simulation arbitrary. As a result, a range of rate constants were tested for each reactions 3.1-3.3 to demonstrate that changing these rate constants over several orders of magnitude still permits formate amplification to proceed (S.7). Base case rate constants for the steps of DHA hydrolysis are described in Table 2.

*Fuel replication kinetics – 'Full Mechanistic System,' non-ideal replicator*

**Table 3 – Simulated 'Full Mechanistic' System**

| Reaction | Forward | Units | Reverse | Units | Ref or SI |
|---|---|---|---|---|---|
| Autocatalytic System | | | | | |
| (1) **$HCO_2^-$** + $CO_2$ + $H_2O$ → $C_2O_4^{2-}$ + $H_3O^+$ | | | | | |
| 1.1 $H_2O$ + hν → $OH^\cdot$ + $H^\cdot$ | 0.00045 | $s^{-1}$ | – | -- | 41 |
| 1.2 $H_2O$ + hν → $e^-_{(aq)}$ + $H_2O^+$ | 0.00045 | $s^{-1}$ | – | -- | 41 |
| 1.3 **$HCO_2^-$** + $H^{\cdot}$ → $CO_2^{\cdot -}$ + $H_2$ | 2.2e8 | $M^{-1}$ $s^{-1}$ | – | -- | 41 |
| 1.4 **$HCO_2^-$** + $OH^{\cdot}$ → $CO_2^{\cdot -}$ + $H_2O$ | 2.5e9 | $M^{-1}$ $s^{-1}$ | – | – | 41 |
| 1.5 **$HCO_2^-$** + $e^-_{(aq)}$ → $CO_2^{\cdot -}$ + $H_2$ + $OH^-$ | 1.0e6 | $M^{-1}$ $s^{-1}$ | – | – | 41 |
| 1.6 $CO_2$ + $OH^{\cdot}$ → $HCO_3^{\cdot}$ | 1.0e6 | $M^{-1}$ $s^{-1}$ | – | – | 41 |
| 1.7 $CO_2$ + $H^{\cdot}$ → $COOH^{\cdot}$ | 8.0e6 | $M^{-1}$ $s^{-1}$ | – | – | 41 |
| 1.8 $COOH^{\cdot}$ + $H_2O$ ↔ $H_3O^+$ + $CO_2^{\cdot -}$ | 3.98e7 | $M^{-1}$ $s^{-1}$ | 1e9 | $M^{-1}$ $s^{-1}$ | 41 |
| 1.9 $CO_2$ + $e^-_{(aq)}$ → $CO_2^{\cdot -}$ | 7.7e9 | $M^{-1}$ $s^{-1}$ | – | – | 41 |
| 1.10 2 $CO_2^{\cdot -}$ → **$C_2O_4^{2-}$** | 1.0e9 | $M^{-1}$ $s^{-1}$ | – | – | 41 |
| ***Reactions 1.11 – 1.148 located in Supporting Information (S.17)* | | | | | 42,43 |
| (2) $C_2O_4^{2-}$ + 2 $H_2O$ → $C_2H_3O_4^-$ + $OH^-$ + 0.5 $O_2$ | | | | | |
| Cathode: $C_2O_4^{2-}$ + 3 $H_2O$ + $2e^-$ → $C_2H_3O_4^-$ + 3 $OH^-$ | | | | | |
| 2.1 **$C_2O_4^{2-}{}_{(aq)}$** + A ↔ A-$C_2O_4^{2-}{}_{(ads)}$ | 2.4 | $s^{-1}$ | 0.8 | $s^{-1}$ | S.2 |
| 2.2 A-$C_2O_4^{2-}{}_{(ads)}$ + 2 $H_2O$ + $2e^-$ ↔ A-$C_2HO_3^-{}_{(ads)}$ + 3 $OH^-$ | 215 | $s^{-1}$ | 0.002 | $s^{-1}$ | S.3 |
| 2.3 A-$C_2HO_3^-{}_{(ads)}$ ↔ $C_2HO_3^-{}_{(aq)}$ + A | 0.8 | $s^{-1}$ | 2.4 | $s^{-1}$ | S.2 |
| 2.4 $C_2HO_3^-{}_{(aq)}$ + $H_2O$ ↔ **$C_2H_3O_4^-{}_{(aq)}$** | 70 | $s^{-1}$ | 1.14 | $s^{-1}$ | 26 |
| Anode: 2 OH- ↔ $H_2O$ + 0.5 $O_2$ + 2e- | 215 | $s^{-1}$ | 0.002 | $s^{-1}$ | S.3 |
| (3) $C_2H_3O_4^-$ + $OH^-$ → 2 $HCO_2^-$ + $H_2O$ | | | | | |
| 3.1 **$C_2H_3O_4^-$** + Cat ↔ Cat-$C_2H_3O_4^-$ | 100 | $M^{-1}$ $s^{-1}$ | 1 | $s^{-1}$ | S.7 |
| 3.2 Cat-$C_2H_3O_4^-$ + $OH^-$ ↔ Cat-DHA* | 1000 | $M^{-1}$ $s^{-1}$ | 1 | $s^{-1}$ | S.7 |
| 3.3 Cat-DHA* → **2** **$HCO_2^-$** + Cat + $H_2O$ | 10 | $s^{-1}$ | -- | -- | S.7 |
| | | | | | |
| Buffering Equilibria | | | | | |
| Phosphate – Refer to reactions and values in Table (2). | | | | | |
| Carbonate – Refer to reactions and values in Table (2). | | | | | |
| | | | | | |
| Solvent Autoionization | | | | | |
| $H_3O^+$ + OH ↔ 2 $H_2O$ | $k_w$ = 1.1e11 | $M^{-1}$ $s^{-1}$ | $k_w'$ = 1.1e-3 | $s^{-1}$ | 22 |

With mechanistic descriptions for reactions 2 and 3, a fully mechanistic description for this autocatalytic system is now explored by decomposing reaction 1 into radiolytic reactions known to perform net formate carboxylation. These are partly detailed in Table 3 and fully described in S.17. The initial step of base-mediated formate carboxylation to yield oxalate represents a thermodynamically challenging reaction, as it requires the C-C coupling of two low-energy reactants, formate and $CO_2$. However, this net reaction has been realized in previous work by Getoff et. al[41], where it was demonstrated that vacuum UV (VUV) photolysis of $CO_2$-saturated formate solutions yield oxalate as the chief carbon product, with a quantum efficiency exceeding 0.6. The proposed mechanism of Getoff's VUV experiments invoked the dimerization of $COO^{\cdot -}$ radical intermediate, formed either through a formate or $CO_2$ precursor, as the rate-limiting step for oxalate synthesis. This would suggest that at long times, equimolar production of $COO^{\cdot -}$ from both formate and $CO_2$ becomes difficult. Given the exponential growth in formate concentration and the similar rate constants for the kinetically-limiting steps for $COO^{\cdot -}$ formation from $CO_2$ and formate, it would be expected that at long times, $COO^{\cdot -}$ intermediates derive almost exclusively from formate, rather than $CO_2$, breaking the exponential scaling of formate accumulation and causing a deviation from exponential replication growth rates. Such a trend is observed in Figure 6.

Radical speciation from the primary products of VUV photolysis in aqueous media forms an incredibly complex chain of reactions, which have been successfully modeled by Pastina et al, and later modified by Severin and coworkers[42,43]. Simulated, steady-state VUV illumination with 124 nm photons for this fully-described autocatalytic system yields a much shorter time base for formate evolution compared to the Basic and Partial simulations. This results from the fast timescale of radical initiation reactions, with formate concentration reaching nearly 1.5 M within roughly 150 seconds (Figure 6a, formate subplot). Formate evolution initiates with an exponential rise that quickly transitions towards limiting, hyperbolic growth within the first 20 seconds of the simulation. Here, formate replication becomes limited by the rate of the radical processes responsible for oxalate evolution, whose concentration reaches a fast plateau within early stages of the reaction. The current density extracted from $O_2$ evolution rapidly stabilizes near 3 A $cm^{-2}$ over this period (Figure 6f). Primary species evolved by VUV illumination of water are hydrogen radicals ($H^{\cdot}$), hydroxide radicals ($OH^{\cdot}$), and solvated electrons, ($e^-_{(aq)}$). These engage in secondary reactions that ultimately yield the $CO_2^{\cdot -}$ intermediate from either $CO_2$ or formate (Table 3, reactions 1.3-1.5, 1.9; S.17)[41,44,45]. The $CO_2^{\cdot}$ intermediate then dimerizes to yield the oxalate intermediate[41,44,45] (Table 3, reaction 1.10) fed into the reaction for oxalate reduction. Incorporating these radical mechanisms into the overall scheme for formate amplification also result in the co-evolution of molecular hydrogen, $H_2$. As with formate and other cycle intermediates, the kinetics of $H_2$ evolution prove to be highly sensitive to rate constants driving these radical reactions, particularly the rate of $OH^{\cdot}$ and $H^{\cdot}$ radical initiation upon VUV irradiation. $H_2$ evolution is observed to follow a weakly exponential trend at short timescales before transitioning to linear growth. Radical initiation under these conditions also results in the rapid accumulation of base (Figure 6p, $OH^-$ subplot). Transient VUV photoillumination was also explored, revealing a sharp, sigmoidal trend for formate evolution that is diagnostic of self-limiting autocataysis, plateauing at ca. 0.35 M formate (Figure 7). Concentrations of the key intermediates (oxalate, glyoxylate/dihydroxyacetic acid and their adsorbates) each exhibit sharp increases before rapidly decaying to zero. Significant base production and negligible proton evolution is observed upon integration of these radical mechanisms for reaction step 1; as a result, there is virtually no consumption of the diprotic alkaline buffer used in these simulations, in contrast to the Basic and Partial simulation results. This is consistent with the large quantities of molecular hydrogen evolved through the radical chemistry. Here, fast sequences of proton-dependent, $H^{\cdot}$ evolution and $H^{\cdot}$-$H^{\cdot}$ annihilation yield $H_2$ provide a sink for protons that would otherwise consume buffer, as observed in the Partial Mechanistic System (Figure 6, $H_2$ subplot). The time-dependent energy efficiency for the Full Mechanistic System is calculated according to:

$$\eta_{real} = \frac{v_{rxn} \sum_i B_{0,i}(t) \Delta G_{c,i}}{\int_0^t P_{VUV}(t) + i(t) \epsilon dt} \quad \text{eq. 6.}$$

Term $B_{0,i}(t)$ gives the time-dependent concentration of the $i^{th}$ fuel component (here, formate, oxalate, glyoxylate/dihydroxyacetic acid and $H_2$ represent the main fuel components in the reaction volume with significant embodied energy content), $v_{rxn}$ is the reaction volume, and $\Delta G_{c,i}$ is the free energy of combustion of the $i^{th}$ fuel component, with values of -200.27, -273.63, -477.41, -624.43 and 237 kJ $mol^{-1}$ for formate, oxalate, glyoxylate, dihydroxyacetate and hydrogen, respectively (S.3). $P_{vuv}$ is the illumination power of the VUV source, $i(t)$ gives the electrolyzer current, and $\epsilon$ is the electrolyzer applied potential. Application of eq. 6 results in very low overall energy efficiencies for the VUV-coupled processes, peaking at 21% and 16% for steady-state and transient illumination cases, respectively. Losses here are dominated by the requirement for using high energy, 124 nm photons to drive water photolysis for radical generation at a high power density (20.2 W $cm^{-2}$, corresponding an effective 10 M photon concentration at steady-state illumination and a rate constant of photon flux $k_{rad} = 0.0045\ s^{-1}$; S.9). These moderate efficiencies highlight a significant opportunity for optimizing this cycle, provided alternative routes for formate carboxylation that do not require high-energy, H-O-H bond homolysis, are realized.

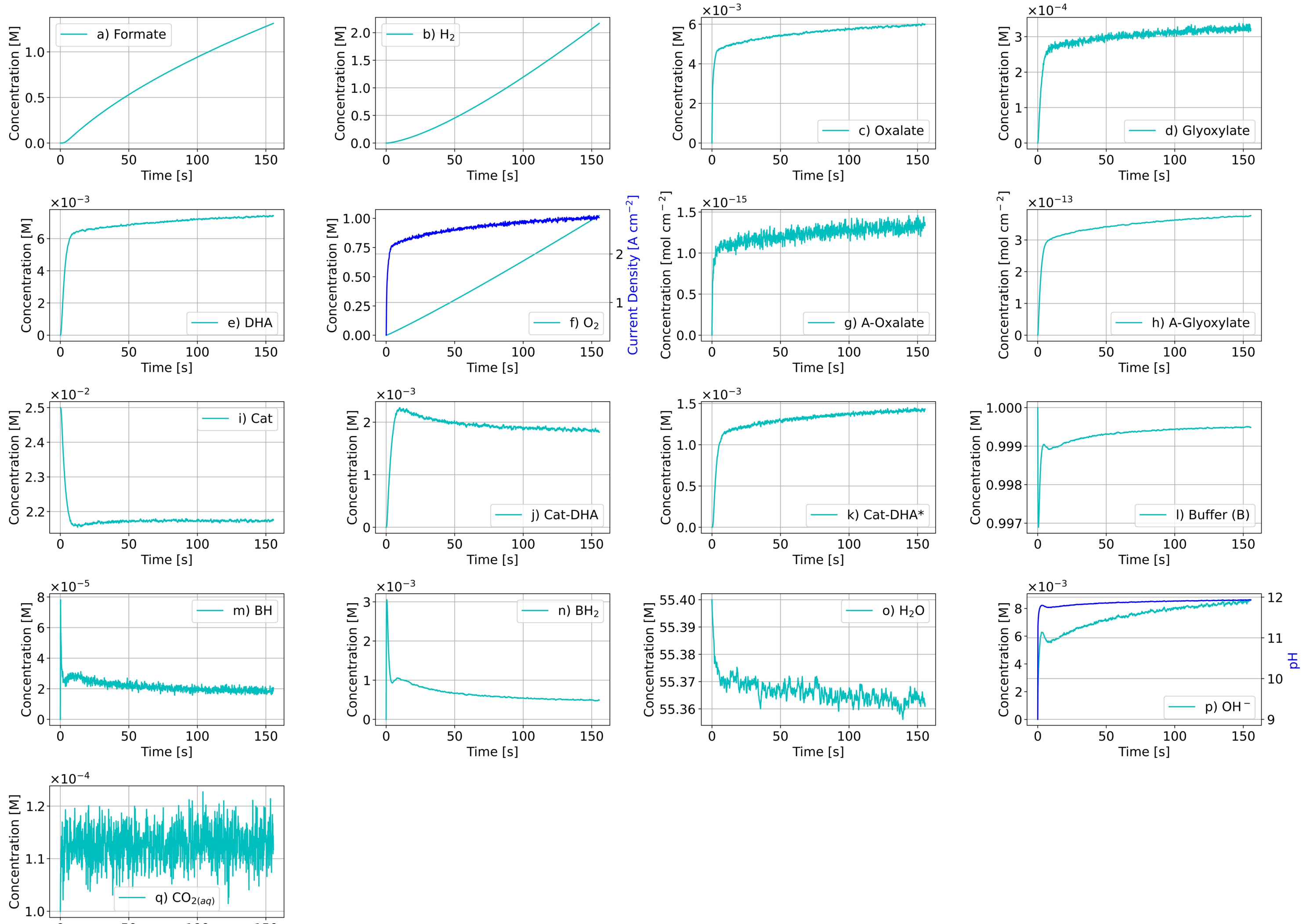

**Figure 6:** Behavior of the system with a mechanistic description for all steps, coupled to anodic OER. $B_0 = 100$ μM; $[CO_2]$ is held constant at 100 μM. The simulation now incorporates radical-initiated, VUV-pumped mechanisms for formate carboxylation previously reported to drive the net conversion of formate to oxalate via formate carboxylation (step 1). VUV irradiation is at steady-state, with water consumption in the simulation mitigated through the incorporation of an $H_2O$ reservoir feeding the reactor.

In addition, more extensive studies on $CO_2$ radical mechanisms conducted by Horne et al[46]. incorporate the full array of potential reactions under ionizing radiation originating from nuclear decays. While the emitted radiation has energies orders of magnitude greater than VUV, their primary radiolysis products in $CO_2$ saturated media are similar to those considered in the VUV cases.

A natural extension of the similarity in primary products evolved using VUV light vs nuclear ionizing radiation is possibility of using ionizing radiochemical emissions as inputs for driving formate carboxylation in the autocatalytic cycle. The reports by Severin[43] and Pastina[42] demonstrate that the radical OH · , H · and solvated electron species invoked by Getoff et al. as the key initiators of formate carboxylation in $CO_2$-sparged aqueous solutions, are also produced through the exposure of water to various nuclear decay sources, including proton and gamma ray emitters. However, autocatalysis here requires that the relative ratios of primary radical species produced by nuclear emissions (G values) fall within a band of radical product ratios similar to those measured by Getoff for VUV irradiation. However, substituting VUV irradiation in the simulation with known G values for primary radical generation in water by either 2 and 10 MeV protons or 10 MeV gamma ray irradiation do not result in nonlinear formate evolution (S.18). The failure is due, in part, to the rapid production of protons generated by irradiation with these energies, leading to a rapid consumption of the alkalinity necessary for the cycle. Furthermore, only a linear component is observed for $H_2$ yields in each of these simulations (S.18). This suggests that using nuclear decay emissions to mediate formate carboxylation would need to be tuned appropriately, such that their G values for radical production approximate those of VUV radiation.

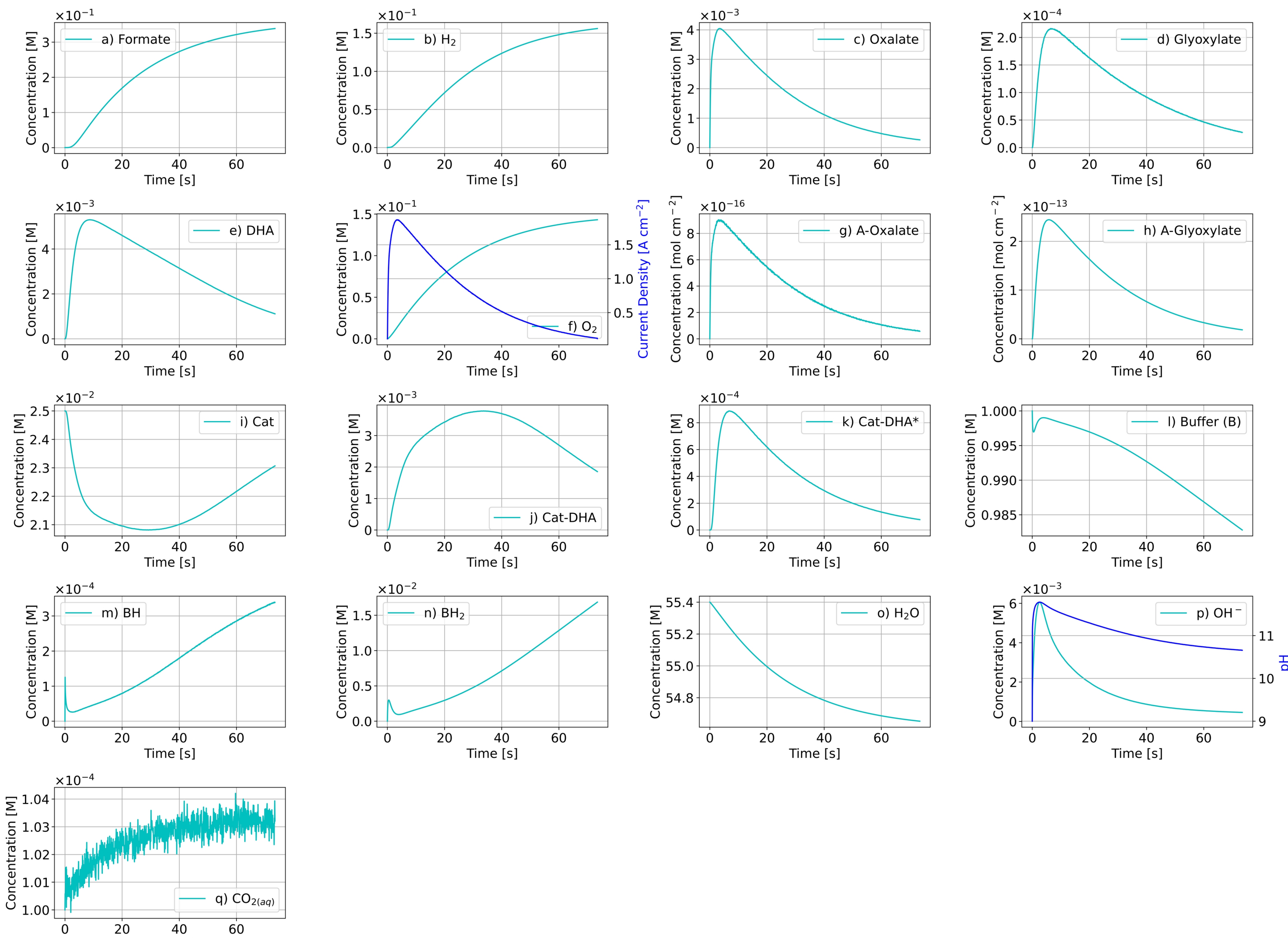

**Figure 7** Behavior of the system with a mechanistic description for all steps, coupled to anodic OER, under transient VUV illumination (124 nm). $B_0$ = 100 μM; [$CO_2$] is held constant at 100 μM. The simulation now incorporates radical-initiated, VUV-pumped mechanisms for formate carboxylation previously reported to drive the net conversion of formate to oxalate via formate carboxylation.

*Fuel replication thermodynamics*

Replication energetics of the net cycle are thermodynamically uphill, requiring energetic inputs to push reaction equilibria in the direction of continuous formate evolution. Steps 1 and 2 are individually endergonic, with reported free energies of +43 kJ $mol^{-1}$ for formate carboxylation[24] and +246 kJ $mol^{-1}$ for the two-electron reduction of oxalate to glyoxylate coupled to water oxidation, for an overall reaction $\Delta E^0$ = 1.275 V ($E^0$ = -0.475 V for the oxalate/glyoxylate half reaction)[24]. Hydrolytic fission of glyoxylate (via DHA) into two formate molecules (step 3) has a reported free energy change of -55 kJ $mol^{-1}$ [24].

Notably, the requirement of an exothermic bond-breakage for completing the cycle is consistent with the heat-dissipative nature of replicating systems[47]. The energetics of a single reaction cycle represents the sum of these serial reaction steps, yielding a total free energy change of $\Delta G^o$ = +234 kJ mol$^{-1}$ per cycle iteration (Figure 8).

The proportionality between energy consumed by product synthesis ($U_{tot}$) and total product yield mandates that the rate of energy consumption also scale exponentially. As a result, in the limit of long times ($t >> \tau$), a fuel replicator will require an energy-dense source ($P$) satisfying the condition:

$$\frac{dU_{tot}}{dt} < P \qquad \text{eq. 7a,}$$

to maintain exponential growth. Absent other depletion conditions (such as supply or mass-transport limited flux of key reactants or intermediates), rates will converge to a constant value limited by the source power density. Preferential kinetics relative to steady-state operation occurs for the general condition where the rate of autocatalysis exceeds that of linear catalysis:

$$k_{linear}\prod_j [A_j]^{m_j} < B_0(1+\Xi(t))^{kt}[kln(1+\Xi(t)) + \frac{kt}{1+\Xi(t)}\frac{d\Xi}{dt}] \qquad \text{eq. 7b.}$$

Here, $k_{linear}$ is the overall rate constant of a comparable linear process; reaction rate is proportional to the steady-state concentrations of reactants $A_j$, of some arbitrary reaction order $m_j$. (i.e., electrochemical reduction of $CO_2$ to formate in 1:1 $CO_2$: formate stoichiometry). Rate advantage for the autocatalytic approach occurs where the conditionality of 7b is met. Graphical representation of this inequality is captured by Figure 9a.

A clear challenge exists for how to pair such a process for fuel production with a suitable energy source, as most power sources − such as sunlight − are only available at steady-state. Kinetic advantage can therefore only occur relative to some linear system, for cases where the time-averaged current density (or equivalently, time-averaged growth rate) of an autocatalytic system exceeds that of some steady-state linear case (eq. 7b). As a result, for all periods where autocatalysis operates below the rate of the power source, energy availability goes underutilized. At long times, for an optimized replication process only limited by energy availability (as opposed to reactants), the rate of fuel growth and energy consumed by autocatalysis will approach the limit allowed by the source's maximum power density:

$$U_{tot} = v_{rxn}\sum_i \Delta G_{c,i} B_{0,i}\left(1+\Xi_i(t)\right)^{k_i t} < \int_0^t P(t)dt \qquad \text{eq. 7c,}$$

where the right-hand side (RHS) of the inequality in eq. 7c is the source power density. The summation of the RHS captures the effect of integrating autocatalytic energy consumption over time for all fuel species $i$ generated by the process.

Upper limits on achievable thermodynamic efficiency for autocatalytic replication can be estimated through a previously established framework for determining the lower bound of heat dissipation in a replicating system described by England[47]. According to this approach, any well-defined system of replication and its surrounding bath will be characterized by heat transfer that is a function of the replicator's durability, defined as the inverse of its decay constant ($1/\delta$). Heat dissipation by the replicator will also be a function of the replicator's internal entropy ($\Delta S_{int}$), and growth rate constant($g_{max}$)[47]. Critically, replicator decay in the case of formate is given by the decomposition of formate into the products of CO and $H_2O$[48], not the recombination of two formate molecules into DHA, a very low probability occurrence due to its vanishingly small equilibrium constant (~2.3x10$^{-10}$ at 298 K). This distinction is mentioned here, as recent work has brought into question the general applicability of England's framework beyond the cases where: 1) second-order reversion of a replicator into its starting materials is much slower than first-order replicator decay into some waste product(s) (formate decays into CO and $H_2$) and 2) there is no relationship between the rates of the two processes[49]. However, in the case examined by this study, both these conditions apply.

The internal entropy of a formate replicator may be calculated according to the Gibbs relation:

$$\Delta S_{\text{int}} = \frac{\Delta H - \Delta G}{T}$$

eq. 7d,

where $\Delta H$ is the heat of formation of formic acid at (-425 kJ mol$^{-1}$ at 25 $^{o}$C), $\Delta G$ is formic acid's free energy of formation (-361 kJ mol$^{-1}$ at 25 $^{o}$C), and $T_{sys}$ is the system temperature, 298 K (S.3). Using these values with eq. 7d gives a value for $\Delta S_{int}$ -0.2 kJ mol$^{-1}$ K$^{-1}$ for a formate replicator at room temperature. Formate durability, $1/\delta$, may be estimated by considering known rates of spontaneous formate decomposition (to either $CO_2$ and hydrogen or carbon monoxide and water) under conditions relevant to those in the proposed autocatalytic cycle. Barham and coworkers have measured the Arrhenius behavior of formic acid decay constants, reporting a first-order decay constant of 700 s$^{-1}$ for decomposition of formic acid into CO and $H_2O$ at 40 $^{o}$C[48]. Using these data, a decay constant $\delta$ = 8.3x10$^{-11}$ s$^{-1}$ at 25 $^{o}$C could be extrapolated (S.10).

**Figure 8**

Thermodynamics of the proposed autocatalytic cycle. Conversion of oxalate to glyoxylate (in equilibrium with its hydrate, dihydroxyacetate, DHA) is shown using alkaline water oxidation as the complementary anodic half reaction. The process results in the net consumption of one water equivalent per cycle turnover.

Applying these values for $\Delta S_{int}$ and $\delta$ to eq. 7e provides solution for $q$, the minimum waste heat transferred from the replicating system to the bath[47], for a given value of $g_{max}$:

$$q \geq k_B T \ln\left(\frac{g_{max}}{\delta}\right) - T \Delta S_{\text{int}}$$

eq. 7e,

Equation 7e estimates for the upper bound on thermodynamic efficiency and lower bound on generated waste heat ( $q$) as a function of $g_{max}$ for a formate replicator, are shown in Figure 9. Critically, the calculated values for replication-dependent heat dissipation based on England's proof[47] assume values with a minimum bound of ca. 89 kJ mol$^{-1}$, within range of the exergonic step of glyoxylate/DHA bond fission (-55 kJ mol$^{-1}$). As replication rate constants increase, dissipative losses increase logarithmically. As a result, the upper bound on energy efficiency for this process, as determined from the free energies of reaction for individual steps of the cycle shown in Figure 8, point to a thermodynamic limit of $\eta$ = 0.81 per cycle (289 kJ mol$^{-1}$ total input per cycle; 234 kJ mol$^{-1}$ maximum available for conversion into formate). Calculating heat dissipation (eq. 7e) allows for approximating time-dependent efficiency (eqs. 7f, 7g) for formate replication. Using the definitions in eqs. 7c and 7e, the energy efficiency for desired products in an autocatalytic fuel cycle may be expressed as:

$$\eta = \frac{U(t)_{tot}}{\int P(t)\,dt}$$

eq. 7f,

with an upper efficiency bound of:

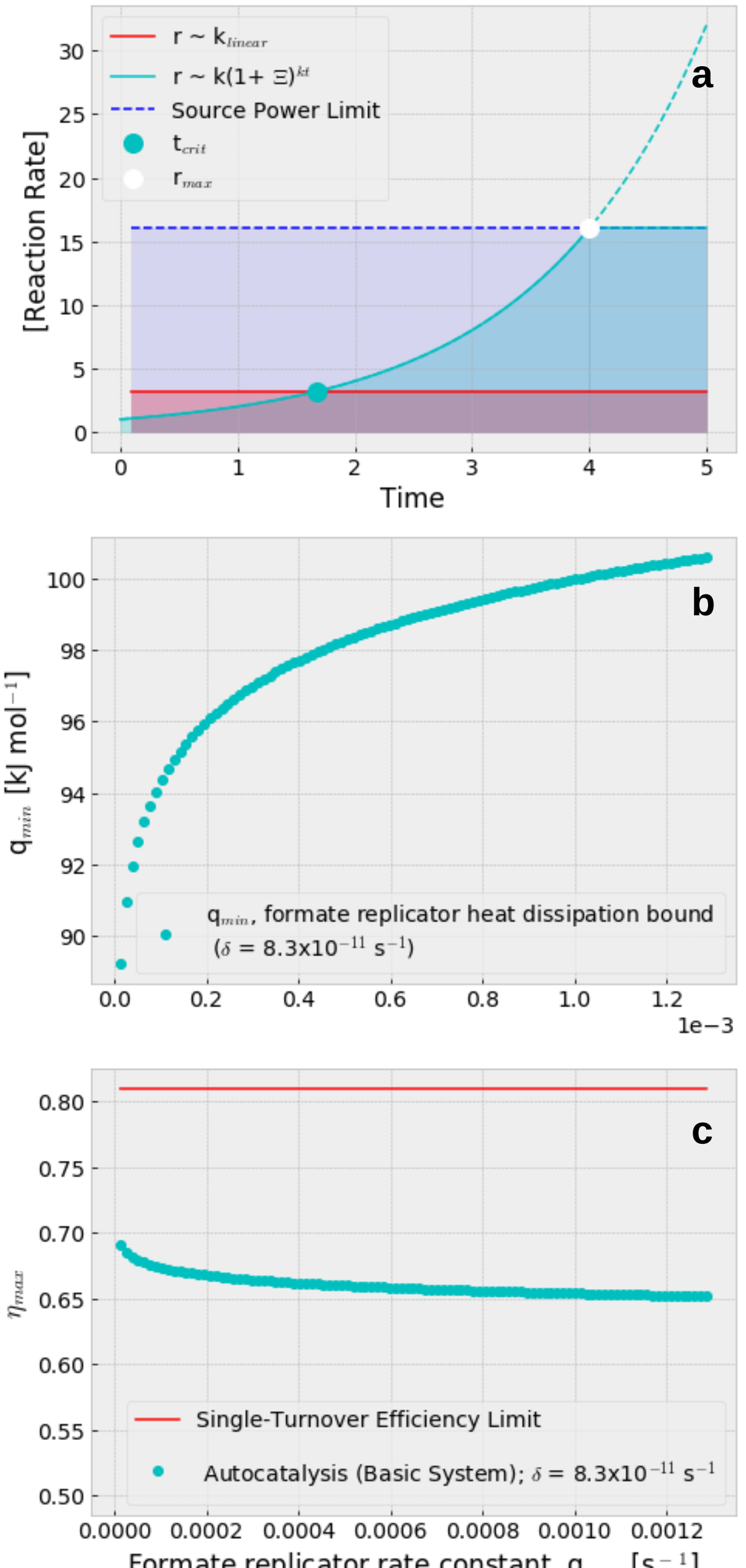

**Figure 9** **(a)** A graphical representation of eqs. 7b & 7c. Rates of product evolution through autocatalysis ($\Xi = 1$) becomes preferential relative to linear catalysis for $t > t_{crit}$, with an intersection (cyan dot) described by equation 7b. **(b)** Lower bounds of replicator heat dissipation as a function of replicator rate constant, $g_{max}$. **(c)** Energy efficiency ($\eta$) scaling of $g_{max}$ for formate replication (Basic System, cyan); T = 298 K. The thermodynamic limit for a single turnover (red) is calculated assuming a perfect cycle where energy losses only occur at the exothermic step of DHA bond fission to yield 2 formate equivalents (minimum input of 289 kJ mol$^{-1}$; maximum possible yield of 234 kJ mol$^{-1}$ as formate; $\eta_{max} = 0.81$).

$$\eta = 1 - \frac{q}{\int P(t)\,dt}$$

eq. 7g.

Using the replication rate constant extracted from formate growth in the 'Basic' system ($g_{max} = 0.0013$ s$^{-1}$) at 298 K, eq. 7g yields an upper-bound efficiency estimate of 65% (Figure 9c; S.9).

## Discussion

*Practical implications*

There are three key advantages offered by the possibility of autocatalytic energy storage. First is the clear opportunity for building very fast systems for fuel generation, with rates that can scale exponentially with time. The doubling of product yield per cycle (in the ideal limit) for a $2^n$ autocatalytic system also means that fast overall rates of catalysis are possible even for systems featuring very slow catalytic rate constants. In most linear systems, this is difficult to achieve as their reaction rates only scale linearly with rate constants and substrate concentrations are typically at steady-state.

Second, autocatalytic systems will see their rate of fuel production (and so power consumption) increase until a plateau is reached at the power limit dictated by the source power density. That is, an optimized system exhibits adaptive behavior, with a peak rate that will continue increasing with time, so long as there is sufficient rate of energy input and substrate availability, to accommodate replicator growth. This property has the potential to be incredibly beneficial to electrolyzer performance, as these devices tend to have peak currents that are governed by mass-transport-limited substrate flux and availability. Contrary to steady-state systems, replicating fuel cycles actually help mitigate such mass-transport limitations, as the concentrations of the primary reactant amplify with time. In the case of a typical electrolyzer, if the peak current of the system is far below that of the power source driving it, total power utilization only becomes possible with a sufficient multiple of electrolyzer units, with a sum power density of equaling *P*. This opens the possibility that for some arbitrarily large power source, two systems with comparable rate constants, one linear, one autocatalytic, it is possible to utilize the power source with a single autocatalytic electrolyzer, whereas multiple linear electrolyzers are required to utilize the source power density. Of course, this also means such a system could only run as a repeat series of transient growth cycles followed by reactant collection to restore the system to its initial state, in preparation for another growth cycle.

Third, since the substrate and product are the same molecule, and an autocatalytic system could be operated in batch (with the exception of a steady-state, $CO_2$ input), the system acts as a concentrator for the target product. That is, it incorporates the functions of a fuel generator (such as an electrolyzer) and concentrator (a solvent still, etc.) into a single unit. The long-studied dependence between "fitness" of biological and chemical replicators and the ability of one replicator to dominate in number at long times, implies a potential role of autocatalysis in designing integrated systems for concentrating products as they are being generated, while minimizing side products (competing, "unfit" replicators)[10,20,47,50,51]. This fact is also evidenced by findings here, with formate becoming a dominant product at long times relative to oxalate and glyoxylate/DHA intermediates, despite the fact that all four exhibit exponential growth behavior during the simulation time.

The decision to design an autocatalytic scheme around formate production was motivated by a number of critical factors. First, as the simplest liquid $CO_2$ reduction product, formate requires the least amount of chain buildup prior to the bond breakage event responsible for formate doubling. The reduced number of intermediate steps required for transforming formate precursors into $C_2$ compounds relative to other liquid fuel candidate schemes (S.12-S.14) means reduced opportunities for efficiency losses and the generation of side products that could interfere with the primary amplification reaction in a real system. Furthermore, formate/formic acid is a candidate hydrogen carrier, featuring a volumetric energy density similar to that of liquid hydrogen compressed to 700 atmospheres, a storage benchmark relevant to hydrogen deployment in automotive applications[17]. The mature state of hydrogen fuel cell technology and the advancements made in catalytic dehydrogenation of formic acid to $H_2$[52,53], suggest formate's potential as a carbon-neutral fuel or hydrogen storage reservoir.

The strategy used to conceive this fuel cycle demonstrates how autocatalytic energy storage schemes may be rationally constructed by integrating existing, linear electrochemistries. Since this process is partly dependent on electrochemical reduction, this step of the fuel cycle may be realized by coupling it to electrochemical oxygen evolution reactions (OER). The possibility of electrocatalytic oxalate reduction to glyoxylate over Pb cathodes was reported by Tafel et al[34,54], along with more recent reports of GaSnOx electrocatalysts capable of reducing oxalate to glyoxylate with over 97% efficiency[36], indicating that reaction (2) is experimentally tractable.

Aside from its more fundamental implications, such nonlinear approaches may reduce barriers to industrial-scale production of fuel or commodity chemicals. Where fuel generation moves as an exponential function of time, such chemistries could potentially relax the capital requirements on plant size and process operation time needed to achieve profitable economies of scale. Finally, combining this autocatalytic approach with preexisting ionizing radiation sources to drive formate carboxylation suggest a possible use case for nuclear wastes in energy storage applications. However, practical application would have to address the significant energy efficiency losses that arise from primary radical generation, and whether the G values for radiation sources could be tuned appropriately.

*Generalizing PCR for chemical energy storage*

Results here validate the underlying premise that general steps of primer annealing, nucleotide polymerization and ligation, and final dimer denaturation in PCR are translatable to a broader set of chemical reactions. It's observed that the chief constraints such schemes are defined by three broad requirements: (1) that the precursor/target can be built up into a molecular intermediate or transition state that represents a dimer of the target product, (2) that the process of chain buildup include a balancing of redox equivalents, where the terminal hydrocarbon prior to some binary fission event (here, DHA* C-C bond breakage) has 2n reducing equivalents relative to the template/product oxidation state, and (3) this approximate dimer species can be decomposed to yield the target product in a 1:2 stoichiometry.

The requirement of chain buildup prior to bond fission is reflected in prior work detailing the replication mechanisms of molecular fiber structures[55,] chimeric nucleobase/amino acid assemblies[56] in origin-of-life research, and definitions of minimal self-replicating systems detailed by Paul et al[57]. Notably, constraints (1) and (2) are implicit statements about the molecular symmetries required of the pre-fission intermediate. Such symmetry requirements are reminiscent of DNA, where the two-fold (anti)symmetry of the double helix was observed by Watson and Crick to suggest the very mechanism by which it could be replicated *in vivo*[58]. These three fundamental constraints of the proposed fuel cycle are relatively slack, suggesting a large theoretical space of target molecules that could be produced via PCR-type schemes of fission-dependent autocatalysis. To emphasize the generality of this approach, other conceivable autocatalytic reactions satisfying these requirements and following the PCR-type framework used here, were also outlined (S.12-S.14). However, reducing such processes to practice will be highly dependent on the nature of the individual reaction cycles under consideration. The mechanistic challenges of realizing formate amplification indicate that engineering such reaction cycles for even marginally more complex molecular targets would be non-trivial.

The foundational logic behind $2^n$ product amplification should be extendable to the higher-order cases of ternary, quaternary, etc. fission, where even greater rates of product formation, scaling respectively as $m^n$ (for m = 3, 4, etc.) would occur in the limits of perfect cycle selectivities. However, developing autocatalytic cycles with large exponent bases would carry even greater requirements for catalytic selectivity and the extent of chain buildup into pre-fission intermediates of either m-fold symmetry or m repeating units. Work here suggests that highest priority should be placed on optimization of reaction selectivity over rates, as even minority side reactions can result in the rapid accumulation of undesirable products, stalling the amplification process.

In addition to autocatalytic cycling proceeding continuously – the general case explored by this manuscript – running the reaction as a sequence of discretized steps (as done in PCR), where the $n^{th}+1$ reaction step does not initiate until the completion of the nth step, represents a secondary configuration by which autocatalytic substrate production may be achieved. This type of discrete implementation would have the benefit of enabling separate control each individual reaction, minimize the possibility of chemical cross-talk. For discrete cycling, the time for one cycle for based on stoichiometric rate laws in the Basic System may be approximated as:

$$\tau = \frac{1}{k_{cycle}} = \lim_{A_1 \to 0} \frac{1}{k_1}\left(\frac{1}{A_1} - \frac{1}{A_{1,0}}\right) + \sum_{2}^{j} \lim_{A_j \to 0} \frac{1}{k_j} \ln\left(\frac{A_{j,0}}{A_j}\right) \qquad \text{eq. 8.}$$

This assumes reactions that can be treated as sequential chemical transients with respect to substrate concentrations, $A_j$. The first limit expresses the time for transient substrate consumption for a $2^{nd}$ order process, the case for the initial formate carboxylation reaction. The summation term generalizes the time for steps two and three to occur as a sequence, with the transient reaction times for each being represented by first-order decay kinetics. Here, terms $k_1$ and $k_j$ give the respective $2^{nd}$ and $1^{st}$ order rate constants describing the relevant reaction step. Terms $A_{1,0}$ and $A_{j,0}$ denote the initial concentration of substrate $A$ at the start of the $j^{th}$ reaction, while terms $A_1$, $A_j$ give the concentration of substrate at time $t$. Evaluating equations 1b and 1c using the calculated $\tau$ for the discontinuous case allows for approximating formate growth in the limit of discrete cycling.

*Technical barriers to implementation*

Despite its conceptual simplicity and the standing precedent set by PCR, significant technical barriers lie in demonstrating the autocatalytic formate cycle described here. One of the key intermediates, glyoxylate, may also undergo decomposition through UV photochemistry[59], although this has been reported at timescales far longer than the reaction cycle times considered here. These challenges are compounded by the difficulty of facilitating reaction (1), the carboxylation of formate. These low-energy compounds are generally difficult to chemically activate; for this reaction, no synthetic route beyond radical chemistries is known to the author. While Getoff et al. has demonstrated net photochemical formate carboxylation, with high quantum efficiencies in the vacuum ultraviolet regime, the absence of any economical source of VUV radiation makes such a strategy impractical[41]. Furthermore, these calculations suggest the high energy VUV photons used for photolysis would cause significant efficiency losses. Achieving autocatalysis with high energy efficiency will require alternatives to photolysis. Radical production by voltammetric methods, such as electrochemical Fenton reactions for $OH^{\cdot}$ generation or using ionizing radiation from waste radionucleides, could provide avenues for facilitating reaction (1) with fewer energetic costs[60]. It is critical to note that this simulation assumes many chemical intermediates and processes do not cross-interact but in practice, unknown chemical side products or reactions could significantly change system behavior. Finally, while not explored here, past work in the field of organometallic $CO_2$ insertion into aldehyde bonds has been demonstrated for higher aldehydes[61]. If such reactivity can be shown to be extendable to C-H insertion in formate, it would provide a direct route for realizing formate carboxylation to oxalate without the complicated speciation and reactivity of radical-driven chemistry.

Development of a selective catalyst for glyoxylate/DHA decomposition into two formate equivalents will be essential. To the author's knowledge, there is currently no underlying reason for investigating this reaction in the first place; such rationale are now provided by this paper. Ongoing experimental work will act as a follow-on to this theoretical manuscript to characterize the kinetics and selectivity of a Mn-cyclam complex for catalyzing reaction (3). This future work will also resolve the exact stoichiometry of formate evolved with respect to hydrolyzed glyoxylate. It is also possible that, as with the Mn-bearing glyoxylate oxidase, Mn-cyclam catalyzes reactions in which glyoyxlate is decomposed into 1 equivalent of $CO_2$ and formate through Mn redox-disproportionation, rather than two formate equivalents.

Notably, the importance of selective constituent reactions is reminiscent of the role played by DNA polymerase III in PCR chemistry. DNA polymerase III catalyzes rapid nucleotide incorporation ($k_{cat}$ ~ 1000 $s^{-1}$) into duplicate strands with an error rate of order 1 in 1 million (excluding the additional fidelity provided by DNAP's exonuclease activity)[62,63]. This high specificity is essential for DNA polymerase III; without it, the exponential nature of DNA replication would lead to the fast accumulation of random errors (mutations) within genetic material, impacting biochemical functions in living organisms. Indeed, the exploitation of low-fidelity DNA polymerases in error-prone PCR are used for the rapid generation of random mutant libraries *in vitro*. Despite the error rates of promiscuous DNA polymerases being on the order of just tenths of a percent, the incorporation rate of nucleotide mutations in error-prone PCR demonstrate how even minor side reactions may cause rapid side product accumulation in real autocatalytic systems[64].

**Conclusions**

The extension of nonlinear chemical kinetics to reaction schemes for sustainable fuel production carries a largely unexplored potential to reshape conceptual approaches to the design of renewable energy systems. In particular, chemical processes exploiting the essential reaction characteristics of PCR – particularly the form of autocatalytic cycling that incorporates a symmetric fission step – may reorient our design focus. Such processes have the attributes necessary for enabling exponential yields of fuel as a function of time, growth relations that dwarf the conventional, linear schemes characterizing the vast majority of work done in the field of renewable chemicals and fuel production. Furthermore, the exponential growth of cycle intermediates open the possibility of generating significant quantities of the more reduced/longer-chain compounds that are generated prior to molecular bond fission, provided that the relative rates of the constituent reactions can be suitably controlled.

Practical application of autocatalytic fuel synthesis will require surmounting several technical challenges. Highly selective catalysts will be needed to drive the individual reaction steps comprising such chemical cycles, which remain a broad challenge in chemistry. While these simulations suggest reactions (2) and (3) demand less selective catalysts than reaction (1) for the entire cycle to proceed, catalyzing efficient oxalate reduction and glyoxylate hydrolysis also mark non-trivial challenges. However, resolving these problems would pave a path towards rethinking chemical energy storage – and chemical reactions more generally – in terms of nonlinear processes and their potential benefits relative to traditional, linear reaction schemes.

**Materials and Methods**

Materials and Methods can be found in the Supporting Information.

**Supporting Information**

Computational methodology, simulation parameterization, kinetic schemes for Basic, Partial and Full systems, calculations of reaction selectivity, influence of rate constants, and reactivity under batch $CO_2$, Arrhenius extrapolation of formic acid decay constants, characterization of a glyoxylate (DHA) hydrolysis catalyst, examples of potential autocatalytic schemes for the production of methanol, dimethyl ether, carbon monoxide, and calculations for radiolysis-mediated formate carboxylation can be found in the Supporting Information.

**Data Availability**

Data involving the setup of the simulations used in this manuscript are provided in the Supporting Information. Kinetiscope simulation packages and python scripts used for data workup will be made available at https://github.com/MEG-LBNL.

**Acknowledgments**

This research was supported by the Early-Career Laboratory Directed Research and Development (LDRD) award through Lawrence Berkeley National Laboratory, funding through an LDRD from the LBNL Carbon Negative Initiative, and by the Clean Energy Manufacturing Program, U.S. Department of Energy, Office of Science, Office of Basic Energy Sciences, CSGB Division under contract no. DE-AC0205CH11231. The Kinetiscope simulation environment (http://hinsberg.net/kinetiscope/index.html) was used for all kinetics simulations presented in this work. The author thanks Joseph Varghese, David Larson, Frances Houle, and Walter Drisdell for their helpful comments.

**ORCID**

Peter Agbo: 0000-0003-3066-4791

**Correspondence**

PAgbo@lbl.gov

**Competing Interests**

This work relates to a Lawrence Berkeley National Laboratory invention disclosure by P.A.

**Dedication**

This work is dedicated to Sophia Agbo and in memory of Samuel Agbo and Lovette Chiezey.

# Self-replicating fuels via autocatalytic molecular bond fission

Peter Agbo[1,2]*

Chemical Sciences Division[1], Molecular Biophysics & Integrated Bioimaging Division[2] Lawrence Berkeley National Laboratory, Berkeley CA 94720

*Corresponding author
PAgbo@LBL.gov

**Supporting Information**

## S.1 Computational & Quantitative Methods

Microkinetics simulations of autocatalytic formate synthesis were executed using the Kinetiscope simulation environment (https://hinsberg.net/kinetiscope/). As a stochastic kinetics simulation engine, Kinetiscope discretizes the problem of microkinetics simulation using the formalism pioneered by Gillespie[1], explicitly tracking of the time evolution of all molecules in a reaction volume using Monte Carlo methods. 0-D simulations were created using the single compartment configuration of Kinetiscope, using the chemical schemes described in S.15-17. All simulations were run on an HP G4 Z4 workstation with 80 gigabytes of RAM and an Intel Xeon Processor. Parameterization details of kinetic and thermodynamic inputs for the Basic, Partial and Full mechanistic models are elaborated upon in sections S.2-S.3.

In this simulations, species held at steady-state - $CO_2$, photons (h$\nu$) and radiochemical particle emissions) are represented by kinetiscope reaction inputs of the form:

A + B => C + B,

for reactions where B is a species held at steady-state.

Reactions are initiated such that each species with a non-zero concentration at t=0 is represented by at least 1 particle in the simulation compartment.

For the Basic (toy) simulation, the system is treated as a well-mixed In the case of the redox processes for oxalate reduction and oxygen evolution, these are not explicitly electrochemical reactions in the Basic simulation. Therefore, they do not scale with an electrode area (which is unspecified.) As a result, reaction fluxes and currents can only be normalized with respect to reaction volume ($cm^3$) rather than a reaction area ($cm^2$). Electrode area is later incorporated in the Partial and Full mechanistic simulations.

VUV irradiation in the full simulation is treated using branch ratios for OH., H. and $e^-_{(aq)}$ values consistent with 124 nm photon illumination, as reported by Getoff and Schenck[2]. Radiochemical irradiation branch ratios used to determine rate constants (fluxes) for irradiation in the simulation are based on radiochemical G values taken from Domenich et al. and Pastina et al[3,4].

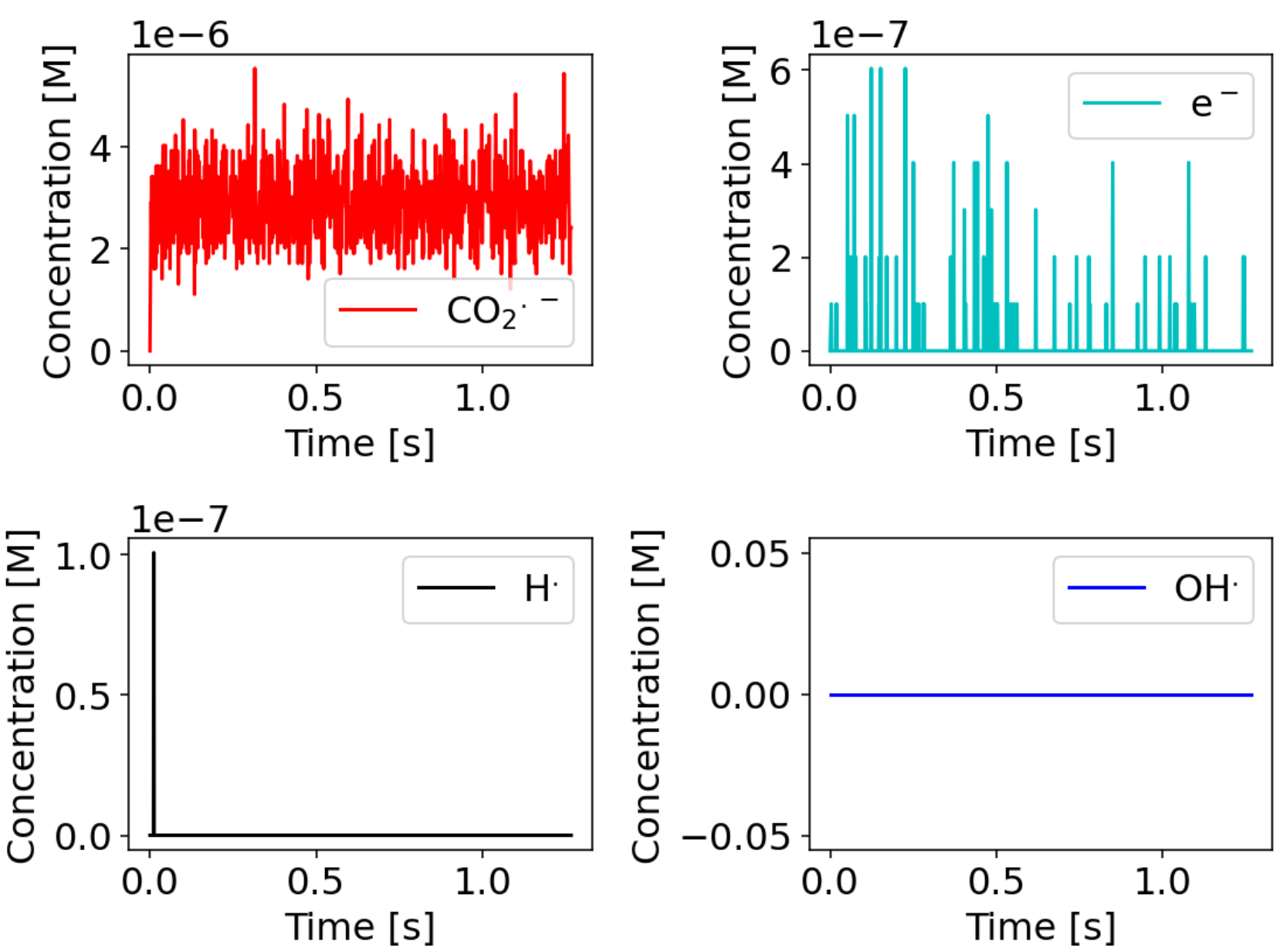

**Figure S1**
Simulation output of $[CO_2^{\cdot-}]$, $[e^{-}_{(aq)}]$, $[H^{\cdot}]$ and $[OH^{\cdot}]$ for the fully-described autocatalytic cycle (Full System), under VUV (124 nm) irradiation at steady-state. $[CO_2]$ = 100 µM, and the reaction is seeded with 100 µM formate. The data demonstrate why formate evolution assumes a linear growth rate at long times. Specifically, the formation of oxalate and all species downstream, are dependent on the $[CO_2^{\cdot-}]$, which is found to rapidly assume a steady-state value within seconds.

* * *

*Application of the unbounded replicator model for non-ideal systems – parameter extraction*

Below, bounded autocatalysis is calculated according to the canonical replicator model for a process A + B → 2 B:

$$[B](t) = \frac{[A]_0 + [B]_0}{1 + \frac{[A]_0}{[B]_0} e^{-([A]_0+[B]_0)kt}} \quad (1)$$

where $k = 0.3$, $[B]_0 = [A]_0 = 100$ μM.

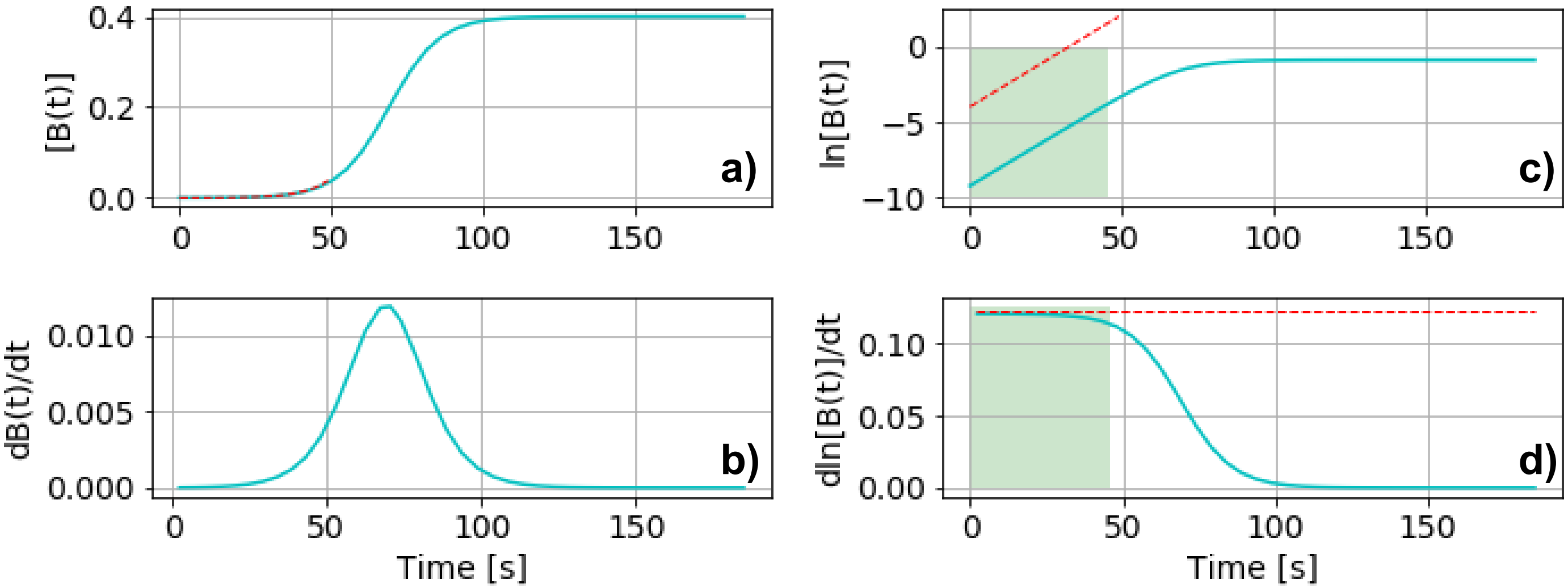

**Figure S2**

**a)**An example of bounded (non-ideal) autocatalytic growth over time, for some arbitrary chemical replicator. Here, the convolution of exponential growth kinetics at early times and the onset of growth-limiting processes at intermediate times, make fitting to an exponential growth model difficult, obscuring the values of the growth rate constant ($k$) and/or replicator selectivity (Ξ). However, **b)** for bounded growth functions, calculation of the 1st derivative of a sigmoid yields a peak, marking times in the growth phase where autocatalysis is now dominated by the growth-limiting processes rather than the kinetics of replicator expansion. This provides a useful guide on the regimes which should be excluded from curve fitting when attempting parameter extraction using an unbounded exponential model. **c)** calculation of ln[B(t)] should yield regions at early times (green shaded region; the pure exponential growth phases) that yield a straight line (red) with a slope of kln(1+Ξ). **d)** Alternatively, calculating dln[B(t)]/dt yields a flat line (red) of constant value equaling kln[(1+Ξ)] over the same period (green shaded region).

*Basic system simulation with phosphate and carbonate buffer equilibria*

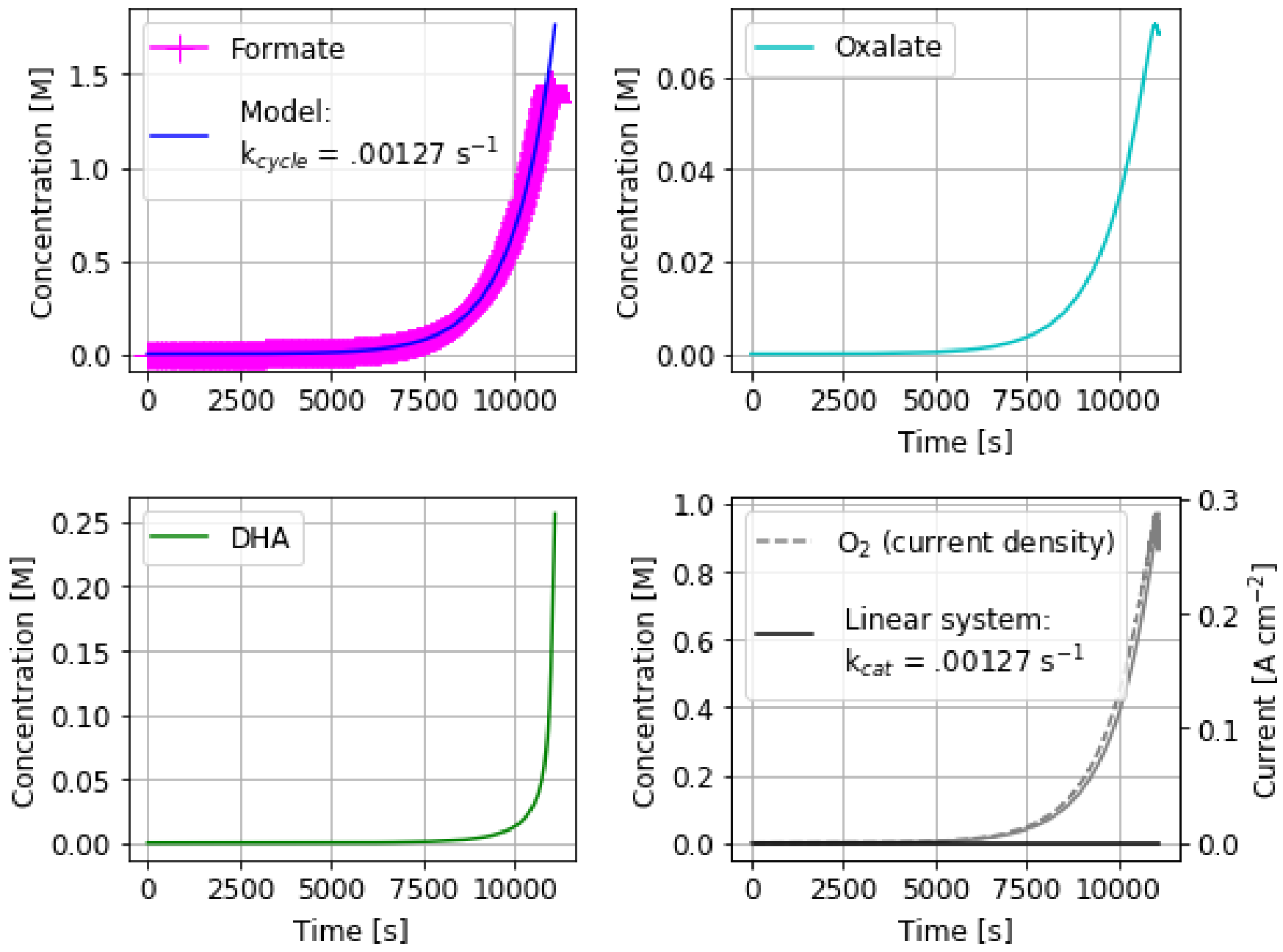

**Figure S3**

Incorporating the additional details of phosphate and carbonate buffer equilibria into the Basic System simulation are found to produce little change in system response, with the autocatalytic rate constant for formate evolution resulting giving a best fit of $k = 0.00127$ $s^{-1}$. As a result, changes in system response going from the Basic to Partial System simulations can be attributed to the mechanistic details of reaction 2 (electrochemical oxalate reduction) and reaction 3 (DHA hydrolysis).

**Table S1**

| **Reaction** | **Forward** | **Units** | **Reverse** | **Units** | **Ref** |
|---|---|---|---|---|---|
| Buffering Equilibria | | | | | |
| Phosphate – Values for $H_3PO_4^-$ (**$BH_2$**)↔ $H_2PO_4^{2-}$ (**BH**) ↔ $HPO_4^{3-}$ (**B:**) equilibrium ($pK_{a1}$ = 7.20, $pK_{a2}$ = 12.32) | | | | | |
| $H_3O^+$ + B: ↔ BH + $H_2O$ | 1.58e7 | $M^{-1}$ $s^{-1}$ | 1 | $s^{-1}$ | 5 |
| $OH^-$ + BH ↔ B: + $H_2O$ | 6.34e6 | $M^{-1}$ $s^{-1}$ | 1 | $s^{-1}$ | 5 |
| BH + $H_3O^+$ ↔ $BH_2$ + $H_2O$ | 2.38e12 | $M^{-1}$ $s^{-1}$ | 1 | $s^{-1}$ | 5 |
| $OH^-$ + $BH_2$ ↔ BH + $H_2O$ | 4.21e1 | $M^{-1}$ $s^{-1}$ | 1 | $s^{-1}$ | 5 |
| Carbonate | | | | | |
| $CO_2$ + $H_2O$ ↔ $H_2CO_3$ | 0.04 | $s^{-1}$ | 12 | $s^{-1}$ | 6,7 |
| $CO_2$ + $OH^-$ ↔ $HCO_3^-$ | 1.21e4 | $M^{-1}$ $s^{-1}$ | 4e-4 | $s^{-1}$ | 6 |
| $H_2CO_3$ ↔ $H^+$ + $HCO_3^-$ | 1e7 | $s^{-1}$ | 5e10 | $M^{-1}$ $s^{-1}$ | 8 |
| $HCO_3^-$ ↔ $H^+$ + $CO_3^{2-}$ | 3 | $s^{-1}$ | 5e10 | $M^{-1}$ $s^{-1}$ | 8 |
| | | | | | |

**S.2 Unit Conversions Between Homogeneous and Heterogeneous Concentration**

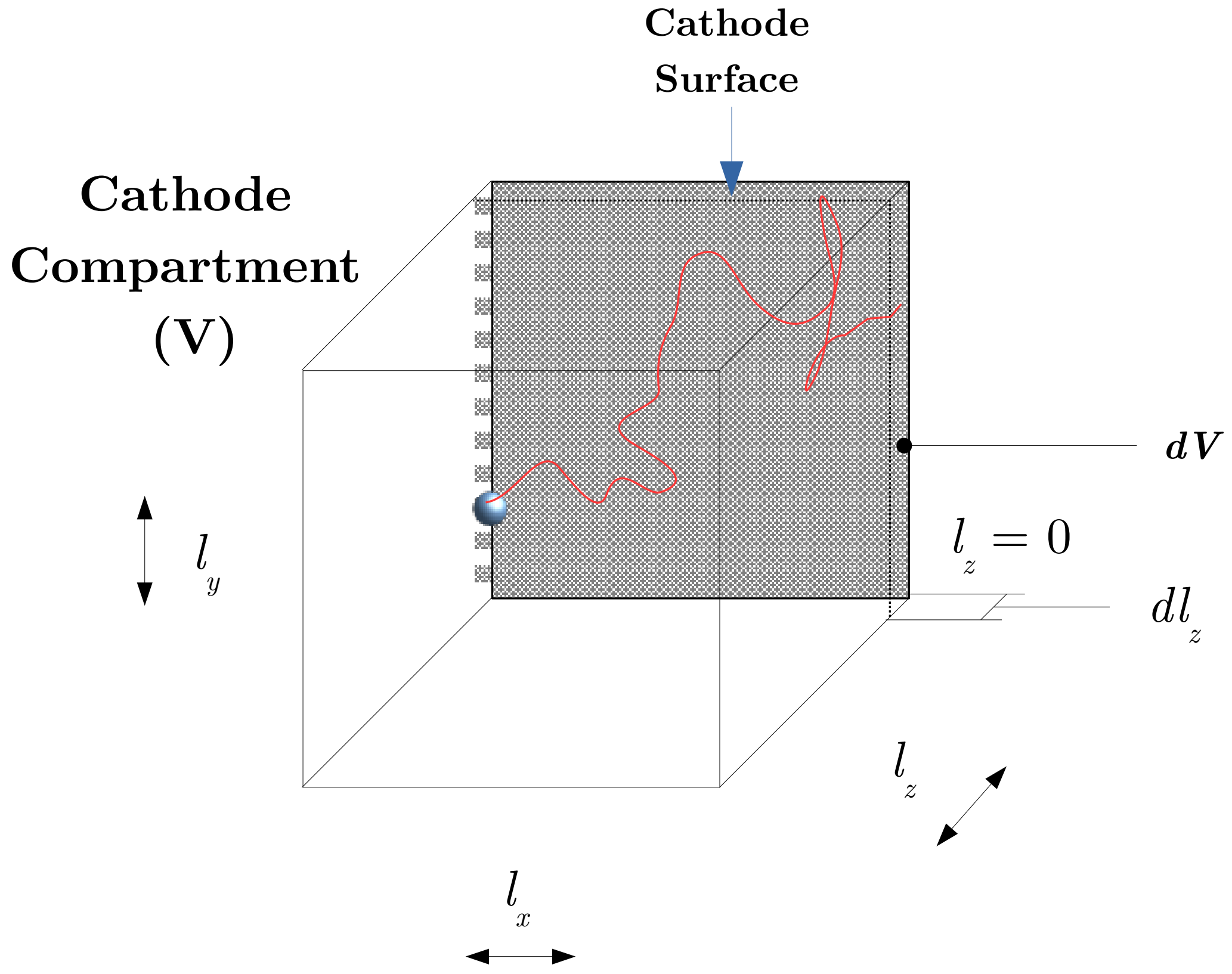

**Figure S4**

Considering dimensionality of diffusion processes in the 0-D simulation model. A diffusive (random walk) trajectory for a molecule of oxalate (or glyoxylate) is shown (red line), for random walks between from the compartment bulk into the region defining the electrode surface ($dl$ = 0.2 nm). Diffusion of oxalate to the cathode surface was modeled by treating the “surface” region – the region as distances from the surface where oxalate is close enough to the surface as to be considered an adsorbate, taken as $dl$ = 0.2 nm. The resulting volume (dV) in which an asorbate can be found, for a 1 $cm^2$ electrode bounding one face of a 1 $cm^3$ volume, is then 2x$10^{-8}$ $cm^3$. = 2x$10^{-11}$ l. Density of active centers are approximated as a metal catalyst ($10^{14}$ $cm^{-2}$) deposited on a high-aspect ratio (~1000x) substrate for an active site density of $10^{17}$ $cm^{-2}$ (1.66x$10^{-7}$ mol $cm^{-2}$). In the single-compartment simulation in Kinetiscope all concentration units must be in molar. Conversion to “molarity” of this active site density distributed over a differential volume comprising the surface region gives an effective active site concentration 1.66x$10^{-7}$ mol / 2x$10^{-11}$ l = 8300 M.

*          *          *

*Diffusion approximations for oxalate adsorption/desorption processes:*

Step 2 in the proposed sequence of reactions may be decomposed such that oxalate reduction to glyoxylate/dihydroxyacetic acid is governed by an initial step of oxalate diffusion from the bulk region to the surface region, followed by adsorption to a cathodic surface:

(1) $C_2O_4^{2-}{}_{(bulk)} \leftrightarrow C_2O_4^{2-}{}_{(surf)}$ ($k_1$, $k_1$')

(2) $C_2O_4^{2-}{}_{(surf)} \overset{A}{\leftrightarrow} A\text{-}C_2O_4^{2-}{}_{(ads)}$. ($k_2$, $k_2$')

Species denoted "bulk," "surf," and "ads," respectively refer to states of oxalate that are in bulk solution, the electrode surface boundary region (taken as any distance within ~2 angstroms of the electrode surface as described by Figure S4, and oxalate adsorbed and bound to the electrode surface. Note that in this approach, we treat the binding of oxalate to an electrode active site (A) as pseudo-first order, as in the surface region, the effective active site concentration is very high (8300 M effective concentration). Oxalate diffusion from the bulk to surface region is handled by considering the following rate equations:

$$\frac{d[C_2O_{4(surf)}^{2-}](t)}{dt} = k_1[C_2O_{4(bulk)}^{2-}](t) + k'_2[A-C_2O_{4(ads)}^{2-}](t) - k_2[C_2O_{4(surf)}^{2-}](t) - k'_1[C_2O_{4(surf)}^{2-}](t) \tag{m}$$

$$\frac{d[C_2O_{4(ads)}^{2-}](t)}{dt} = k_2[C_2O_{4(surf)}^{2-}](t) + k'_2[A-C_2O_{4(ads)}^{2-}](t) \tag{n}$$

It is difficult to provide meaningful rate constants for oxalate adsorption/desorption. However, interconversion between oxalate in the surface region and oxalate adsorbate is assumed to be rapid relative to oxalate diffusion from the bulk into the surface region. As a result, this gives a rate expression controlled by the first step:

$$r = k_1[C_2O_{4(bulk)}^{2-}](t) - k'_1[C_2O_{4(surf)}^{2-}](t) \tag{o}$$

Over timescales much shorter than the replication cycle time, we approximate these processes as steady-state. Solving for $C_2O_4^{2-}{}_{(surf)}$ in terms of $C_2O_4^{2-}{}_{(ads)}$, the following rate description for oxalate conversion to an adsorbate results:

$$r = k_1[C_2O_{4(bulk)}^{2-}](t) - \frac{k'_1 k'_2}{k_2}[C_2O_{4(ads)}^{2-}](t) = k_1[C_2O_{4(bulk)}^{2-}](t) - \frac{k'_1}{K_2}[C_2O_{4(ads)}^{2-}](t) \tag{p}$$

a unimolecular rate constant for glyoxylate diffusion between the bulk and surface region is calculated using the oxalate diffusion coefficient, $D_{oxa} = 1.03\text{x}10^{-5}$ $cm^2$ $s^{-1}$, as discussed in the manuscript[9]. The ratio $k_2/k'_2 = K_2$, the equilibrium constant for oxalate adsorption/desorption within the electrode surface region. With oxalate adsorption/desoprtion equilibria unclear, we take the value of these fast processes to be similar, such that $K_2 \sim 1$. To check against this assumption, we perform a sensitivity analysis for this system for $K_2 = 1\text{x}10^{-3}$, 1, and 1000.

(Partial System response shown). As seen in Figure S5, values of $K_2$ spanning even six orders of magnitude do not appreciably change the formate doubling time, suggesting validity in the overriding assumption that oxalate adsorption/desorption rates do not significantly influence process kinetics.

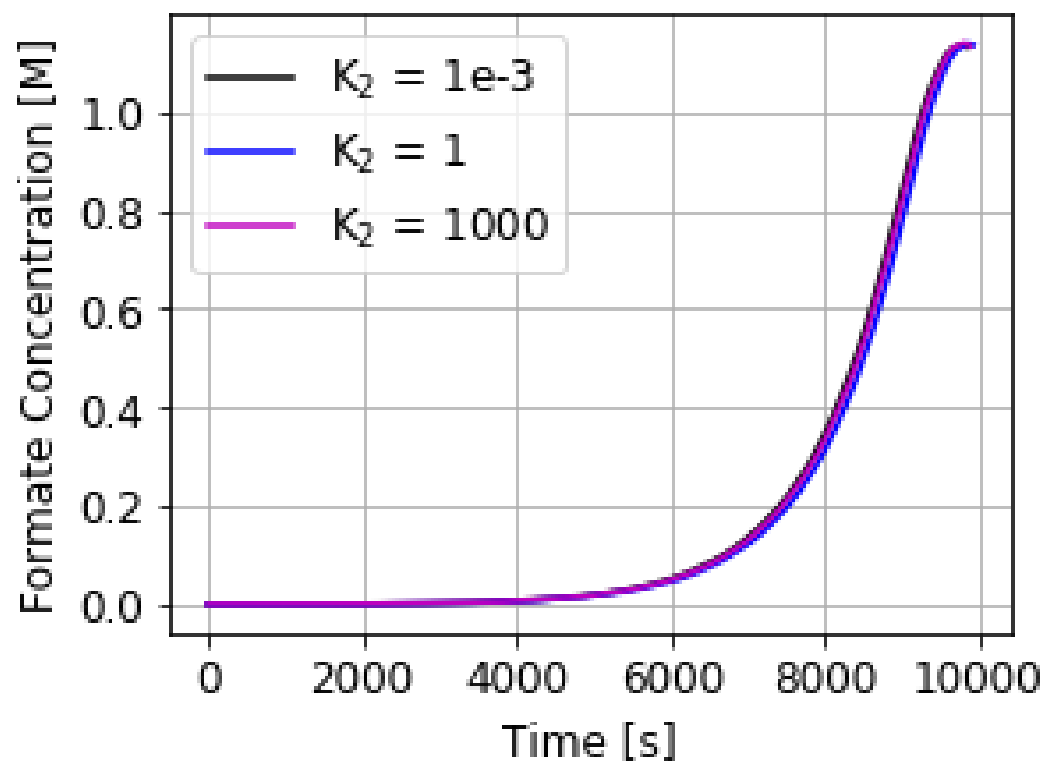

**Figure S5**

Partial System formate replication as a function of changing oxalate adsorption/desorption equilibrium constants. Changing values of $K_2$ over six orders of magnitude ($K_2 = 1x10^{-3}$, 1, and 1000) reveals that formate replication displays very low sensitivity to the value of this equilibrium constant.

**S.3 Model Parameterization – Determination of Rate Constants**

*Step 1 rate constants*

Rate constants used for simulation of radical reactions constituting step 1 were taken from Pastina et al[4]., Severin et al[3]., and and Getoff et al[2,10]. These values are detailed in section S.17.

*　*　*

*Step 2 rate constants*

Reaction 2 is decomposed according to the following mechanism:

$$\mathrm{A} + \mathrm{oxa}_{(aq)} \overset{(k_{2a},\, k_{2a}')}{\leftrightarrow} [\mathrm{A\text{-}oxa}_{(ads)}] \overset{(k_{ET},\, k_{ET}')}{\leftrightarrow} [\mathrm{A\text{-}gly}_{(ads)}] \overset{(k_{2b})}{\rightarrow} \mathrm{A} + \mathrm{gly}_{(aq)}$$

In this mechanism, $A_T$ is the total active site density, oxa is the concentration of the cathode substrate, oxalate, A-oxa$_{(ads)}$ is the density of active sites with adsorbed substrate (oxalate), A-gly$_{(ads)}$ is the density of active sites with adsorbed product (glyoxylate). Gly gives the concentration of free glyoxylate.

For oxalate reduction, take electrochemical reduction of the adsorbed product species to be the rate-determining process in the conversion to glyoxylate. This is consistent with reports that oxalate reduction over graphite and mercury cathodes[11] is rate-controlled by ET to oxalate adsorbates, and assumes that substrate interconversion between substrate-free and electrode-adsorbed states are characterized by rapid equilibria. With ET to the substrate as our rate-determining step, we start from the rate law:

$$r_2(t,V) = k_{ET}[A-oxa_{(ads)}](t) - k'_{ET}[A-gly_{(ads)}](t) \quad \text{(q).}$$

Here, rate constants $k_{ET}$ and $k'_{ET}$ are the interfacial electron transfer rate constants governing the respective, electrode-mediated reduction ($k_{ET}$) and oxidation ($k'_{ET}$) of the oxalate/glyoxylate adsorbates:

$$k_{ET}(\eta) = k_0\left[\exp\left(\frac{nF\alpha\eta}{RT}\right)\right] \;;\; k'_{ET}(\eta) = k_0\left[\exp\left(-\frac{nF\eta(1-\alpha)}{RT}\right)\right] \quad \text{(r).}$$

Using values of $n$ = 0.3 V, $\alpha$ = 0.5 and $T$ = 298 k, for a prototypical exchange current density of $10^{-5}$ A $cm^{-2}$, multiplied by a 1000x aspect ratio (total surface coverage = $10^{17}$ $cm^{-2}$), yields rate constants of $k_0$ = 0.624 $s^{-1}$, $k_{ET}$ = 215 $s^{-1}$, $k'_{ET}$ = 0.002 $s^{-1}$.

*　*　*

*Step 3 rate constants*

A usable expression for the catalyst-DHA* complex is determined according to Mechanism A:

$$\mathrm{DHA} + \mathrm{Cat} \leftrightarrow \mathrm{Cat\text{-}DHA} \leftrightarrow \mathrm{Cat\text{-}DHA^*} \rightarrow \mathrm{Cat} + 2\,\mathrm{HCO_2^-}$$

Here, Cat is a homogeneous catalyst, Cat-DHA denotes a transient catalyst-dihydroxyacetate complex, and Cat-DHA* is the transient catalyst-pre-fission Intermediate (DHA*) complex, which decays to two formate molecules while regenerating the catalyst. Because there is no available data on the kinetics of this reaction, rate constants were chosen to be conservative, with low values being used to parameterize DHA-catalyst binding equilibria, DHA conversion to form the DHA*, along with assumptions of slow decay of the DHA* into two formate equivalents. In addition, sensitivity analyses were conducted for this reaction, to show the impact of a range of rate constants on overall cycle behavior (S.7).

*    *    *

*Buffering equilibria*

Rate constants for buffering equilibria were calculated according to the general expression

$$K_a = k_f/k_r,$$

for a process

$$\text{B:} + H_3O^+ \xleftrightarrow{(k_f,\ k_r)} \text{BH} + H_2O$$

where $K_a$ is the acid-base equilibrium constant for a buffer of some specified $pK_a$, $k_f$ is the forward rate constant for protonation and $k_r$ represents the reverse rate constant for deprotonation. $K_b$ values were calculated according to the relation.

$$K_b = 1x10^{14}/K_a.$$

### S.3 Model Parameterization – Thermochemical Data

Values for the free energies of combustion for all species in Table S1 were found by applying Hess' Law to combustion reactions for formic acid, oxalic acid, and dihydroxyacetic acid and their conjugate bases. The free energies of formation for all components of each combustion reaction were then used to determine the overall free energy change for combustion. Values for hydrogen were determined from the relation:

$$\Delta G^0 = -nF\Delta E^0 \quad \text{(s).}$$

**Table S2**

$\Delta G_c^0$ values in bold (conjugate base forms) were used in the calculation of replication energy efficiencies in S9.

| Compound | $\Delta_f G^0$ [kJ mol$^{-1}$] | $\Delta H_c^0$ [kJ mol$^{-1}$] (acid) | $\Delta E^0$ [V] | $\Delta G_c^0$ [kJ mol$^{-1}$] | Reference |
|---|---|---|---|---|---|
| **$H_2O$** | -237.19 | - | - | - | 12 |
| **$OH^-$** | -157.27 | - | - | - | 12 |
| **$O_2$** | 0 | - | - | - | 12 |
| **$CO_2$** | -394.37 | | | | 12 |
| **Formate/ Formic Acid** | -351.00/-371.79 | -254.3 | - | **-200.27**/-259.77 | 12–14 |
| **Oxalate/ Oxalic Acid** | -673.00 | -242.9 | - | **-273.63**/-352.93 | 13 |
| **Dihydroxyacetate/ Dihydroxyacetic acid** | -478.85[a] | -524.8 | - | **-624.43**/-784.27 | 12,15 |
| **Glyoxylate/Glyoxylic Acid** | -468.60 | -523.00 | | **-477.41**/-557.33 | 12 |
| **Hydrogen** | - | - | 1.2 | -237.3 | 16 |

*a.* Value is calculated starting from the literature value for the free energy of formation of glyoxylate ($\Delta G_c$ =-468.60 kJ/mol) and adjusted using the glyoxylate/dihydroxyacetate equilibrium constant, K = 0.016, from Eggins and McMullan.[15]

Calculations of $\Delta G_c^0$ for various replication fuel products. Terms in red denote free energies of formation for conjugate bases being used in the calculation of acid form products. As a result, the resulting $\Delta G_c^0$s for the acid form products are only estimates within 10-20 kJ/mol of their real values.

Formic Acid:

$HCO_2H + ½ O_2 \rightarrow CO_2 + H_2O$

$\Delta G_c^0 = [(1)(-394.37) + (1)(-237.19)] - [(1)(-371.79) + (0.5)(0)] = -259.77$ kJ/mol

Formate:

$HCO_2^- + ½ O_2 \rightarrow CO_2 + OH^-$

$\Delta G_c^0 = [(1)(-394.37) + (1)(-157.27)] - [(1)(-351.00) + (0.5)(0)] = -200.27$ kJ/mol

Oxalic Acid:

$HC_2O_4^- + ½ O_2 \rightarrow 2\ CO_2 + OH^-$

$\Delta G_c^0$ = [(2)(-394.37) + (1)(-237.19)] – [(1)(-673.00) + (0.5)(0)] = -352.93 kJ/mol

Oxalate:

$HC_2O_4^- + ½ O_2 \rightarrow 2 CO_2 + OH^-$

$\Delta G_c^0$ = [(2)(-394.37) + (1)(-157.27)] – [(1)(-673.00) + (0.5)(0)] = -273.63 kJ/mol

Dihydroxyacetate:

$H_3C_2O_4^- + O_2 \rightarrow 2 CO_2 + 2 OH^-$

$\Delta G_c^0$ = [(2)(-394.37) + (2)(-157.27)] – [(1)(-478.85) + (1)(0)] = -624.43 kJ/mol

Dihydroxyacetic Acid:

$H_4C_2O_4 + O_2 \rightarrow 2 CO_2 + 2 H_2O$

$\Delta G_c^0$ = [(2)(-394.37) + (2)(-237.19)] – [(1)(-478.85) + (1)(0)] = -784.27 kJ/mol

Glyoxylate:

$HC_2O_3^- + O_2 \rightarrow 2 CO_2 + OH^-$

$\Delta G_c^0$ = [(2)(-394.37) + (1)(-157.27)] – [(1)(-468.60) + (1)(0)] = -477.41 kJ/mol

Glyoxylic Acid:

$H_2C_2O_3 + O_2 \rightarrow 2 CO_2 + H_2O$

$\Delta G_c^0$ = [(2)(-394.37) + (1)(-237.19)] – [(1)(-468.60) + (1)(0)] = -557.33 kJ/mol

### S.4 Product Selection as a Function of Rate Constants – Reaction 1 Sensitivity

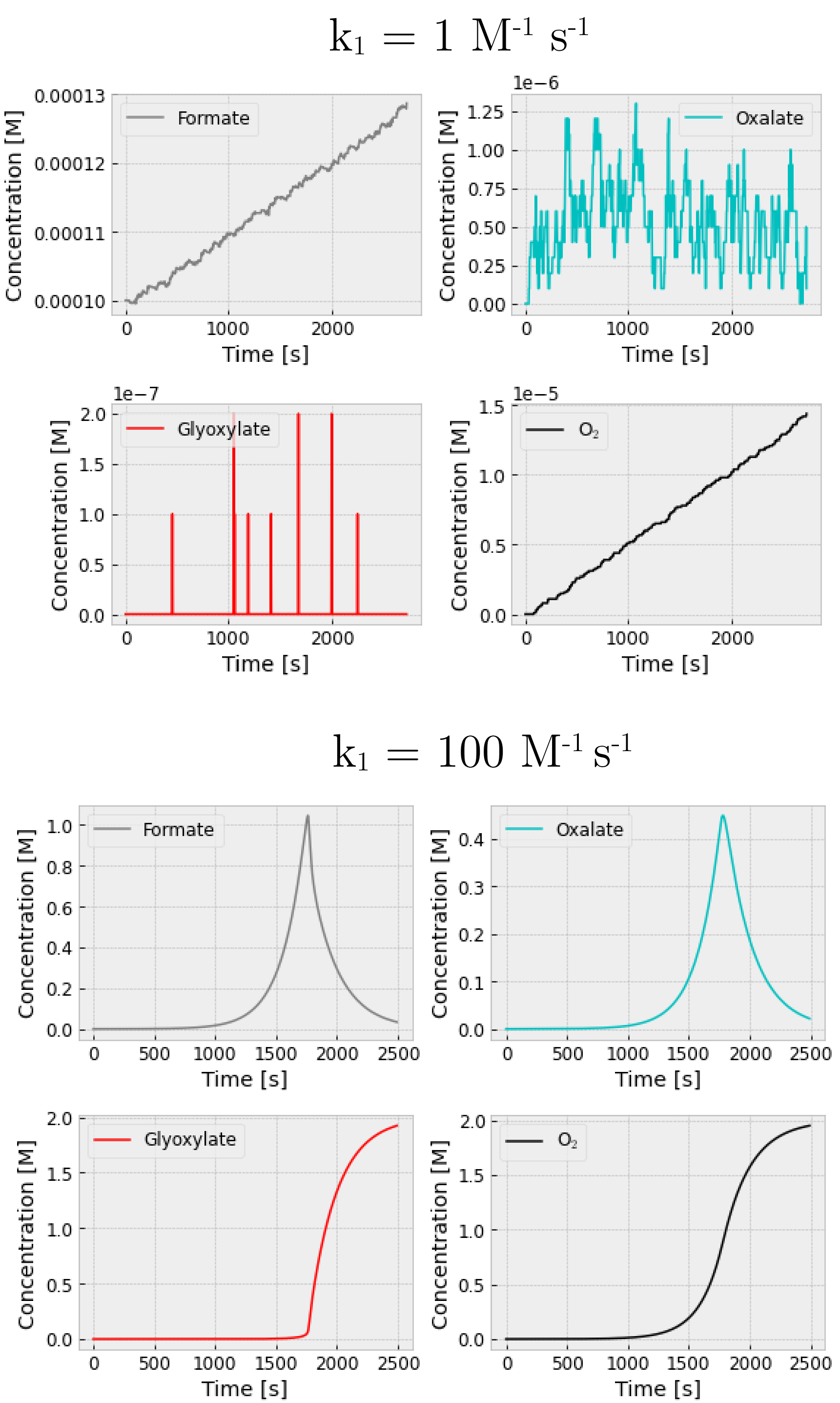

**Figure S6 – Basic System**

Influence of rate constant $k_1$ on the evolution of autocatalytic intermediates formate, oxalate and glyoxylate, along with $O_2$ evolution. $k_2 = 0.01$ s$^{-1}$, $k_3 = 1000$ M$^{-1}$ s$^{-1}$. Spikes and hard transitions observed in these reactions occur from the total consumption of buffer (simulated as a batch species) during the simulation time.

### S.4 Product Selection as a Function of Rate Constants – Reaction 2 Sensitivity

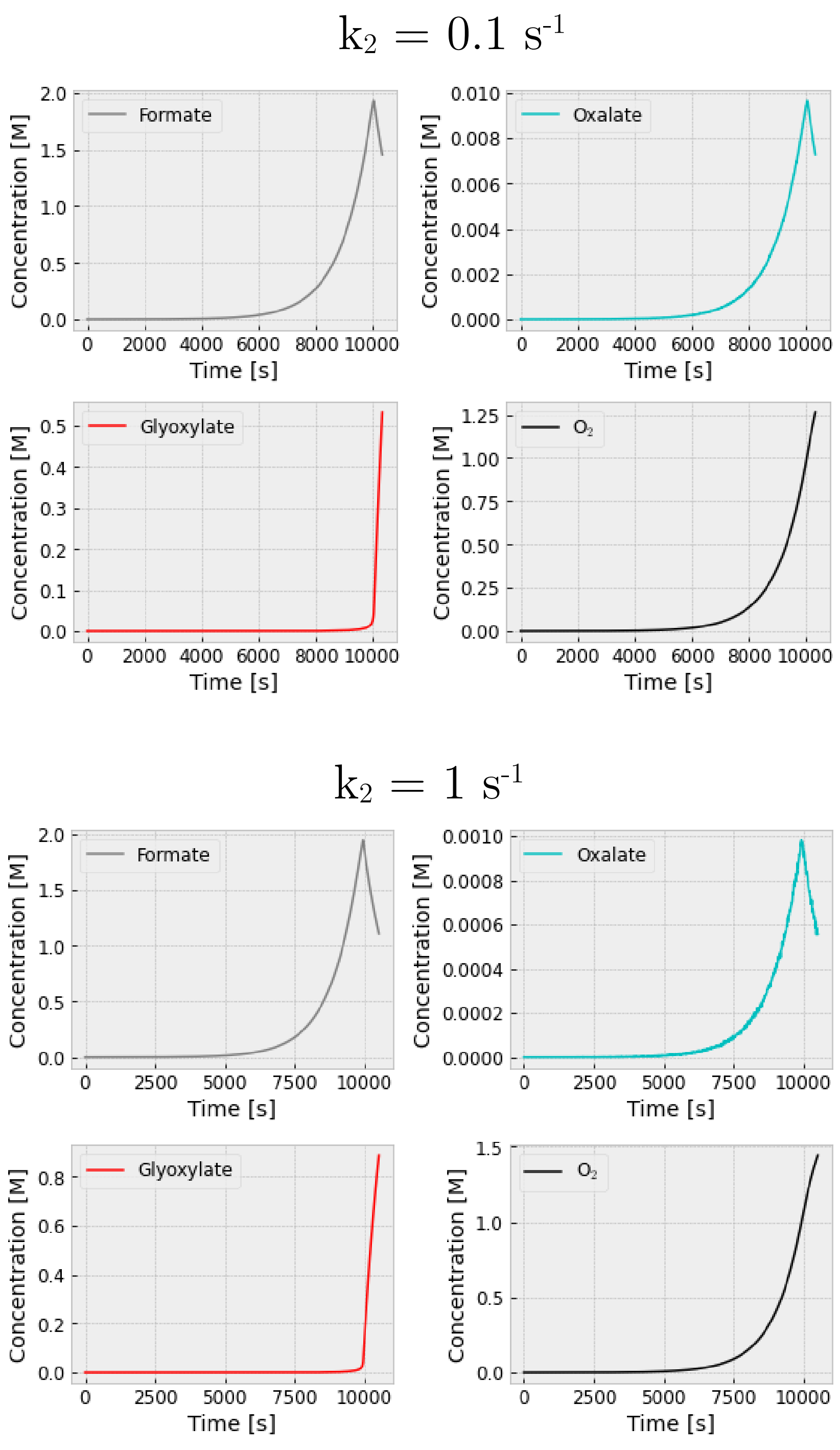

**Figure S7 – Basic System**

Influence of rate constant $k_2$ on the evolution of autocatalytic intermediates formate, oxalate and glyoxylate, along with $O_2$ evolution. $k_1 = 10\ M^{-1}\ s^{-1}$, $k_3 = 1000\ M^{-1}\ s^{-1}$. Spikes and hard transitions observed in these reactions occur from the total consumption of buffer (simulated as a batch species) during the simulation time.

### S.4 Product Selection as a Function of Rate Constants – Reaction 3 Sensitivity

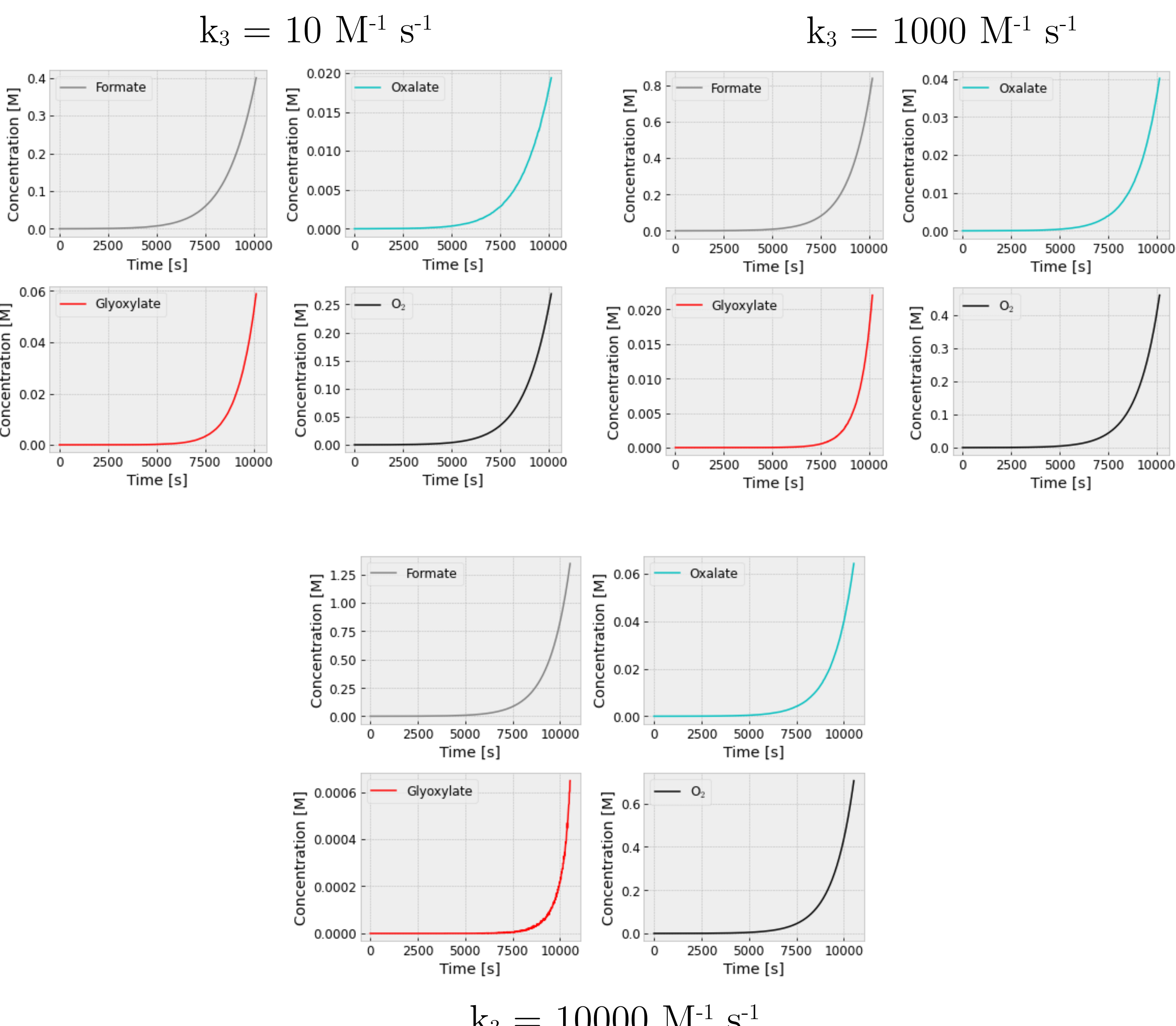

**Figure S8 – Basic System**

Influence of rate constant $k_3$ on the evolution of autocatalytic intermediates formate, oxalate and glyoxylate, along with $O_2$ evolution. $k_1 = 10$ $M^{-1}$ $s^{-1}$, $k_2 = 0.1$ $s^{-1}$.

**S.5 Effects of Side Reactions**

*Defining selectivity for individual reactions*

Using the proposed scheme outlined in Figure 1b, the product selectivity ($\Xi$) for any particular $j^{th}$ step of the fuel cycle may be generally expressed as:

$$\Xi = \frac{(r_f - r_b)}{(r_f - r_b) + \sum (r_{s,f} - r_{s,b})} \quad \text{(s)}$$

where the summation in the denominator gives the total rate of all back and forward side reactions ($r_{s,f}$ , $r_{s,b}$), and terms $r_f$, $r_b$ give the rates of the primary forward and back reactions, respectively. Applying this formula to the reaction scheme outlined in Table 1, and assuming the same stoichiometric rate laws used for the basic system, the selectivity for step 1 is then:

$$\Xi_1(t) = \frac{k_1[HCO_2^-](t)[CO_2](t) - k'_1[oxa](t)}{k_1[HCO_2^-](t)[CO_2](t) - k'_1[oxa](t) + k_{S1}[HCO_2^-](t) - \sum_i k'_{S1i}[S_{1i}](t)} \quad \text{(t).}$$

Terms $k_1$ and $k'_1$ give the rate constants of the primary forward and back reactions for the first step of the cycle, $ks_{1i}$, $k'_{S1i}$ are the forward and back rate constants for the $i^{th}$ side reaction, [$HCO_2^-$] is the formate concentration and [oxa] refers to the concentration of oxalate. The selectivity for step two, the two-electron reduction of oxalate to dihydroxyacetate (DHA, the equilibrium hydrate of glyoxylate) is expressed as:

$$\Xi_2(t) = \frac{k_2[oxa](t) - k_2'[DHA](t) + k_{S2}[oxa](t)}{k_2[oxa](t) - k_2'[DHA](t) + k_{S2}[oxa](t) - \sum_i k'_{S2i}[S_{2i}](t)} \quad \text{(u).}$$

Here, $k_2$ and $k'_2$ give the rates of the primary forward and back reactions for the second cycle step, while $k_{S2}$ and $k'_{s2i}$ are the rate constants for the forward and $i^{th}$ reverse side reactions of the second step. The term [DHA] denotes the concentration of dihydroxyacetate (the hydrate of glyoxylate); [$OH^-$] refers to the hydroxide ion concentration. For the third step, a similar expression for reaction selectivity results, yielding eq. 2d:

$$\Xi_3(t) = \frac{k_3[DHA][OH^-](t)}{k_3[DHA][OH^-](t) + k_{S3}[DHA](t) - \sum_i k'_{S3i}[S_{3i}](t)} \quad \text{(v).}$$

Terms $k_3$, and $k'_3$ give the rates of the primary forward and back reactions for the third step of the reaction cycle, DHA hydrolysis to two molecules of formate. Terms $k_{S3}$ and $k'_{S3i}$ are the rate constants for the forward and $i^{th}$ reverse side reactions of the third step. The product of $\Xi_1 \Xi_2 \Xi_3 = \Xi$, the selectivity for an entire cycle (eq. 2a).

**S.5 Effects of Side Reactions – Selectivity of Reaction 1**

Changes to relative rates of rate constants $k_1$, $k_2$, $k_3$ and their associated side reactions, $k_{s1}$, $k_{s2}$ *and* $k_{s3}$, cause drastic changes to replication cycle behavior, highlighting the necessity of appropriately tuning the rates of formation of each intermediate to realize nonlinear replicator growth. Calculating the selectivity for individual reactions and the overall cycle as these rate constants are changed, show how suboptimal reaction rate constants may cause the cycle to veer between genuine replication behavior, to zero-growth, to net formate consumption cases. In this latter instance, no replication behavior exists, and total cycle selectivity for formate replication is therefore undefined.

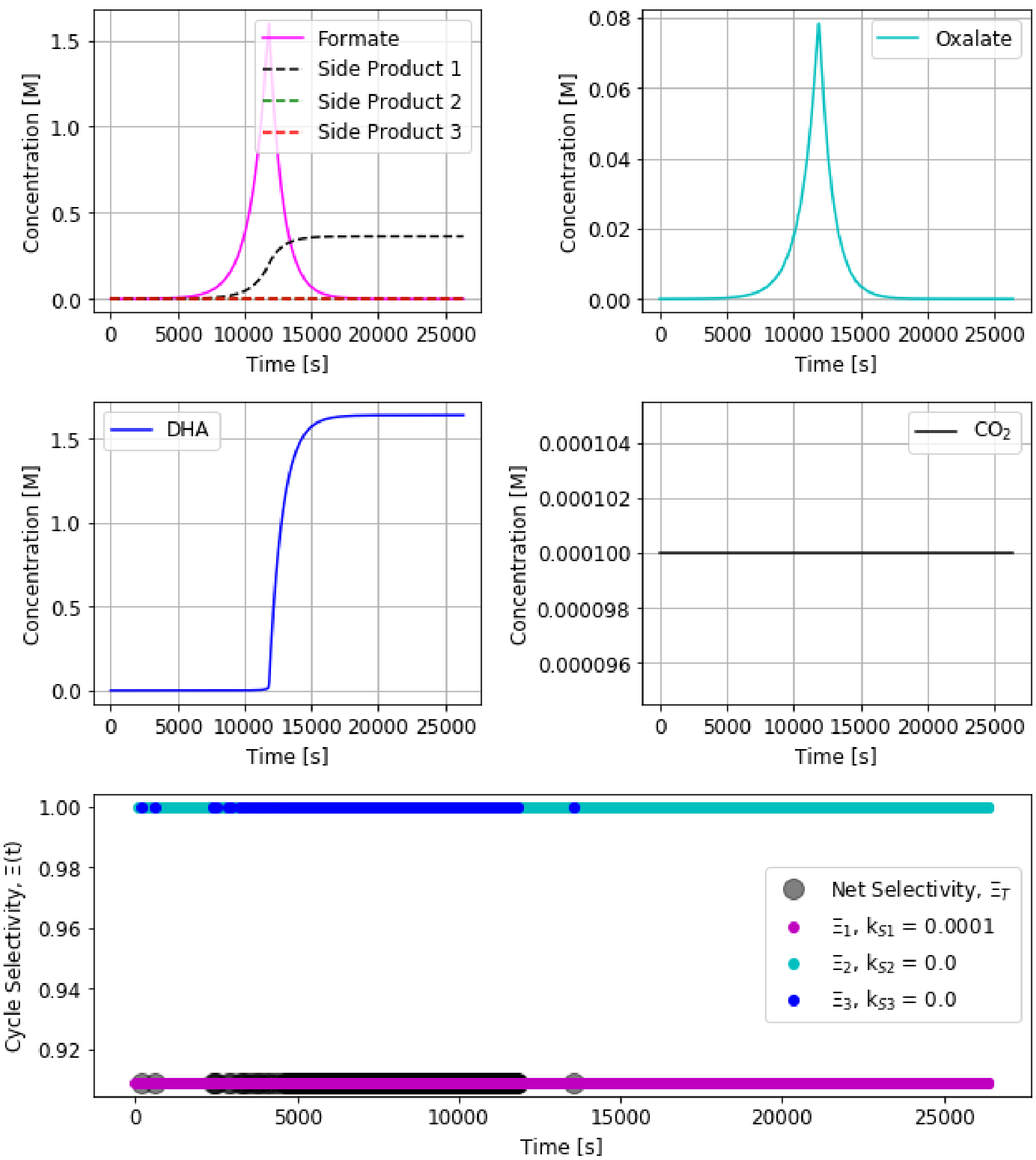

**Figure S9 – Basic System**

$k_{s1}$ = 0.0001 $s^{-1}$, $k_{s2}$ = 0.0 $s^{-1}$, $k_{s3}$ = 0.0 $s^{-1}$

Influence of side reactions in step 1 (formate carboxylation). Side reactions are generalized as unimolecular decays following the rate law $r_{s1} = k_{s1}$[formate]. Spikes and hard transitions observed in these reactions occur from the total consumption of buffer (simulated as a batch species) during the simulation time. Net selectivity is undefined for times where replication has been arrested, and subsequent reactions only result in the consumption of formate.

### S.5 Effects of Side Reactions – Selectivity of Reaction 1

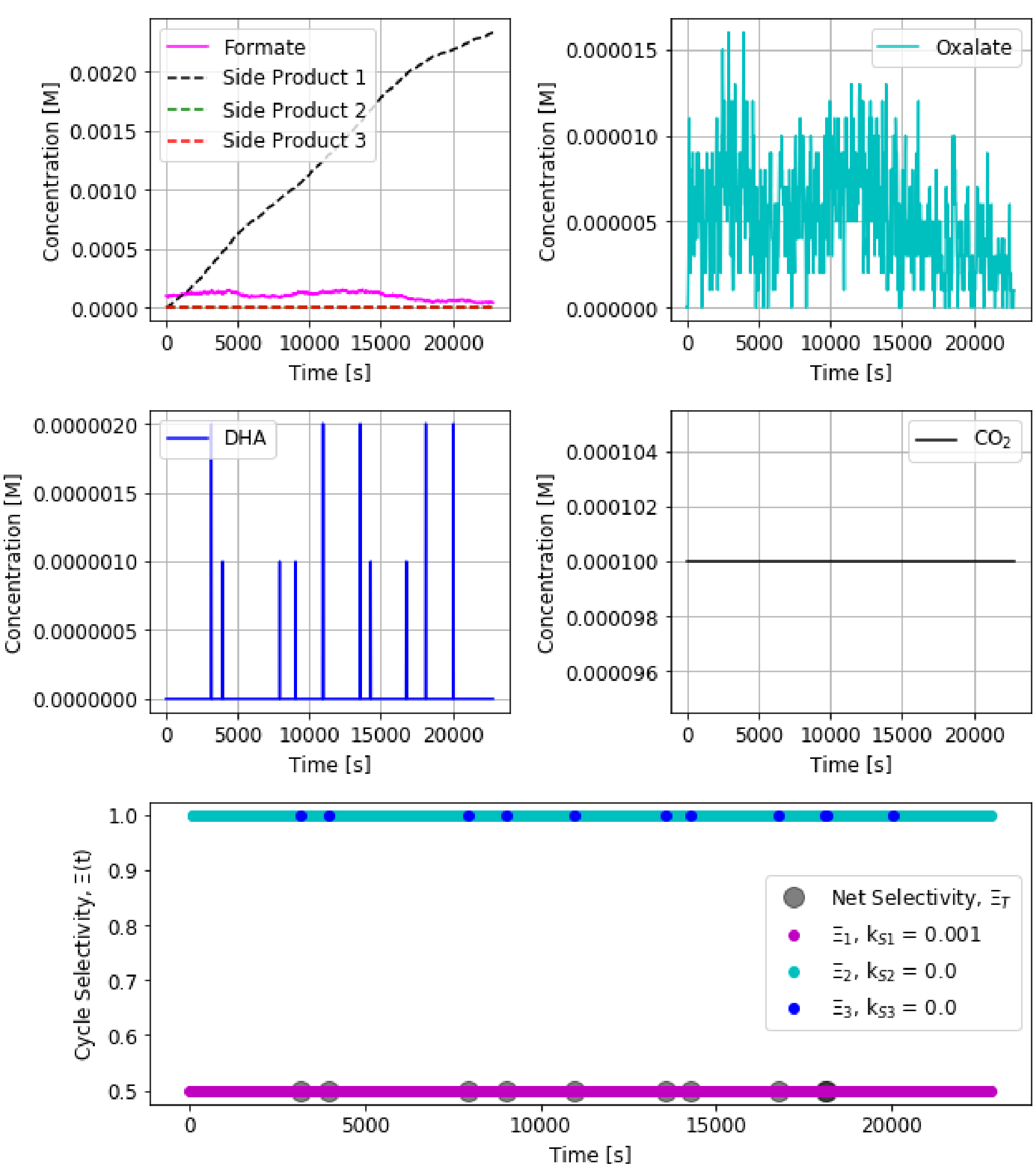

**Figure S10 – Basic System**

$k_{s1} = 0.001$ s$^{-1}$, $k_{s2} = 0.0$ s$^{-1}$, $k_{s3} = 0.0$ s$^{-1}$

Influence of side reactions in step 1 (formate carboxylation). Side reactions are generalized as unimolecular decays following the rate law $r_{s1} = k_{s1}$[formate]. Net selectivity is undefined for times where replication has been arrested, and subsequent reactions only result in the consumption of formate.

### S.5 Effects of Side Reactions – Selectivity of Reaction 1

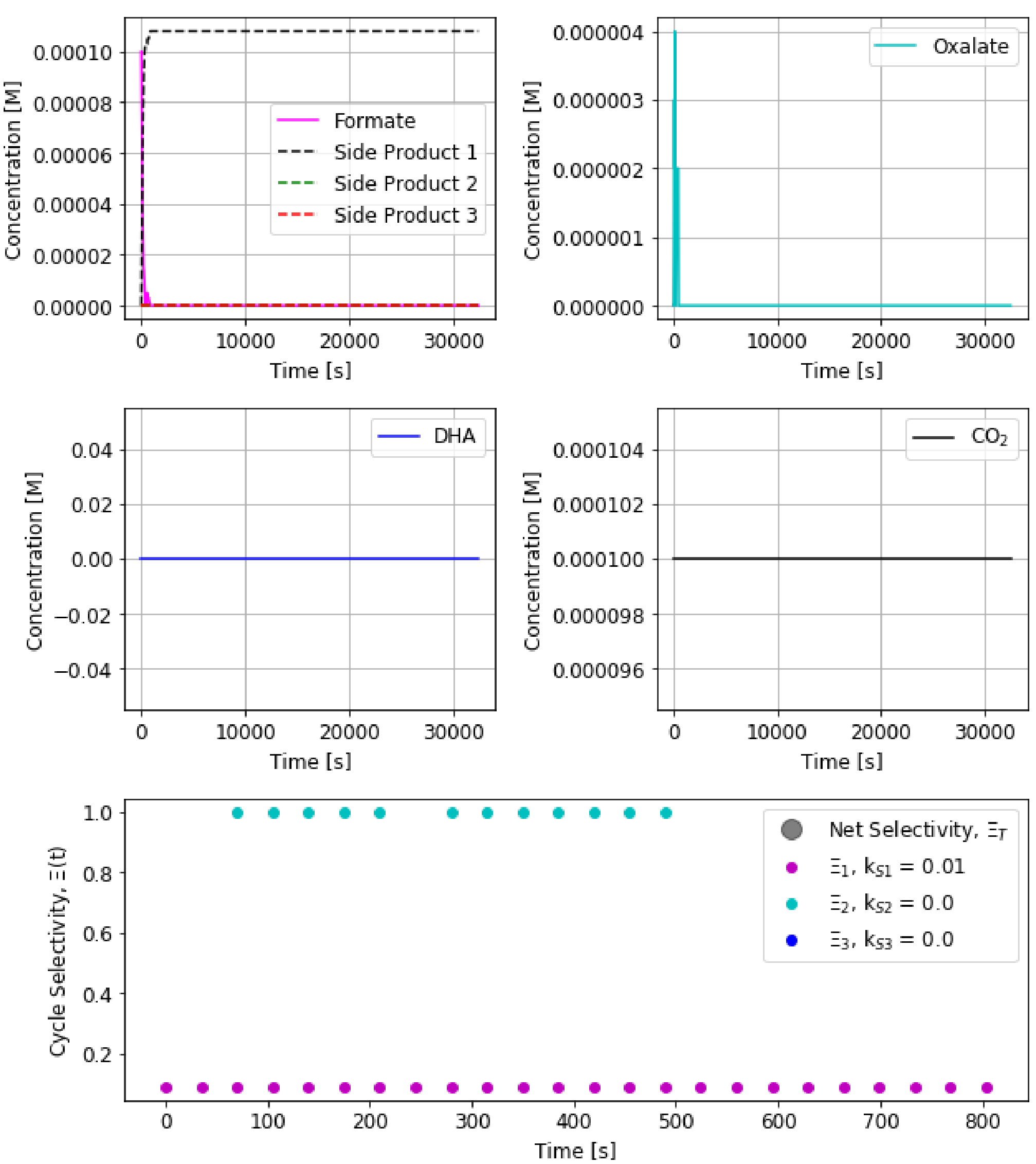

**Figure S11 – Basic System**

$k_{s1} = 0.01\ \text{s}^{-1}$, $k_{s2} = 0.0\ \text{s}^{-1}$, $k_{s3} = 0.0\ \text{s}^{-1}$

Influence of side reactions in step 1 (formate carboxylation). Side reactions are generalized as unimolecular decays following the rate law $r_{s1} = k_{s1}[\text{formate}]$. At this value of $k_{s1}$, the rate of side product formation completely arrests formate turnover into oxalate at early times in the reaction, halting formate growth. This results in an undefined net reaction selectivity for the overall cycle. Net selectivity is undefined for times where replication has been arrested, and subsequent reactions only result in the consumption of formate.

**S.5 Effects of Side Reactions – Selectivity of Reaction 1**

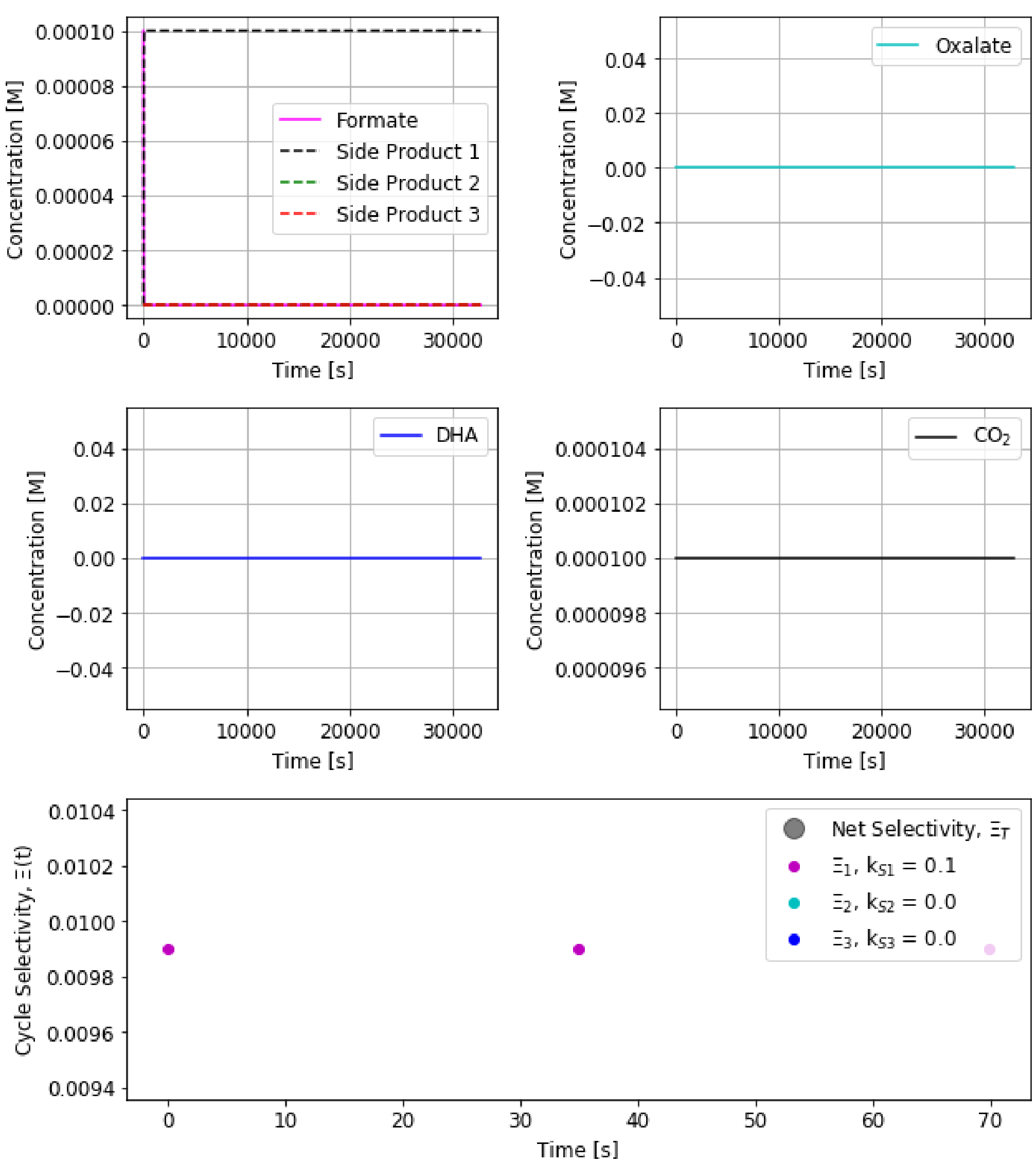

**Figure S12 – Basic System**

$k_{s1}$ = 0.1 $s^{-1}$, $k_{s2}$ = 0.0 $s^{-1}$, $k_{s3}$ = 0.0 $s^{-1}$

Influence of side reactions in step 1 (formate carboxylation). Side reactions are generalized as unimolecular decays following the rate law $r_{s1} = k_{s1}$[formate]. At this value of $k_{s1}$, the rate of side product formation completely arrests autocatalysis, resulting in an undefined net reaction selectivity.

### S.5 Effects of Side Reactions – Selectivity of Reaction 2

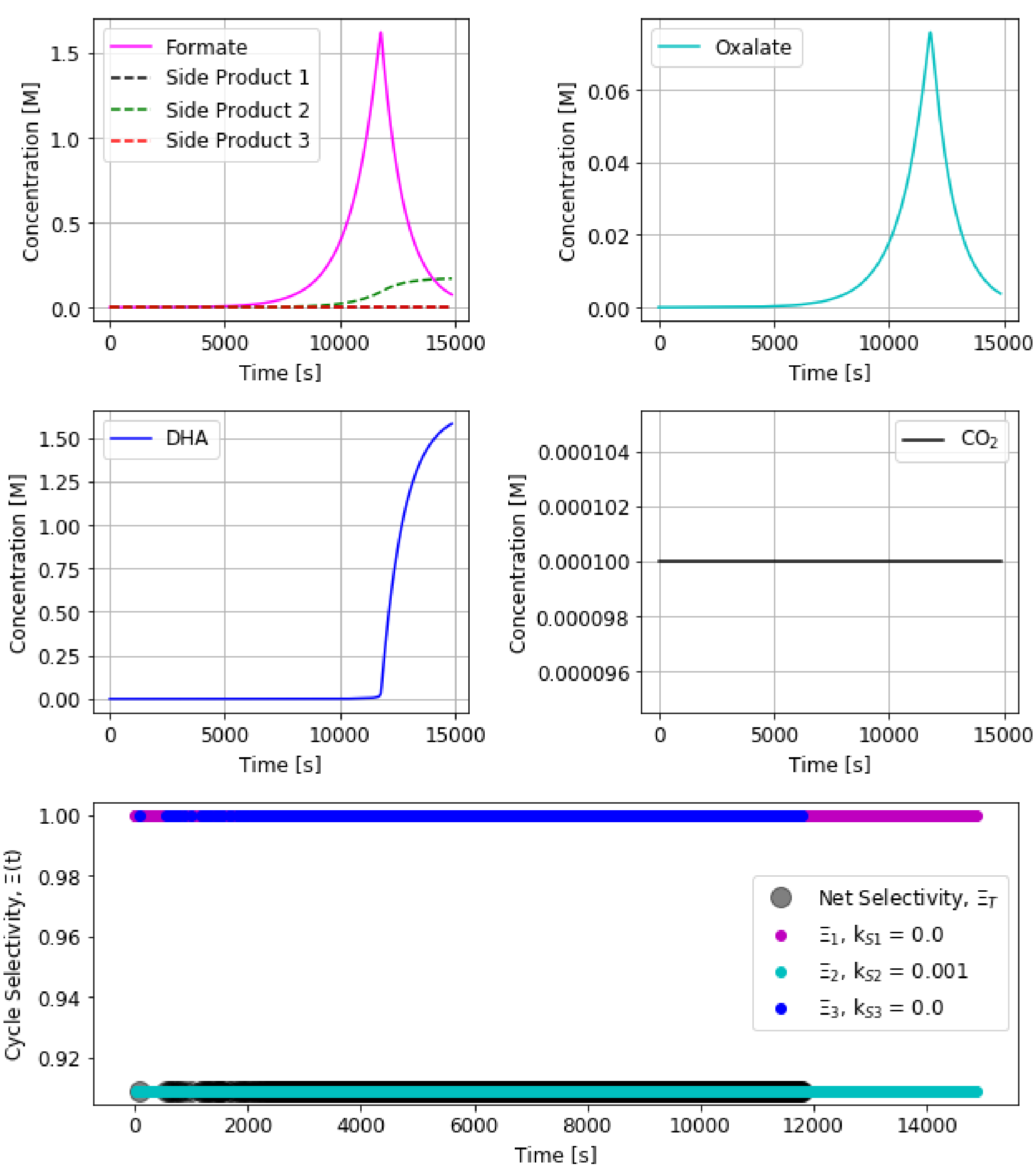

**Figure S13 – Basic System**

$k_{s1} = 0.0$ s$^{-1}$, $k_{s2} = 0.001$ s$^{-1}$, $k_{s3} = 0.0$ s$^{-1}$

Influence of side reactions in step 2 (oxalate reduction). Side reactions are generalized as unimolecular decays following the rate law $r_{s2} = k_{s2}$[oxalate]. Spikes and hard transitions observed in these reactions occur from the total consumption of buffer (simulated as a batch species) during the simulation time. Net selectivity is undefined for times where replication has been arrested, and subsequent reactions only result in the consumption of formate.

### S.5 Effects of Side Reactions – Selectivity of Reaction 2

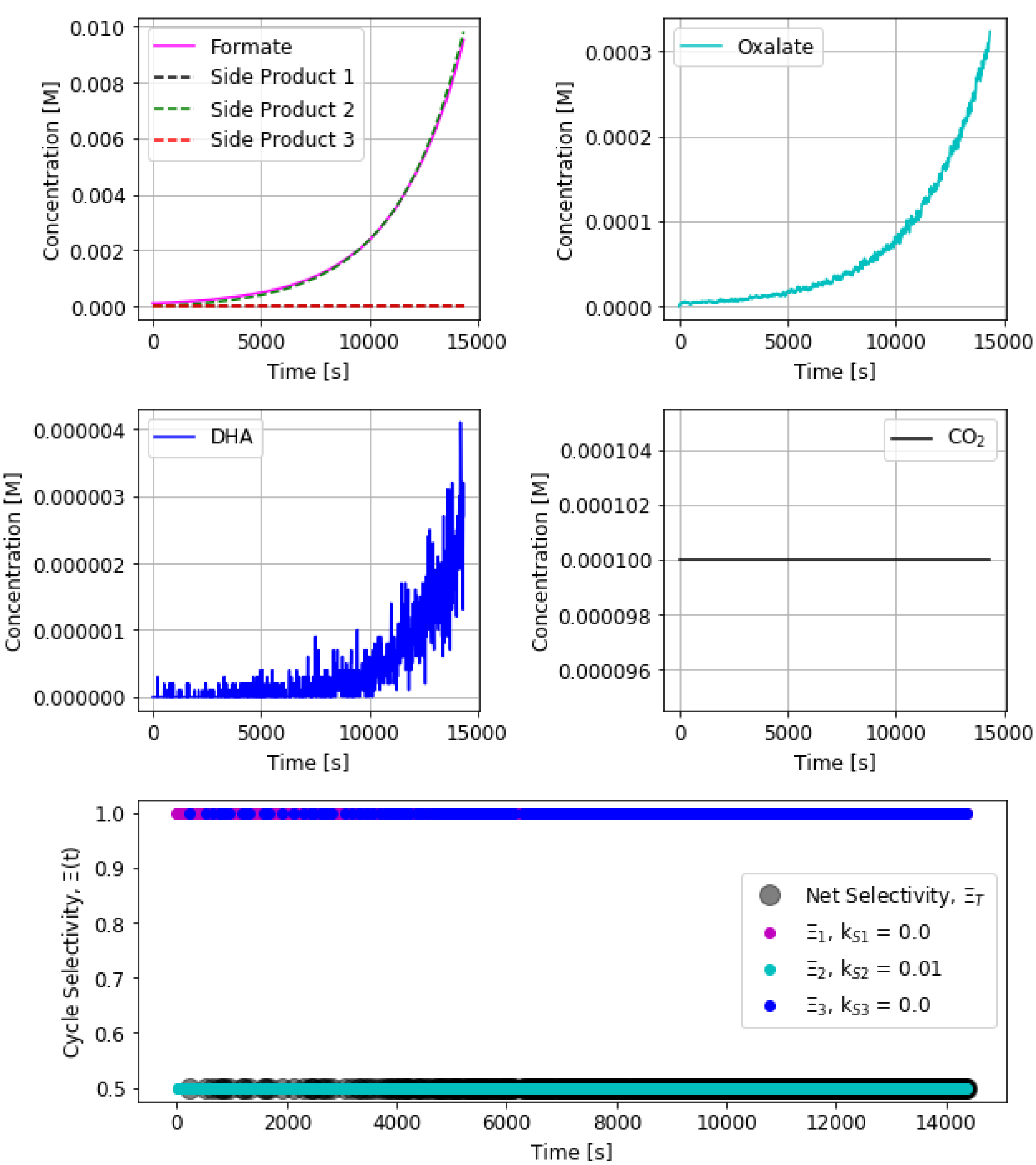

**Figure S14 – Basic System**

$k_{s1} = 0.0$ s$^{-1}$, $k_{s2} = 0.01$ s$^{-1}$, $k_{s3} = 0.0$ s$^{-1}$

Influence of side reactions in step 2 (oxalate reduction). Side reactions are treated as a generic reaction following the rate law $r_{S2} = k_{S2}$[oxalate].

## S.5 Effects of Side Reactions – Selectivity of Reaction 3

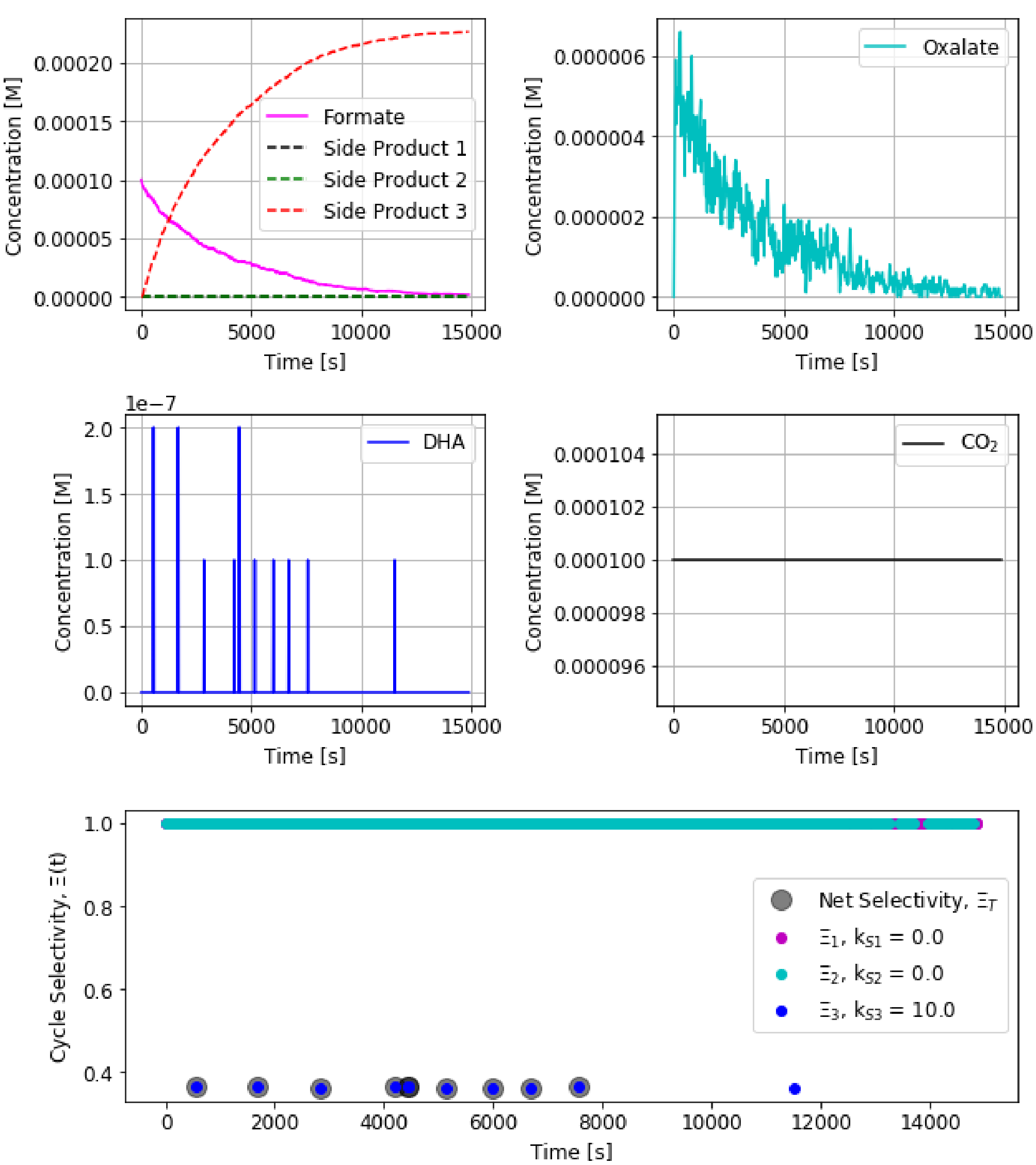

**Figure S15 – Basic System**

$k_{s1}$ = 0.0 s$^{-1}$, $k_{s2}$ = 0.0 s$^{-1}$, $k_{s3}$ = 10.0 s$^{-1}$

Influence of side reactions in step 3 (formate carboxylation). Side reactions are treated as a generic reaction following the rate law $r_{S3} = k_{S3}[\mathrm{DHA}]$. Net selectivity is undefined for times where replication has been arrested, and subsequent reactions only result in the consumption of formate.

## S.5 Effects of Side Reactions – Selectivity of Reaction 3

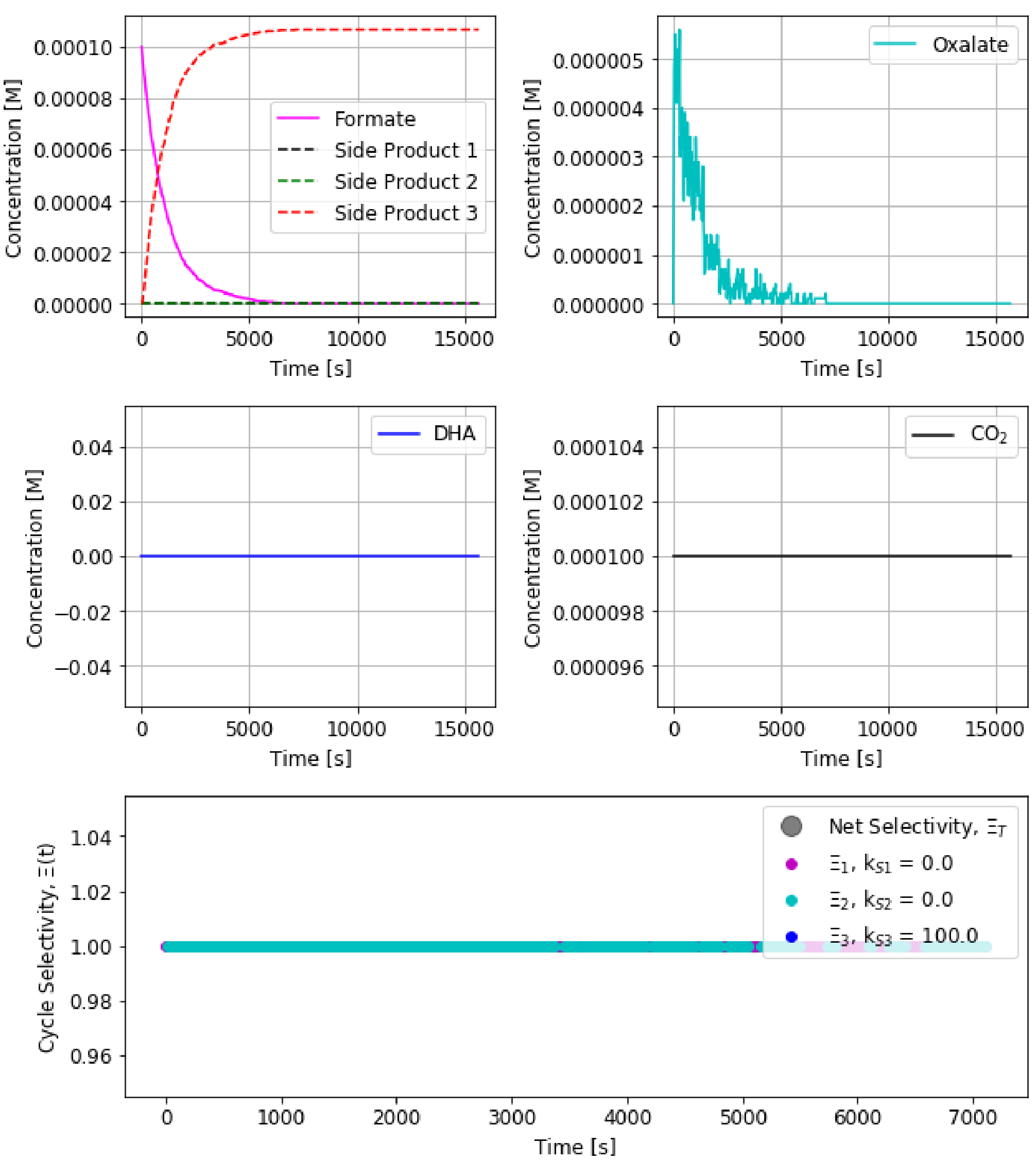

**Figure S16 – Basic System**

$k_{s1}$ = 0.0 s$^{-1}$, $k_{s2}$ = 0.0 s$^{-1}$, $k_{s3}$ = 100.0 s$^{-1}$

Influence of side reactions in step 3 (formate carboxylation). Side reactions are treated as a generic reaction following the rate law $r_{S3} = k_{S3}$[DHA]. Net selectivity is undefined for times where replication has been arrested, and subsequent reactions only result in the consumption of formate.

### S.5 Effects of Side Reactions – Selectivity of Reaction 3

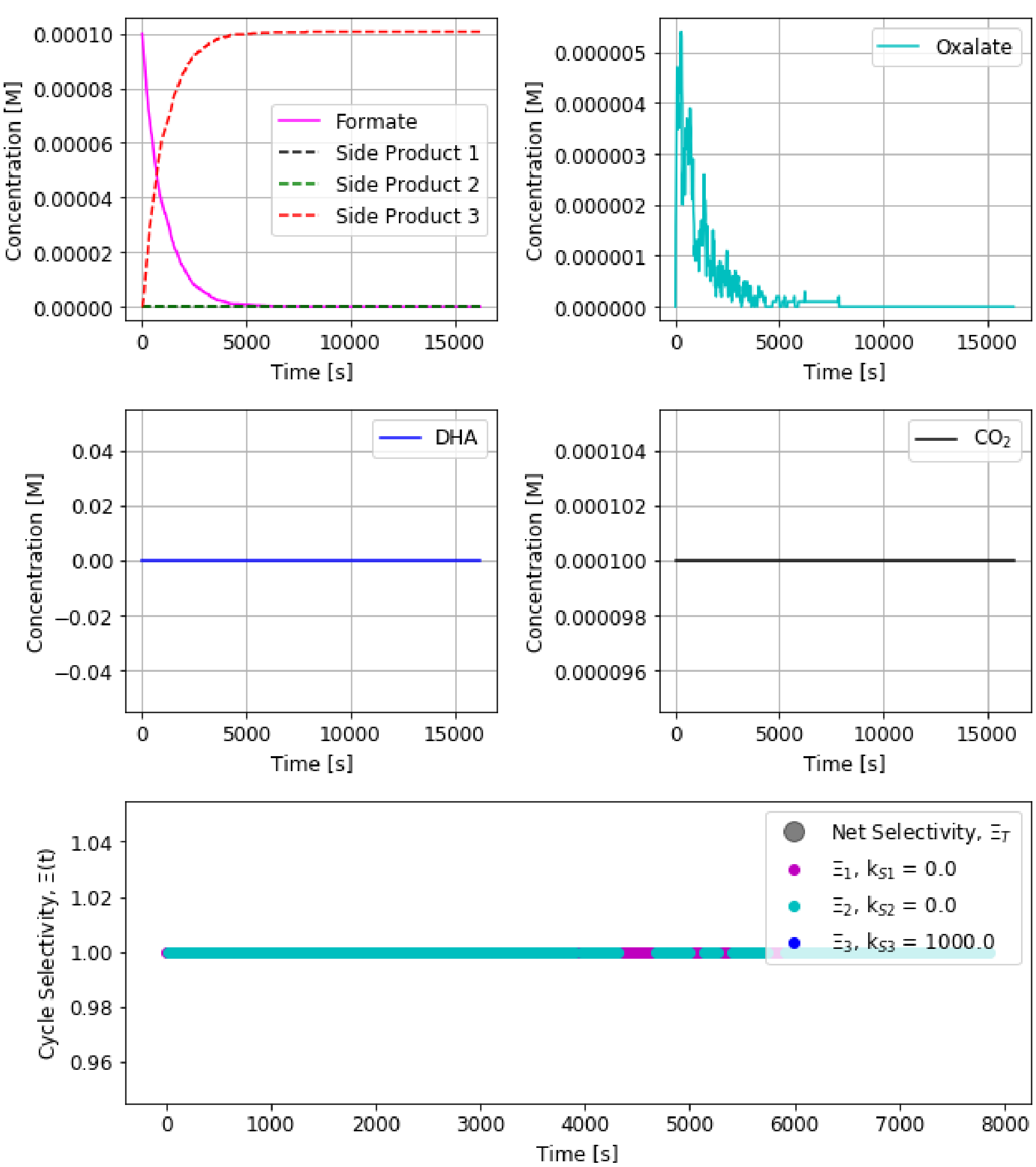

**Figure S17 – Basic System**

$k_{s1} = 0.0$ s$^{-1}$, $k_{s2} = 0.0$ s$^{-1}$, $k_{s3} = 1000.0$ s$^{-1}$

Influence of side reactions in step 3 (formate carboxylation). Side reactions are treated as a generic reaction following the rate law $r_{S3} = k_{S3}$[DHA]. Net selectivity is undefined for times where replication has been arrested, and subsequent reactions only result in the consumption of formate.

## S.6 Effects of Batch $CO_2$ supply

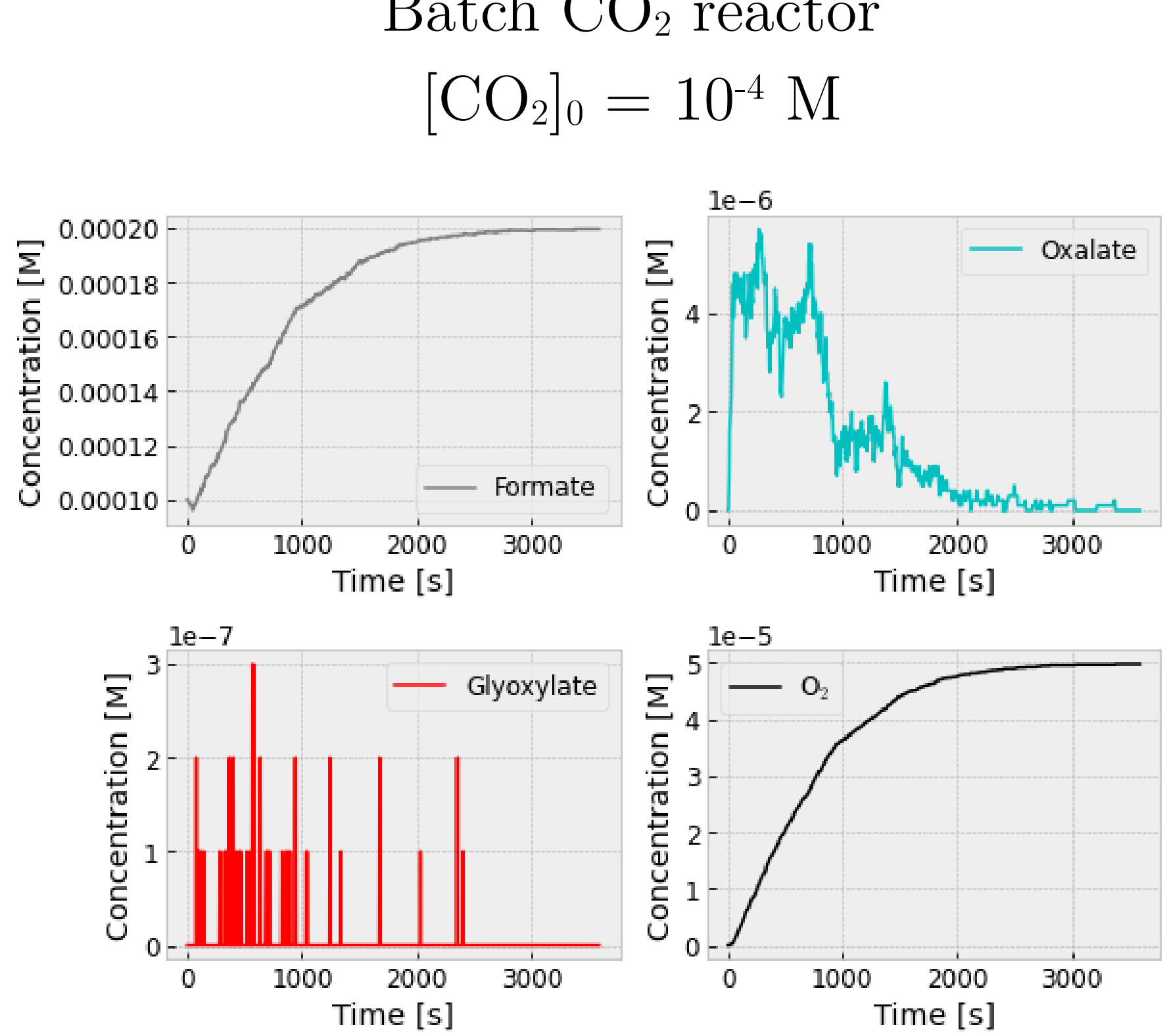

**Figure S18 – Basic System**

Influence of a batch $CO_2$ supply on autocatalysis. Initial formate and $CO_2$ concentrations are 100 μM.

**S.7 Rate Constant Sensitivity Analysis of Glyoxylate/DHA Bond Fission (Partial System)**

Sensitivities tested (base cases in bold):

| | |
|---|---|
| Reaction 3a, DHA + Cat ↔ Cat-DHA: | k = 10, **100**, 1000 $M^{-1}$ $s^{-1}$ |
| Reaction 3b, Cat-DHA + $OH^-$ ↔ Cat-DHA*: | k = 100, **1000**, 10000 $M^{-1}$ $s^{-1}$ |
| Reaction 3c, Cat-DHA* ↔ Cat + 2 $HCO_2^-$ *: | k = 1, **10**, 100 $s^{-1}$ |

Reaction 3a (#14): DHA + Cat ↔ [Cat-DHA]

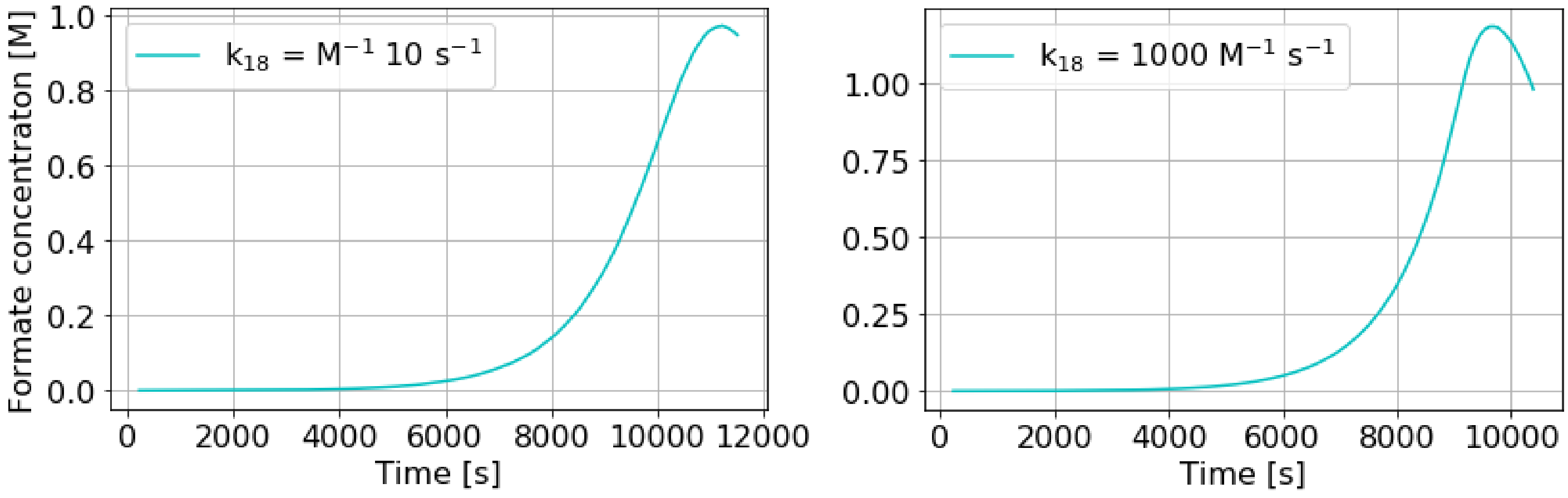

**Figure S19**

Variance of the forward rate constant for DHA binding to catalyst is shown to yield exponential growth over several orders of magnitude (base case: $k_{18}$ = 100 $M^{-1}$ $s^{-1}$), with increases in this rate constant leading to overall decreases in the characteristic time for the replication cycle. The values modeled represent very slow rates and weak equilibrium constants (reverse rate constant is k = 1 $s^{-1}$) for binding, suggesting the possibility of this replication scheme succeeding even when using a very slow and weakly-binding catalyst.

## S.7 Rate Constant Sensitivity Analysis of Glyoxylate/DHA Bond Fission (Partial System)

Reaction 3b (#15): [Cat-DHA] + $OH^-$ ↔ [Cat-DHA*]

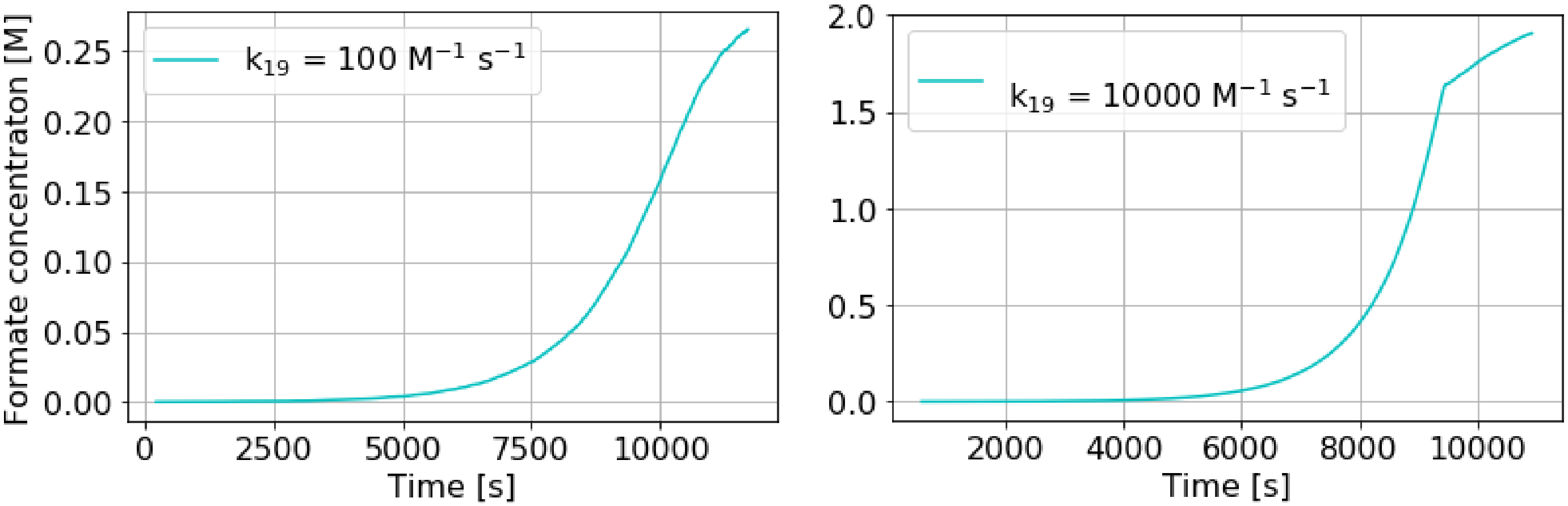

**Figure S20**

Variance of the forward rate constant for formation of the pre-fission intermediate (DHA*) binding to catalyst; base case: $k_{19}$ = 1000 $M^{-1}$ $s^{-1}$). The range of values modeled represent relatively slow rates of bimolecular reaction, nevertheless suggesting the possibility of this replication scheme succeeding even when the kinetics of this step are slow.

Reaction 3c (#16): [Cat-DHA*] → Cat + 2 $HCO_2^-$ + $H^+$

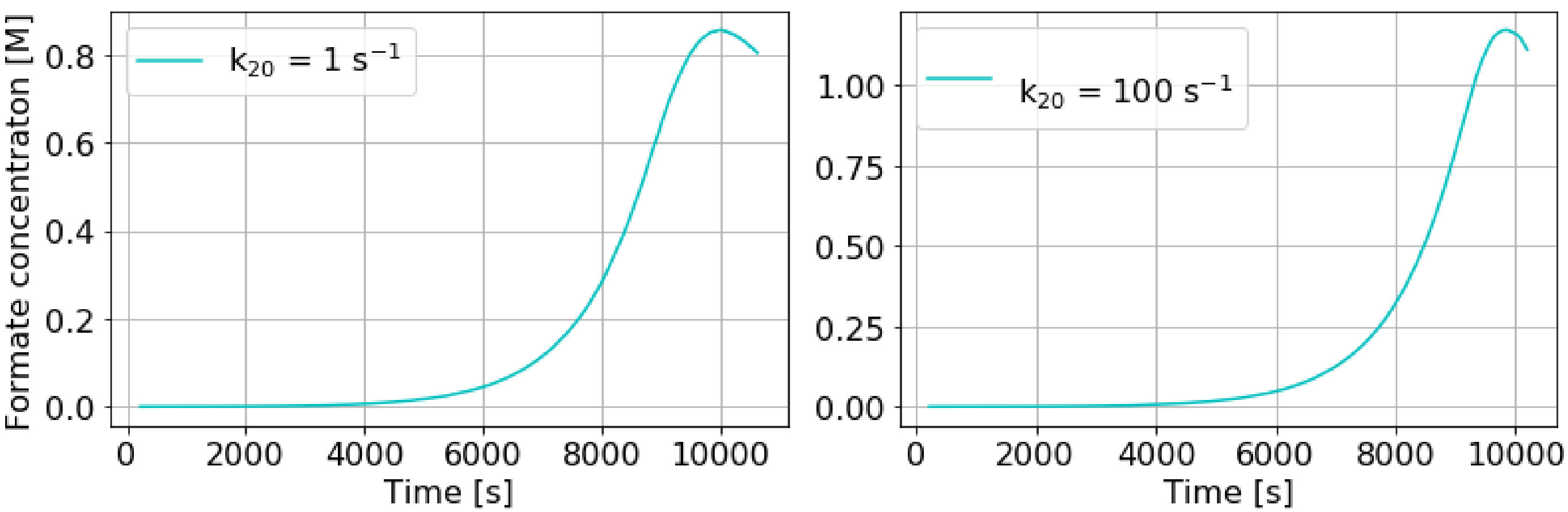

**Figure S21**

Variance of the forward rate constant for decay of DHA* into two molecules of formate; base case: $k_{20}$ = 10 $s^{-1}$). The range of values modeled represent relatively slow rates of first order decays. Despite this, the reaction still exhibits exponential growth kinetics reaction, suggesting the possibility of this replication scheme succeeding even when the rate constants for this step are small.

## S.8 Species Evolution Under VUV Pumping – Steady State Illumination (Full Mechanistic System)

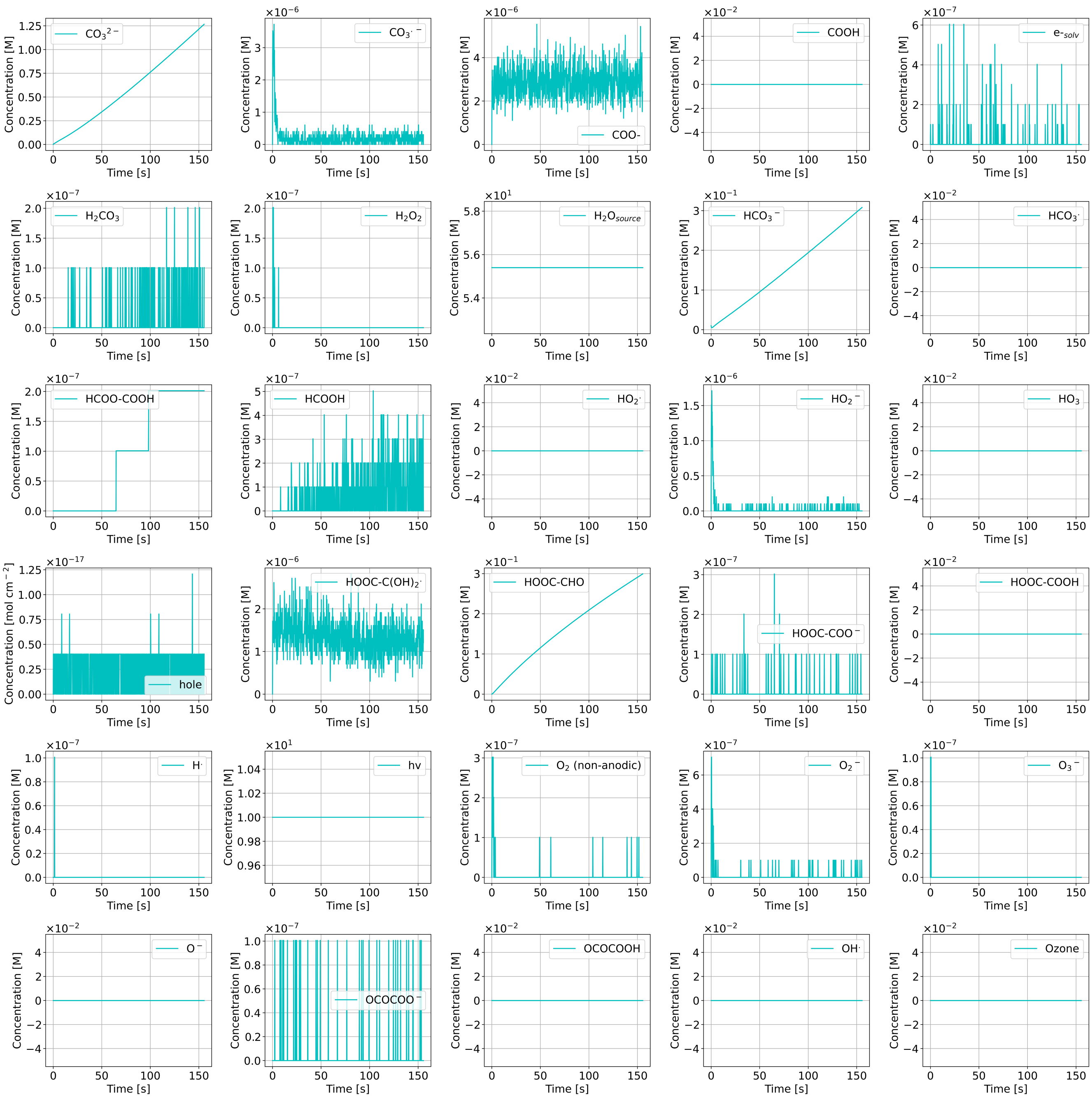

**Figure S22**

Time evolution of additional species in the Full System under steady-state VUV illumination (remaining species of Figure 6). Behavior of the system with a mechanistic description for all steps, coupled to anodic OER. $[\text{Formate}]_0$ = 100 μM; $[CO_2]$ is held constant at 100 μM. The cycle now incorporates VUV-pumped radical mechanisms reported to drive net formate carboxylation to oxalate (step 1).

### S.8 Species Evolution Under VUV Pumping – Transient Illumination (Full Mechanistic System)

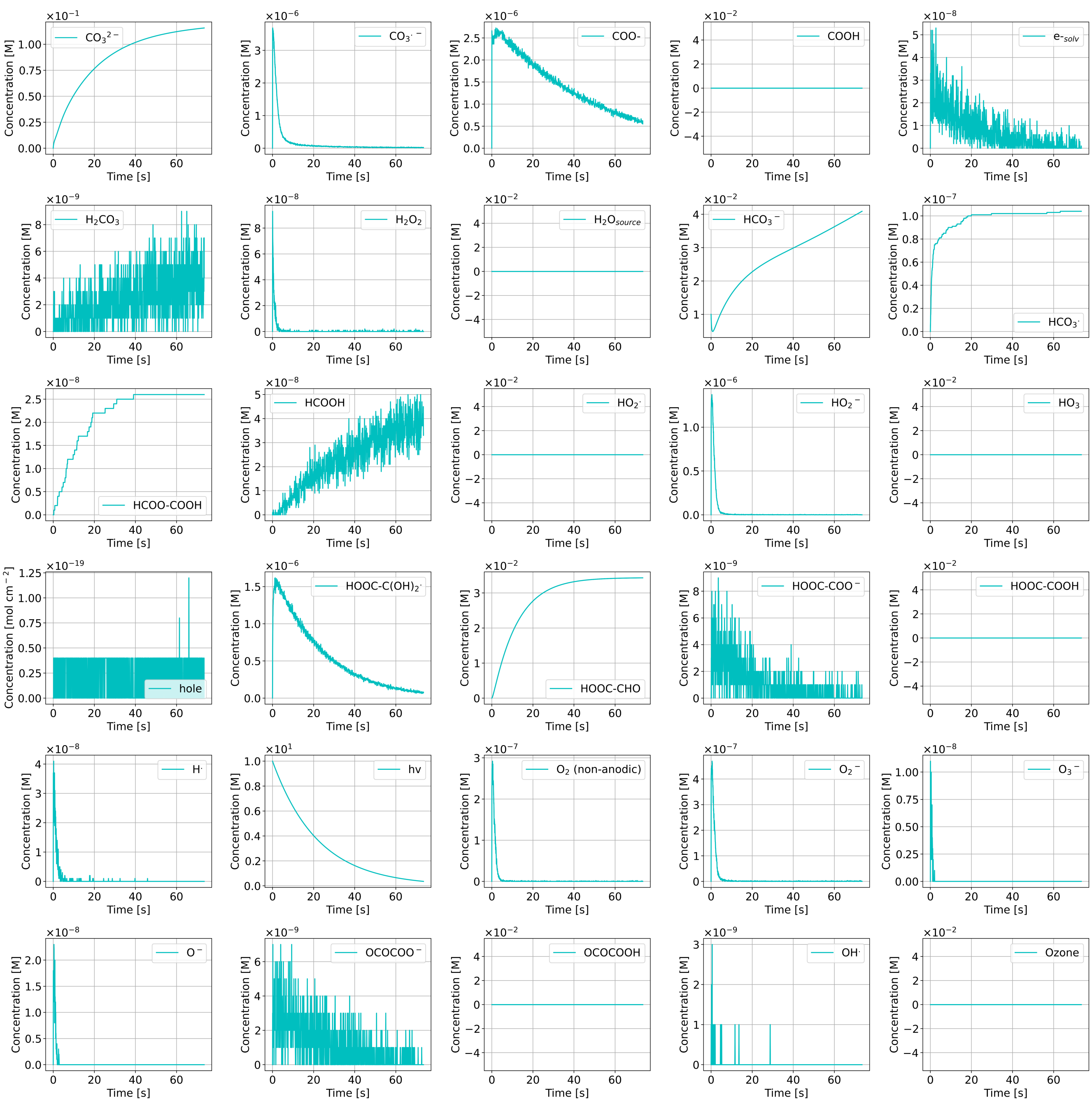

**Figure S23**

Time evolution of additional species in the Full System modeled under transient VUV illumination (remaining species of Figure 7). Behavior of the system with a mechanistic description for all steps, coupled to anodic OER. $[\text{Formate}]_0 = 100\ \mu\text{M}$; $[CO_2]$ is held constant at 100 μM. The cycle now incorporates VUV-pumped radical mechanisms reported to drive net formate carboxylation to oxalate (step 1).

### S.9 Energy Efficiency of Full Mechanistic System - Steady-State and Transient VUV Pumping

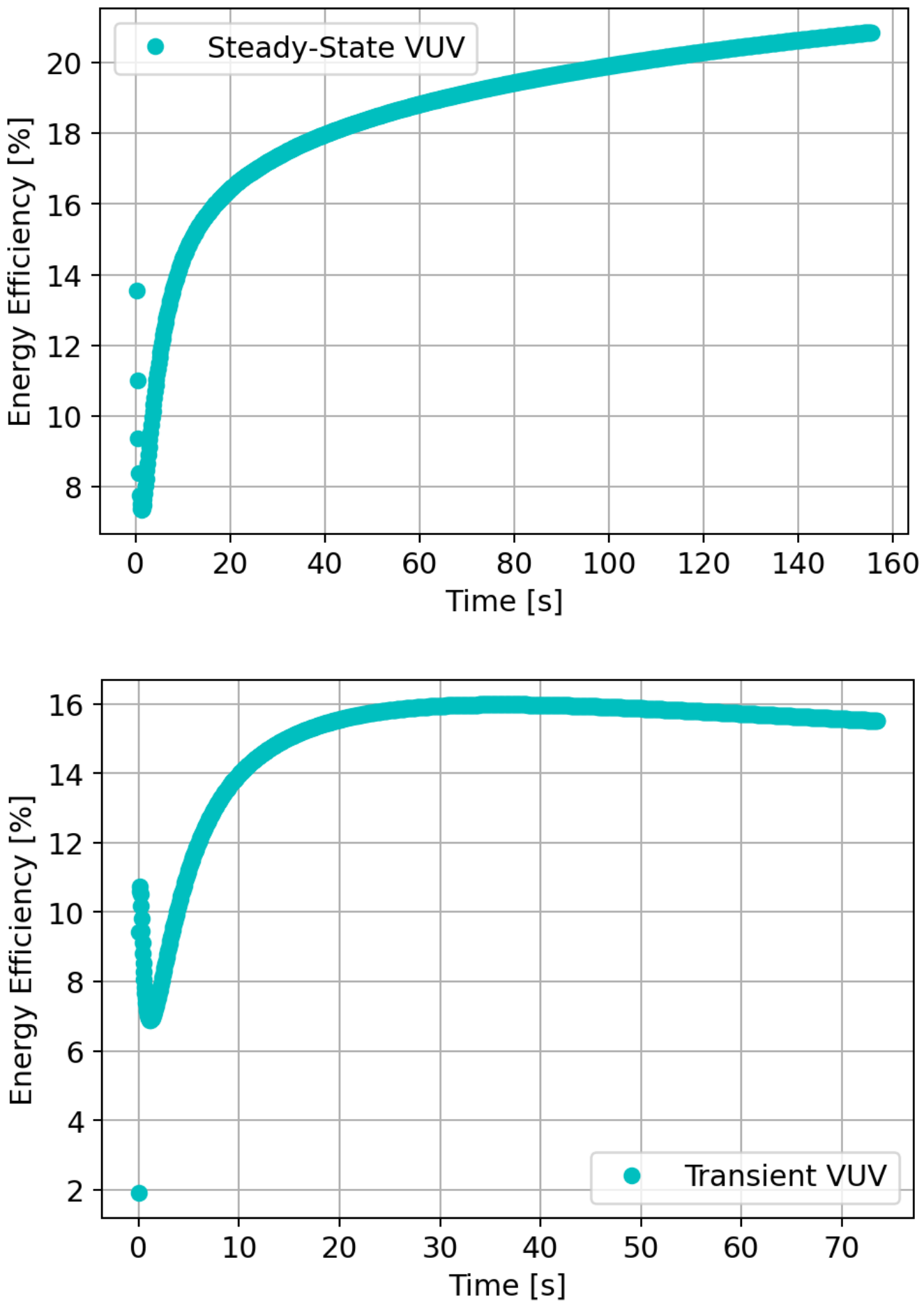

**Figure S24**

Energy efficiency of autocatalytic fuel conversion under VUV illumination at 124 nm, in the case of steady-state (top) and transient (bottom) illumination. The second point of energy input into the cycle, electrochemical conversion of oxalate to glyoxylate/dihydroxyacetate, is treated as an electrochemical electrolyzer operating at 1.575 V (representing a ~ 300 mV overpotential for oxalate reduction to glyoxylate coupled to water oxidation). Oxalate electrolysis is at steady-state for both illumination cases. Efficiencies are calculated using the free energies of combustion of the main fuel products of the cycle, formate, oxalate, glyoxylate, and hydrogen. The low efficiencies are primarily driven by the high energy photons and their high intensity required for sustaining the formate carboxylation radical reactions used in the fuel replication scheme. Photon concentrations used in these simulations (10 M) correspond to an illumination intensity $n$ = of 0.046 mol photons $cm^{-2}$. The rate constant of photon injection ($k_{h\nu}$ = 4.5x$10^{-4}$ $s^{-1}$) sets the initial photon flux (2.07x$10^{-5}$ $mol^{-1}$ $cm^{-2}$ $s^{-1}$) at this illumination intensity. Using these values, The VUV illumination power density, P, is calculated according to:

$$P = nN_A k_{h\nu} \frac{hc}{\lambda} \quad \text{(w)},$$

Where h is Plank's constant and c is the speed of light. For λ = 124 nm, this yields an illumination source power density of 20.2 W $cm^{-2}$.

## S.10 Arrhenius Extrapolation of Measured Formic Acid Decay

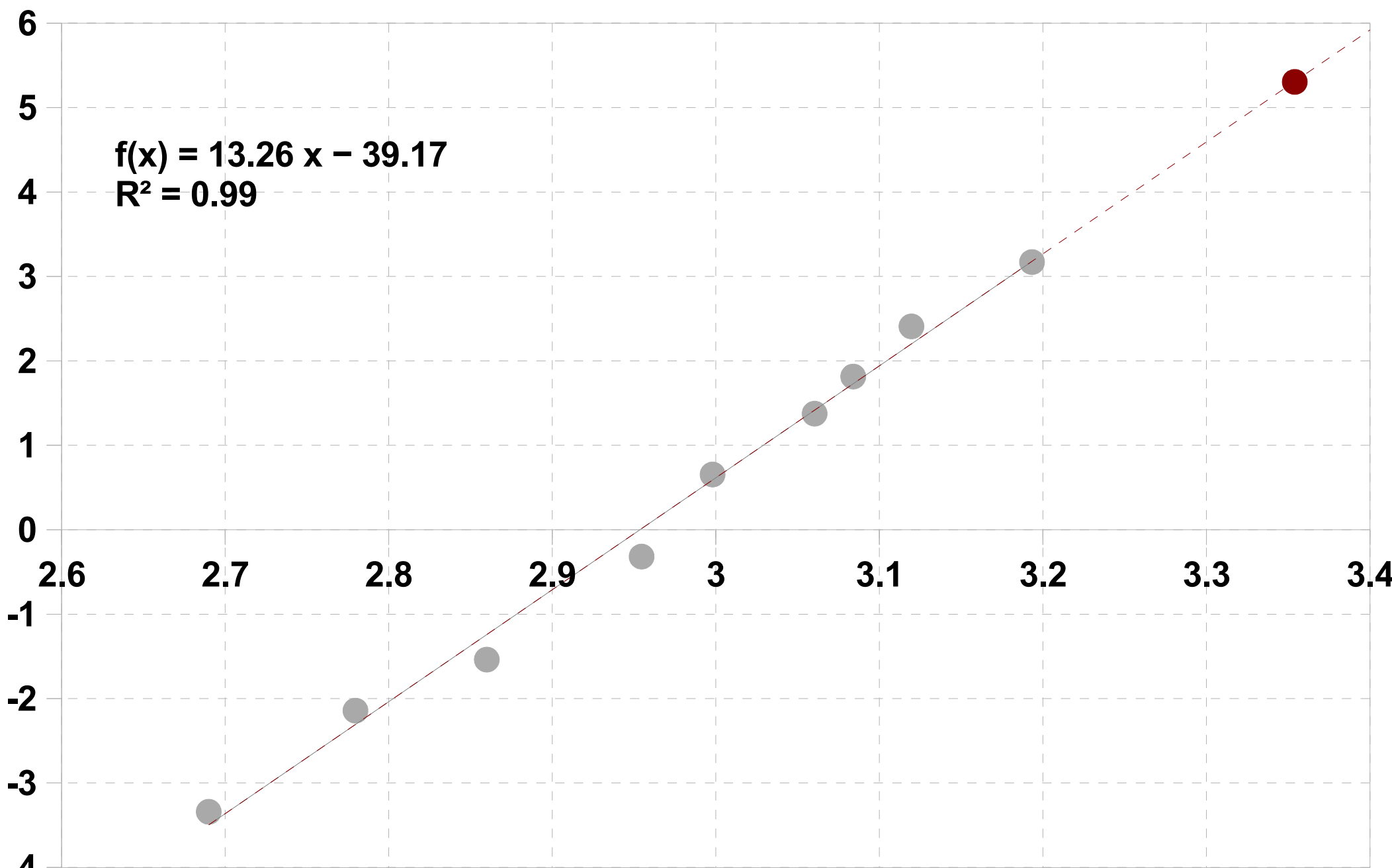

**Figure S25**

Arrhenius plot of measured formic acid decay into CO and $H_2O$ between 40 and 98.6 °C, adapted from Barham et al[17]. Original data are shown as gray dots, with the linear fit shown as a dark line. Linear extrapolation of the Arrhenius relationship to 25 °C is shown as a red dot ($-\ln[k] = 23$; $k_d = 8.3\text{x}10^{-11}$ $s^{-1}$).

## S.11 Glyoxylate Hydrolysis Catalyst Characterization - HPLC

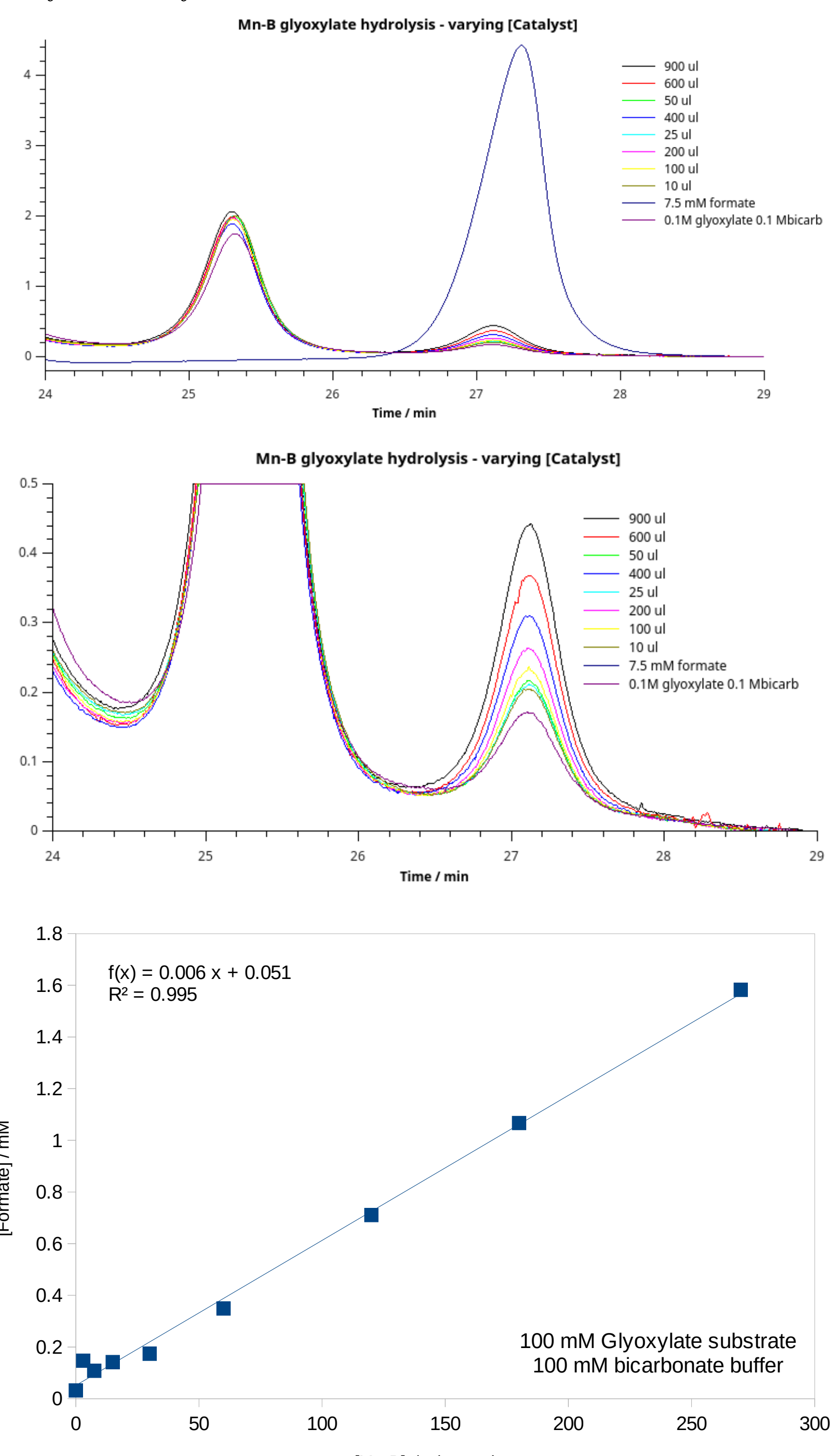

**Figure S26**

HPLC analysis of mixtures of Mn(Tetraazacyclotetradecane) (Mn-B) complex in a solution of 0.1 M glyoxylate in 0.1 M bicarbonate. Extent of formate evolution displays a direct correlation with the amount of Mn-B catalyst added to the glyoxylate solution, suggestive of Mn-B catalyzing the hydrolytic decomposition of glyoxylate into formate. Future work will characterize the kinetics of this reaction and resolve the stoichiometry of formate evolved with respect to hydrolyzed glyoxylate, as it is also possible that reactions in which glyoyxlate is decomposed into 1 equivalent of $CO_2$ and formate each are occuring.

**S.11 Glyoxylate Hydrolysis Catalyst Characterization - UV-Vis**

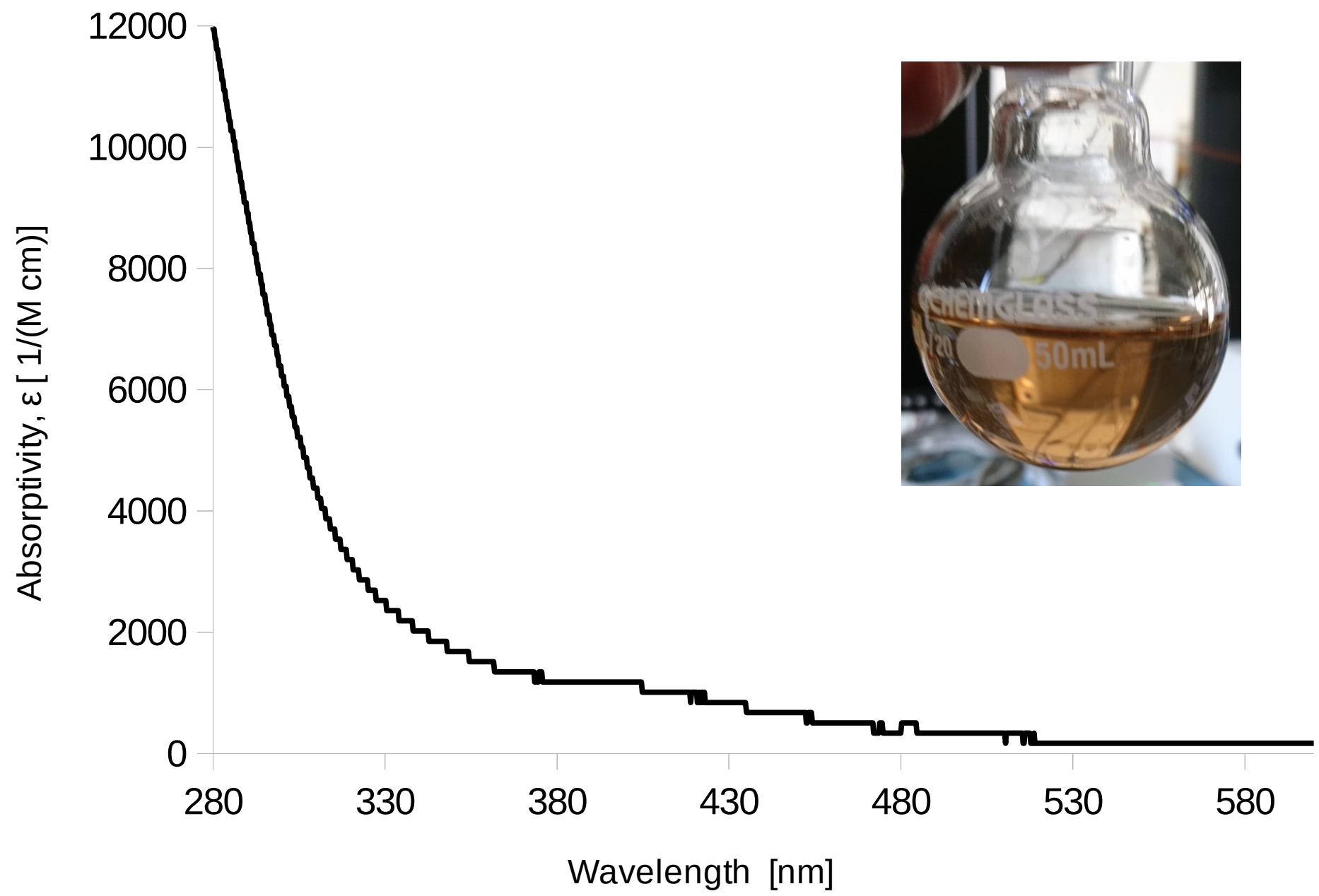

**Figure S27**
UV-Vis absorption spectroscopy of a 330 μM solution of Mn(Tetraazacyclotetradecane) complex.

#### S.12 Putative CO Replication Scheme

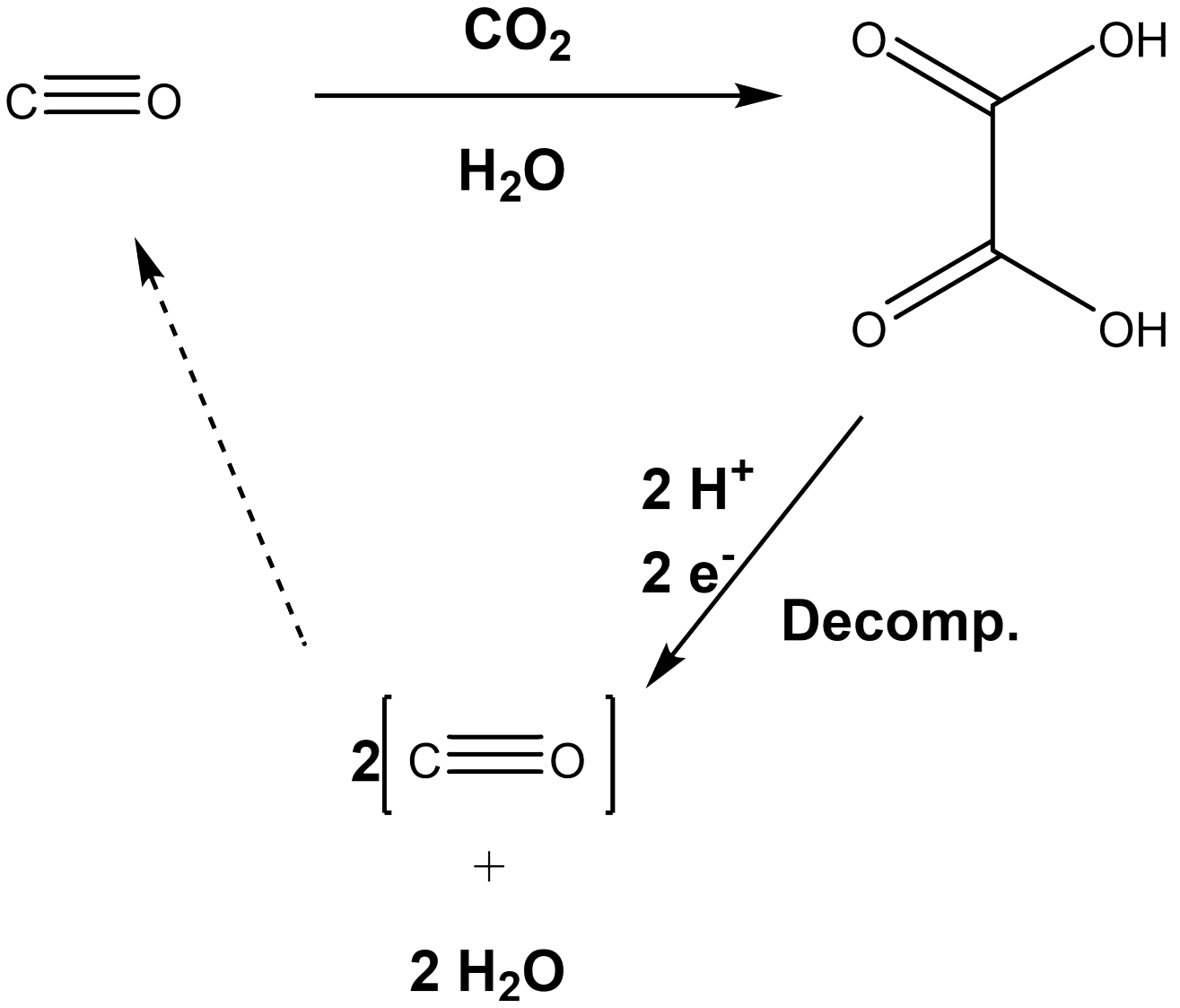

**Figure S28**
The scheme below represents an example of a potential synthetic route for autocatalytic generation of carbon monoxide, with an exponent base of 2. The reaction as shown is for reactions occurring in acid/neutral conditions, rather than base. Ultimately, a process to yield 2 CO + 2 $H_2O$ through the pathway shown will likely proceed through some metastable intermediate, which is implicit in the unspecified decomposition pathway here. Reaction stoichiometries are only reflected here; mechanistic pathways for the efficient decomposition of oxalate/oxalic acid into CO and $H_2O$ would have to be determined for implementing such a cycle.

### S.13 Putative Syngas Replication Scheme

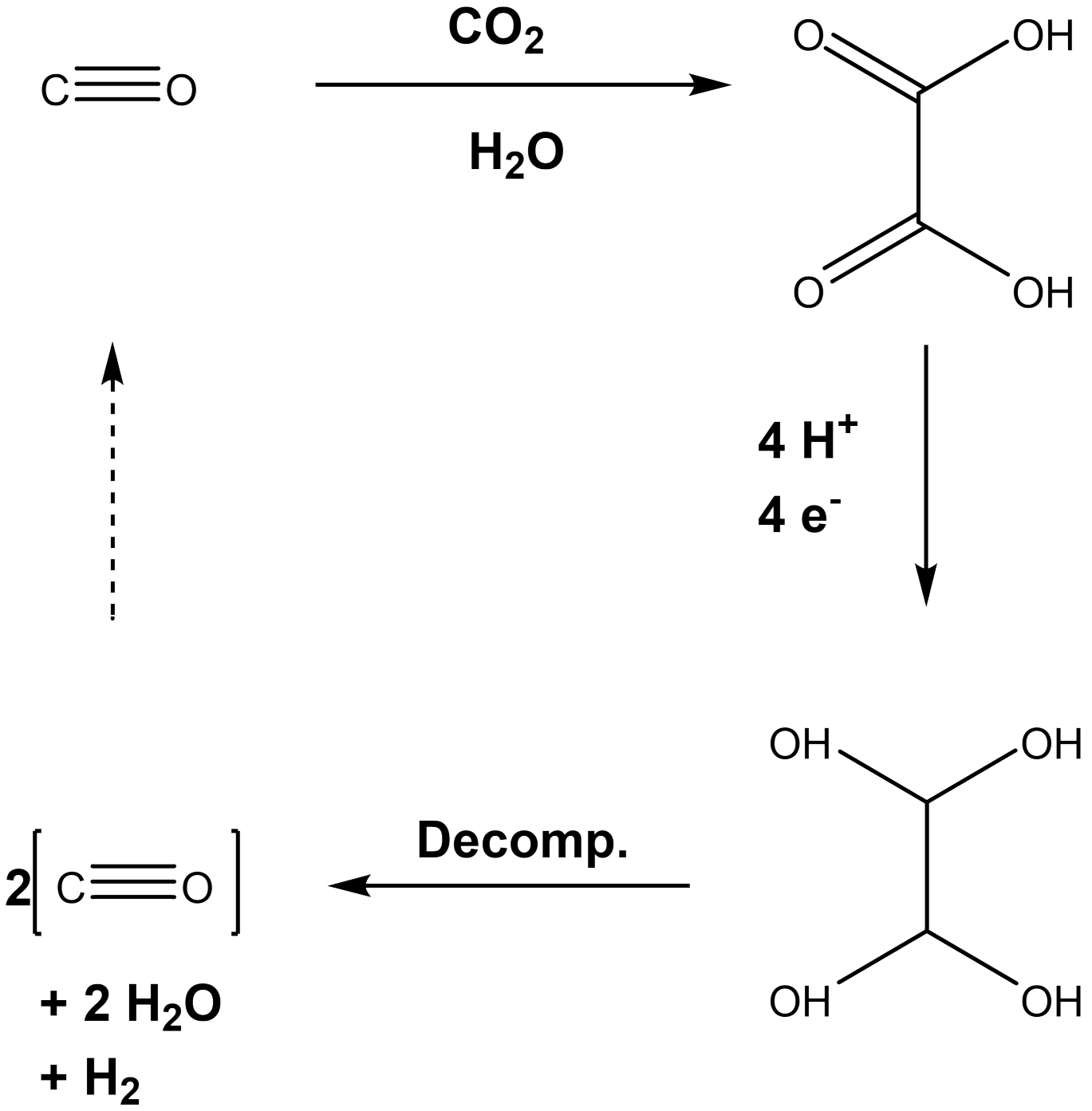

**Figure S29**

The scheme below represents an example of a potential synthetic route for autocatalytic generation of syngas, with a 2:1 CO:$H_2$ ratio yielded at the termination of each cycle (exponent base of 2). The reaction as shown is for reactions occurring in acid/neutral conditions, rather than base. Decomposition of 1,1,2,2-tetrayhydroxyethane (THE) to yield 2 CO + $H_2$ + 2 $H_2O$ will be required. However the details of how to achieve uniform reduction of oxalate to THE and even more challenging, uniform decomposition of THE to CO, $H_2$ and $H_2O$, would require further investigation for implementing this cycle.

## S.14 Putative Methanol Replication Scheme

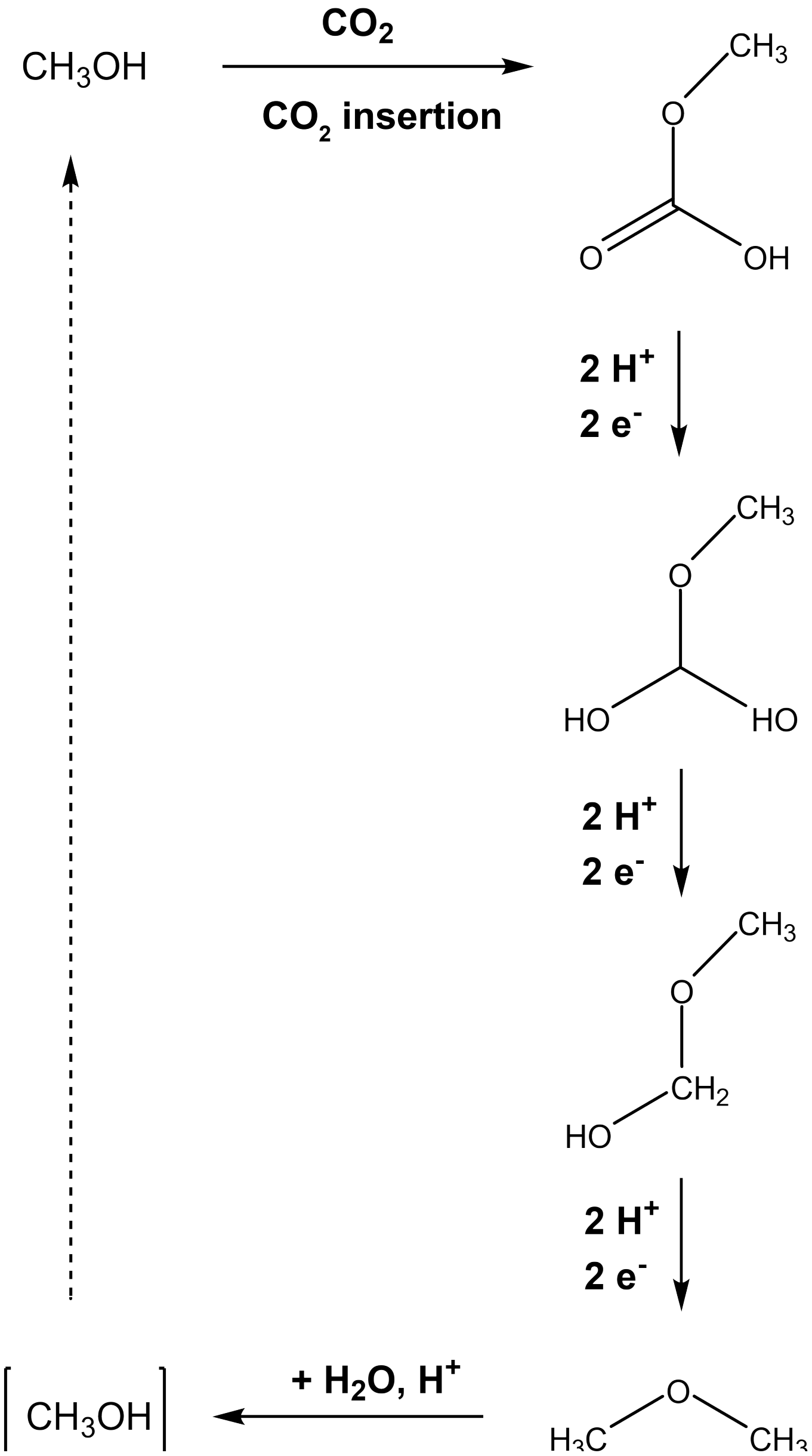

**Figure S30**
The scheme below represents an example of a potential synthetic route for autocatalytic generation of methanol (exponent base of 2). As a result of methanol being a far more reduced product than CO or formate, the scheme requires three successive, 2-electron reductions before generating an intermediate (dimethyl ether) of the oxidation state and molecular symmetry needed to yield two methanol molecules upon hydrolytic cleavage of the ether bond[18,19] (acidic conditions shown here). The significant number of steps required in this instance would likely make autocatalytic methanol production far more challenging. Interestingly, depending on the relative rate constants of the cycle, as demonstrated for formate autocatalysis, selection of dimethyl ether (a potential diesel replacement) as a target product, would also be possible, provided that appropriate reaction rates for the individual steps could be realized.

**S.15. Kinetiscope Simulation Setup and Parameters – Basic System**

```
System5_basic_reactionscheme_03062024+polyprotic_buffer_100uM_CO2.rxn

Time Units: sec
Temperature Units: °K
Energy Units: kJ
Concentration Units: mole/liter
Pressure Units: atm
Length Units: m

Temperature Option: Constant Temperature
Pressure Option: Constant Pressure
Volume Option: Constant Volume
Temperature: 293.15

This Reaction Scheme Contains the Following Compartments:

Compartment Name: single

This Compartment Contains the Following Reaction Steps:

Equation: Formate + CO2 + H2O => Oxalate + CO2 + H3O
Forward k: 10
This Reaction uses a Nonstandard Rate Law

Equation: Buffer + H3O <=> BufferH1 + H2O
Forward k: 1.0e8
Reverse k: 1.0
This Reaction uses a Nonstandard Rate Law

Equation: BufferH1 + H3O <=> BufferH2 + H2O
Forward k: 1.0e13
Reverse k: 1.0
This Reaction uses a Nonstandard Rate Law

Equation: BufferH1 + OH <=> Buffer + H2O
Forward k: 1.0e6
Reverse k: 1.0
This Reaction uses a Nonstandard Rate Law

Equation: BufferH2 + OH <=> BufferH1 + H2O
Forward k: 1.0e1
Reverse k: 1.0
This Reaction uses a Nonstandard Rate Law

Equation: H3O + OH <=> 2 H2O
Forward k: 1.12e11
Reverse k: 1e-3
This Reaction uses a Nonstandard Rate Law

Equation: Formate => side_product_0
Forward k: 1e-30
This Reaction uses a Standard Rate Law

Equation: 2 Oxalate + 4 H2O => 2 DHA + 2 OH + O2
```

Forward k: 0.01
This Reaction uses a Nonstandard Rate Law

Equation: Oxalate  => side_product_1
Forward k: 1e-30
This Reaction uses a Standard Rate Law

Equation: OH + DHA => 2 Formate + H2O
Forward k: 1000
This Reaction uses a Nonstandard Rate Law

Equation: DHA => side_product_2
Forward k: 1e-30
This Reaction uses a Standard Rate Law

This Compartment Uses the Following Species:

Buffer (initial concentration  1)
BufferH1 (initial concentration  0)
BufferH2 (initial concentration  0)
CO2 (initial concentration  1e-4)
DHA (initial concentration  0)
Formate (initial concentration  1e-4)
H2O (initial concentration  55.4)
H3O (initial concentration  1e-9)
O2 (initial concentration  0)
OH (initial concentration  1e-5)
Oxalate (initial concentration  0)
side_product_0 (initial concentration  0)
side_product_1 (initial concentration  0)
side_product_2 (initial concentration  0)

* * *

Basic System with phosphate and carbonate buffering equilibria

System5_basic_reactionscheme_082724+phosphate+carbonate_buffer_100uM_CO2.rxn

Time Units: sec
Temperature Units: °K
Energy Units: kJ
Concentration Units: mole/liter
Pressure Units: atm
Length Units: m

Temperature Option: Constant Temperature
Pressure Option: Constant Pressure
Volume Option: Constant Volume
Temperature: 293.15

This Reaction Scheme Contains the Following Compartments:

Compartment Name: single

This Compartment Contains the Following Reaction Steps:

Equation: Formate + CO2 + H2O => Oxalate + CO2 + H3O

```
Forward k: 10
This Reaction uses a Nonstandard Rate Law

Equation: Buffer + H3O <=> BufferH1 + H2O
Forward k: 1.58e7
Reverse k: 1.0
This Reaction uses a Nonstandard Rate Law

Equation: BufferH1 + H3O <=> BufferH2 + H2O
Forward k: 2.38e12
Reverse k: 1.0
This Reaction uses a Nonstandard Rate Law

Equation: BufferH1 + OH <=> Buffer + H2O
Forward k: 6.34e6
Reverse k: 1.0
This Reaction uses a Nonstandard Rate Law

Equation: BufferH2 + OH <=> BufferH1 + H2O
Forward k: 4.21e1
Reverse k: 1.0
This Reaction uses a Nonstandard Rate Law

Equation: CO2 + H2O <=> H2CO3 + CO2
Forward k: 0.04
Reverse k: 12
This Reaction uses a Nonstandard Rate Law

Equation: CO2 + OH <=> HCO3 + CO2
Forward k: 12.1e3
Reverse k: 40e-5
This Reaction uses a Nonstandard Rate Law

Equation: H2CO3 <=> HCO3 + H3O
Forward k: 1e7
Reverse k: 5e10
This Reaction uses a Standard Rate Law

Equation: HCO3 <=> CO32 + H3O
Forward k: 3
Reverse k: 5e10
This Reaction uses a Nonstandard Rate Law

Equation: H3O + OH <=> 2 H2O
Forward k: 1.12e11
Reverse k: 1e-3
This Reaction uses a Nonstandard Rate Law

Equation: Formate => side_product_0
Forward k: 1e-30
This Reaction uses a Standard Rate Law

Equation: 2 Oxalate + 4 H2O => 2 DHA + 2 OH + O2
Forward k: 0.01
This Reaction uses a Nonstandard Rate Law

Equation: Oxalate  => side_product_1
```

```
Forward k: 1e-30
This Reaction uses a Standard Rate Law

Equation: OH + DHA => 2 Formate + H2O
Forward k: 1000
This Reaction uses a Nonstandard Rate Law

Equation: DHA => side_product_2
Forward k: 1e-30
This Reaction uses a Standard Rate Law

This Compartment Uses the Following Species:

Buffer (initial concentration  1)
BufferH1 (initial concentration  0)
BufferH2 (initial concentration  0)
CO2 (initial concentration  1e-4)
CO32 (initial concentration  0)
DHA (initial concentration  0)
Formate (initial concentration  1e-4)
H2CO3 (initial concentration  0)
H2O (initial concentration  55.4)
H3O (initial concentration  1e-9)
HCO3 (initial concentration  0.01)
O2 (initial concentration  0)
OH (initial concentration  1e-5)
Oxalate (initial concentration  0)
side_product_0 (initial concentration  0)
side_product_1 (initial concentration  0)
side_product_2 (initial concentration  0)
```

### S.16 Simulation 2 – Partial System – Calculation Setup

```
Reaction Scheme Name:
09032024_Partial_mechanistic_scheme_Final[base_model]_(100uM_formate+CO2+25mM_catal
yst)_phosphate+carbonate_buffering_Final.rxn

Time Units: sec
Temperature Units: °K
Energy Units: kJ
Concentration Units: mole/liter
Pressure Units: atm
Length Units: m

Temperature Option: Constant Temperature
Pressure Option: Constant Pressure
Volume Option: Constant Volume
Temperature: 293.15

This Reaction Scheme Contains the Following Compartments:

Compartment Name: single

This Compartment Contains the Following Reaction Steps:

Equation: Formate + CO2 + H2O => Oxalate + CO2 + H3O
Forward k: 10
This Reaction uses a Nonstandard Rate Law

Equation: Buffer + H3O <=> BufferH1 + H2O
Forward k: 1.58e7
Reverse k: 1.0
This Reaction uses a Nonstandard Rate Law

Equation: BufferH1 + H3O <=> BufferH2 + H2O
Forward k: 2.38e12
Reverse k: 1.0
This Reaction uses a Nonstandard Rate Law

Equation: BufferH1 + OH <=> Buffer + H2O
Forward k: 6.34e6
Reverse k: 1.0
This Reaction uses a Nonstandard Rate Law

Equation: BufferH2 + OH <=> BufferH1 + H2O
Forward k: 4.21e1
Reverse k: 1.0
This Reaction uses a Nonstandard Rate Law

Equation: CO2 + H2O <=> H2CO3 + CO2
Forward k: 0.04
Reverse k: 12
This Reaction uses a Nonstandard Rate Law

Equation: CO2 + OH <=> HCO3 + CO2
Forward k: 12.1e3
```

```
Reverse k: 40e-5
This Reaction uses a Nonstandard Rate Law

Equation: H2CO3 <=> HCO3 + H3O
Forward k: 1e7
Reverse k: 5e10
This Reaction uses a Standard Rate Law

Equation: HCO3 <=> CO32 + H3O
Forward k: 3
Reverse k: 5e10
This Reaction uses a Nonstandard Rate Law

Equation: H3O + OH <=> 2 H2O
Forward k: 1.12e11
Reverse k: 1.12e-3
This Reaction uses a Nonstandard Rate Law

Equation: Formate => side_product_0
Forward k: 1e-30
This Reaction uses a Standard Rate Law

Equation: Oxalate + A <=> A-Oxalate_ads
Forward k: 2.4
Reverse k: 0.8
This Reaction uses a Nonstandard Rate Law

Equation: A-Oxalate_ads + 2 H2O <=> A-GLY_ads + 3 OH + 2 hole
Forward k: 215
Reverse k: 0.002
This Reaction uses a Nonstandard Rate Law

Equation: A-GLY_ads <=> GLY + A
Forward k: 0.8
Reverse k: 2.4
This Reaction uses a Nonstandard Rate Law

Equation: GLY + H2O <=> DHA
Forward k: 70
Reverse k: 1.14
This Reaction uses a Nonstandard Rate Law

Equation: 4 OH + 4 hole <=> 2 H2O + O2
Forward k: 215
Reverse k: 0.002
This Reaction uses a Nonstandard Rate Law

Equation: Oxalate  => side_product_1
Forward k: 1e-30
This Reaction uses a Standard Rate Law

Equation: DHA + Cat <=> Cat-DHA
Forward k: 100
Reverse k: 1.0
This Reaction uses a Standard Rate Law

Equation: Cat-DHA + OH <=> Cat-PFI
```

```
Forward k: 1000
Reverse k: 1.0
This Reaction uses a Standard Rate Law

Equation: Cat-PFI => 2 Formate + H2O + Cat
Forward k: 10
This Reaction uses a Standard Rate Law

Equation: DHA => side_product_2
Forward k: 1e-30
This Reaction uses a Standard Rate Law

This Compartment Uses the Following Species:

A (initial concentration  8300)
A-GLY_ads (initial concentration  0)
A-Oxalate_ads (initial concentration  0)
Buffer (initial concentration  1)
BufferH1 (initial concentration  0)
BufferH2 (initial concentration  0)
Cat (initial concentration  2.5e-2)
Cat-DHA (initial concentration  0)
Cat-PFI (initial concentration  0)
CO2 (initial concentration  1e-4)
CO32 (initial concentration  0)
DHA (initial concentration  0)
Formate (initial concentration  1e-4)
GLY (initial concentration  0)
H2CO3 (initial concentration  0)
H2O (initial concentration  55.4)
H3O (initial concentration  1e-9)
HCO3 (initial concentration  0.01)
hole (initial concentration  0)
O2 (initial concentration  0)
OH (initial concentration  1e-5)
Oxalate (initial concentration  0)
side_product_0 (initial concentration  0)
side_product_1 (initial concentration  0)
side_product_2 (initial concentration  0)

This Reaction Scheme Contains the Following Species/Property Definitions:

Species Name: A
Species Name: A-GLY_ads
Species Name: A-Oxalate_ads
Species Name: Buffer
Species Name: BufferH1
Species Name: BufferH2
Species Name: Cat
Species Name: Cat-DHA
Species Name: Cat-PFI
Species Name: CO2
Species Name: CO32
Species Name: DHA
Species Name: Formate
Species Name: GLY
```

```
Species Name: H2CO3
Species Name: H2O
Species Name: H3O
Species Name: HCO3
Species Name: hole
Species Name: O2
Species Name: OH
Species Name: Oxalate
Species Name: side_product_0
Species Name: side_product_1
Species Name: side_product_2
```

### S.17 Simulation 3 – Full System (Steady-state) – Calculation Setup

```
Reaction Scheme Name:
Full_mechanistic_scheme_082224+Pastina_model_VUV_initiation(124nm)_100uM_CO2_v2_ste
ady-state_light+H2O_phosphate+carbonate_buffering.rxn

Time Units: sec
Temperature Units: °K
Energy Units: kJ
Concentration Units: mole/liter
Pressure Units: atm
Length Units: m

Temperature Option: Constant Temperature
Pressure Option: Constant Pressure
Volume Option: Constant Volume
Temperature: 293.15

This Reaction Scheme Contains the Following Compartments:

Compartment Name: single

This Compartment Contains the Following Reaction Steps:

Equation: CO2g <=> CO2 + CO2g
Forward k: 7.6e5
Reverse k: 1.0e4
This Reaction uses a Nonstandard Rate Law

Equation: H2O_source <=> H2O + H2O_source
Forward k: 1.0
Reverse k: 1.0
This Reaction uses a Nonstandard Rate Law

Equation: H2O + hv => Hrad + OHrad + hv
Forward k: 0.045e-2
This Reaction uses a Nonstandard Rate Law

Equation: 6 H2O + 100 hv => 6 esolv + 100 hv
Forward k: 0.045e-2
This Reaction uses a Nonstandard Rate Law

Equation: Formate + Hrad  => COO + H2
Forward k: 2.2e8
This Reaction uses a Standard Rate Law

Equation: Formate + OHrad  => COO + H2O
Forward k: 2.5e9
This Reaction uses a Nonstandard Rate Law

Equation: Formate + esolv + H2O  => COO + H2 + OH
Forward k: 1.0e6
This Reaction uses a Nonstandard Rate Law

Equation: CO2 + OHrad  => HCO3rad
Forward k: 1.0e6
This Reaction uses a Standard Rate Law
```

```
Equation: CO2 + Hrad  => COOH
Forward k: 8.0e6
This Reaction uses a Standard Rate Law

Equation: CO2 + esolv  => COO
Forward k: 7.7e9
This Reaction uses a Standard Rate Law

Equation: 2 COO => Oxalate
Forward k: 1.0e9
This Reaction uses a Standard Rate Law

Equation: Buffer + H3O <=> BufferH1 + H2O
Forward k: 1.58e7
Reverse k: 1.0
This Reaction uses a Nonstandard Rate Law

Equation: BufferH1 + H3O <=> BufferH2 + H2O
Forward k: 2.38e12
Reverse k: 1.0
This Reaction uses a Nonstandard Rate Law

Equation: BufferH1 + OH <=> Buffer + H2O
Forward k: 6.34e6
Reverse k: 1.0
This Reaction uses a Nonstandard Rate Law

Equation: BufferH2 + OH <=> BufferH1 + H2O
Forward k: 4.21e1
Reverse k: 1.0
This Reaction uses a Nonstandard Rate Law

Equation: H3O + OH <=> 2 H2O
Forward k: 1.12e11
Reverse k: 1.12e-3
This Reaction uses a Nonstandard Rate Law

Equation: Formate => side_product_0
Forward k: 1e-30
This Reaction uses a Standard Rate Law

Equation: Oxalate + A <=> A-Oxalate_ads
Forward k: 2.4
Reverse k: 0.8
This Reaction uses a Nonstandard Rate Law

Equation: A-Oxalate_ads + 2 H2O <=> A-GLY_ads + 3 OH + 2 hole
Forward k: 215
Reverse k: .002
This Reaction uses a Nonstandard Rate Law

Equation: A-GLY_ads <=> GLY + A
Forward k: 0.8
Reverse k: 2.4
This Reaction uses a Nonstandard Rate Law

Equation: GLY + H2O <=> DHA
```

```
Forward k: 70
Reverse k: 1.14
This Reaction uses a Nonstandard Rate Law

Equation: 4 OH + 4 hole <=> 2 H2O + O2_anode
Forward k: 215
Reverse k: .002
This Reaction uses a Nonstandard Rate Law

Equation: Oxalate  => side_product_1
Forward k: 1e-30
This Reaction uses a Standard Rate Law

Equation: DHA + Cat <=> Cat-DHA
Forward k: 100
Reverse k: 1.0
This Reaction uses a Standard Rate Law

Equation: Cat-DHA + OH <=> Cat-PFI
Forward k: 1000
Reverse k: 1.0
This Reaction uses a Standard Rate Law

Equation: Cat-PFI => 2 Formate + H2O + Cat
Forward k: 10
This Reaction uses a Standard Rate Law

Equation: DHA => side_product_2
Forward k: 1e-30
This Reaction uses a Standard Rate Law
```

*****The steps below mark the detailed VUV-initiated radical processes for driving net formate carboxylation to oxalate, taken from Domanich et al., and Pastina et al[3,4].**

```
Equation: H2O2 + H2O => H3O + HO2_minus
Forward k: 9.43e-2
This Reaction uses a Nonstandard Rate Law

Equation: H3O + HO2_minus => H2O2 + H2O
Forward k: 5.02e10
This Reaction uses a Standard Rate Law

Equation: H2O2 + OH => HO2_minus + H2O
Forward k: 1.33e10
This Reaction uses a Standard Rate Law

Equation: HO2_minus + H2O => H2O2 + OH
Forward k: 1.27e6
This Reaction uses a Nonstandard Rate Law

Equation: esolv + H2O => Hrad + OH
Forward k: 15.75
This Reaction uses a Nonstandard Rate Law

Equation: Hrad + OH => esolv + H2O
Forward k: 2.44e7
```

```
This Reaction uses a Standard Rate Law

Equation: Hrad + H2O => esolv + H3O
Forward k: 5.83
This Reaction uses a Nonstandard Rate Law

Equation: esolv + H3O => Hrad + H2O
Forward k: 2.09e10
This Reaction uses a Standard Rate Law

Equation: OHrad + OH => O_minus + H2O
Forward k: 1.33e10
This Reaction uses a Standard Rate Law

Equation: O_minus + H2O => OHrad + OH
Forward k: 1.27e6
This Reaction uses a Nonstandard Rate Law

Equation: OHrad + H2O => O_minus + H3O
Forward k: 9.43e-2
This Reaction uses a Nonstandard Rate Law

Equation: O_minus + H3O => OHrad + H2O
Forward k: 5.02e10
This Reaction uses a Standard Rate Law

Equation: HO2 + H2O => O2_minus + H3O
Forward k: 7.73e5
This Reaction uses a Nonstandard Rate Law

Equation: O2_minus + H3O => HO2 + H2O
Forward k: 5.02e10
This Reaction uses a Standard Rate Law

Equation: HO2 + OH => O2_minus + H2O
Forward k: 1.33e10
This Reaction uses a Standard Rate Law

Equation: O2_minus + H2O => HO2 + OH
Forward k: 1.55e-1
This Reaction uses a Nonstandard Rate Law

Equation: esolv + OHrad => OH
Forward k: 3.55e10
This Reaction uses a Standard Rate Law

Equation: esolv + H2O2 => OHrad + OH
Forward k: 1.36e10
This Reaction uses a Standard Rate Law

Equation: esolv  + O2_minus + H2O => HO2_minus + OH
Forward k: 1.30e10
This Reaction uses a Nonstandard Rate Law

Equation: esolv + HO2 => HO2_minus
Forward k: 1.30e10
This Reaction uses a Standard Rate Law
```

```
Equation: esolv + O2 => O2_minus
Forward k: 2.29e10
This Reaction uses a Standard Rate Law

Equation: 2 esolv + 2 H2O => H2 + 2 OH
Forward k: 7.26e9
This Reaction uses a Nonstandard Rate Law

Equation: esolv + Hrad + H2O => H2 + OH
Forward k: 2.76e10
This Reaction uses a Nonstandard Rate Law

Equation: esolv + HO2_minus => O_minus + OH
Forward k: 3.50e9
This Reaction uses a Standard Rate Law

Equation: esolv + O_minus + H2O => 2 OH
Forward k: 2.20e10
This Reaction uses a Nonstandard Rate Law

Equation: esolv + O3_minus + H2O => O2 + 2 OH
Forward k: 1.60e10
This Reaction uses a Nonstandard Rate Law

Equation: esolv + Ozone => O3_minus
Forward k: 3.60e10
This Reaction uses a Standard Rate Law

Equation: Hrad + H2O => H2 + OHrad
Forward k: 4.58e-5
This Reaction uses a Nonstandard Rate Law

Equation: Hrad + O_minus  => OH
Forward k: 1e10
This Reaction uses a Standard Rate Law

Equation: Hrad + HO2_minus => OHrad + OH
Forward k: 9.00e7
This Reaction uses a Standard Rate Law

Equation: Hrad + O3_minus => OH + O2
Forward k: 1.0e10
This Reaction uses a Standard Rate Law

Equation: 2 Hrad => H2
Forward k: 5.14e9
This Reaction uses a Standard Rate Law

Equation: Hrad + OHrad => H2O
Forward k: 1.09e10
This Reaction uses a Standard Rate Law

Equation: Hrad + H2O2 => OHrad + H2O
Forward k: 3.65e7
This Reaction uses a Standard Rate Law
```

```
Equation: Hrad + O2 => HO2
Forward k: 1.31e10
This Reaction uses a Standard Rate Law

Equation: Hrad + HO2 => H2O2
Forward k: 1.14e10
This Reaction uses a Standard Rate Law

Equation: Hrad + O2_minus => HO2_minus
Forward k: 1.14e10
This Reaction uses a Standard Rate Law

Equation: Hrad + Ozone => HO3
Forward k: 3.80e10
This Reaction uses a Standard Rate Law

Equation: 2 OHrad => H2O2
Forward k: 4.81e9
This Reaction uses a Standard Rate Law

Equation: OHrad + HO2 => H2O + O2
Forward k: 8.84e9
This Reaction uses a Standard Rate Law

Equation: OHrad + O2_minus => OH + O2
Forward k: 1.1e10
This Reaction uses a Standard Rate Law

Equation: OHrad + H2 => Hrad + H2O
Forward k: 3.95e7
This Reaction uses a Standard Rate Law

Equation: OHrad + H2O2 => HO2 + H2O
Forward k: 2.92e7
This Reaction uses a Standard Rate Law

Equation: OHrad + O_minus => HO2_minus
Forward k: 2.50e10
This Reaction uses a Standard Rate Law

Equation: OHrad + HO2_minus => HO2 + OH
Forward k: 7.5e9
This Reaction uses a Standard Rate Law

Equation: OHrad + O3_minus => Ozone + OH
Forward k: 2.60e9
This Reaction uses a Standard Rate Law

Equation: OHrad + O3_minus + H2O => 2 O2_minus + H3O
Forward k: 6.0e9
This Reaction uses a Nonstandard Rate Law

Equation: OHrad + Ozone => HO2 + O2
Forward k: 1.1e8
This Reaction uses a Standard Rate Law

Equation: HO2 + O2_minus => HO2_minus + O2
```

```
Forward k: 8.0e7
This Reaction uses a Standard Rate Law

Equation: 2 HO2 => H2O2 + O2
Forward k: 8.4e5
This Reaction uses a Standard Rate Law

Equation: HO2 + O_minus =>  O2 + OH
Forward k: 6.0e9
This Reaction uses a Standard Rate Law

Equation: HO2 + H2O2 => OHrad + O2 + H2O
Forward k: 0.5
This Reaction uses a Standard Rate Law

Equation: HO2 + HO2_minus => OHrad + O2 + OH
Forward k: 0.5
This Reaction uses a Standard Rate Law

Equation: HO2 + O3_minus => 2 O2 + OH
Forward k: 6.0e9
This Reaction uses a Standard Rate Law

Equation: HO2 + Ozone => HO3 + O2
Forward k: 5.0e8
This Reaction uses a Standard Rate Law

Equation: 2 O2_minus + 2 H2O => H2O2 + O2 + 2 OH
Forward k: 0.3
This Reaction uses a Nonstandard Rate Law

Equation: O2_minus + O_minus + H2O => O2 + 2 OH
Forward k: 6.0e8
This Reaction uses a Nonstandard Rate Law

Equation: O2_minus + H2O2 => OHrad + O2 + OH
Forward k: 0.13
This Reaction uses a Standard Rate Law

Equation: O2_minus + HO2_minus => O_minus + O2 + OH
Forward k: 0.13
This Reaction uses a Standard Rate Law

Equation: O2_minus + O3_minus + H2O => 2 O2 + 2 OH
Forward k: 1.0e4
This Reaction uses a Nonstandard Rate Law

Equation: O2_minus + Ozone => O3_minus + O2
Forward k: 1.5e9
This Reaction uses a Standard Rate Law

Equation: 2 O_minus + H2O => HO2_minus + OH
Forward k: 1.0e9
This Reaction uses a Nonstandard Rate Law

Equation: O_minus + O2 => O3_minus
Forward k: 3.75e9
```

```
This Reaction uses a Standard Rate Law

Equation: O_minus + H2 => Hrad + OH
Forward k: 1.28e8
This Reaction uses a Standard Rate Law

Equation: O_minus + H2O2 => O2_minus + H2O
Forward k: 5.0e8
This Reaction uses a Standard Rate Law

Equation: O_minus + HO2_minus => O2_minus + OH
Forward k: 7.86e8
This Reaction uses a Standard Rate Law

Equation: O_minus + O3_minus => 2 O2_minus
Forward k: 7.0e8
This Reaction uses a Standard Rate Law

Equation: O_minus + Ozone => O2_minus + O2
Forward k: 5.0e9
This Reaction uses a Standard Rate Law

Equation: O3_minus => O2 + O_minus
Forward k: 2.62e3
This Reaction uses a Standard Rate Law

Equation: O3_minus + H3O => O2 + OHrad + H2O
Forward k: 9.0e10
This Reaction uses a Standard Rate Law

Equation: HO3 => O2 + OHrad
Forward k: 1.10e5
This Reaction uses a Standard Rate Law
```

*****The steps below incorporate the use of an H2O reservoir input and reactions for carbonate buffering equilibria.**

```
Equation: CO2 + H2O <=> H2CO3 + CO2
Forward k: 0.04
Reverse k: 12
This Reaction uses a Nonstandard Rate Law

Equation: CO2 + OH <=> HCO3 + CO2
Forward k: 12.1e3
Reverse k: 40e-5
This Reaction uses a Nonstandard Rate Law

Equation: H2CO3 <=> HCO3 + H3O
Forward k: 1e7
Reverse k: 5e10
This Reaction uses a Standard Rate Law

Equation: HCO3 <=> CO3 + H3O
Forward k: 3
Reverse k: 5e10
This Reaction uses a Nonstandard Rate Law
```

***The steps below mark the detailed mechanisms for radical reactions of CO2-contaning solutions as described by Horne et al[20].

```
Equation: COO + esolv + H2O => formate + OH
Forward k: 9000000000
This Reaction uses a Nonstandard Rate Law

Equation: COO + Hrad => formate
Forward k: 9000000000
This Reaction uses a Standard Rate Law

Equation: formate + O_minus => COO + OH
Forward k: 1400000000
This Reaction uses a Standard Rate Law

Equation: COO + H2O2 => CO2 + OHrad + OH
Forward k: 200000
This Reaction uses a Standard Rate Law

Equation: COO + HCO3 => CO3rad + formate
Forward k: 20000
This Reaction uses a Standard Rate Law

Equation: COO + O2_minus => CO3 + O2
Forward k: 650000000
This Reaction uses a Standard Rate Law

Equation: H2O2 + esolv => OHrad + OH
Forward k: 12000000000
This Reaction uses a Standard Rate Law

Equation: HCOOH + H2O => formate + H3O
Forward k: 3600000
This Reaction uses a Nonstandard Rate Law

Equation: formate + H3O => HCOOH + H2O
Forward k: 20000000000
This Reaction uses a Standard Rate Law

Equation: HCOOH + esolv => formate + Hrad
Forward k: 140000000
This Reaction uses a Standard Rate Law

Equation: HCOOH + Hrad => COOH + H2
Forward k: 680000
This Reaction uses a Standard Rate Law

Equation: HCOOH + OHrad => COOH + H2O
Forward k: 140000000
This Reaction uses a Standard Rate Law

Equation: formate + O2_minus => COO + HO2_minus
Forward k: 0.01
This Reaction uses a Standard Rate Law

Equation: COOH + H2O <=>  H3O + COO
```

```
Forward k: 100000000
Reverse k: 20000000000
This Reaction uses a Nonstandard Rate Law

Equation: COO + H3O => COOH + H2O
Forward k: 20000000000
This Reaction uses a Standard Rate Law

Equation: COOH + esolv + H2O => HCOOH + OH
Forward k: 90000000000
This Reaction uses a Nonstandard Rate Law

Equation: COOH + Hrad => HCOOH
Forward k: 9000000000
This Reaction uses a Standard Rate Law

Equation: COOH + H2O2 => CO2 + OHrad + H2O
Forward k: 50000000
This Reaction uses a Standard Rate Law

Equation: COOH + COO => CO2 + formate
Forward k: 1000000000
This Reaction uses a Standard Rate Law

Equation: 2 COOH => CO2 + HCOOH
Forward k: 1700000000
This Reaction uses a Standard Rate Law

Equation: COO + esolv + H2O => formate + OH
Forward k: 9000000000
This Reaction uses a Nonstandard Rate Law

Equation: COO + Hrad => HCOOH
Forward k: 9000000000
This Reaction uses a Standard Rate Law

Equation: COO + O2 => CO2 + O2_minus
Forward k: 2400000000
This Reaction uses a Standard Rate Law

Equation: COO + H2O2 => CO2 + OHrad + OH
Forward k: 600000
This Reaction uses a Standard Rate Law

Equation: 2 COO => OCOCOO
Forward k: 7.5e8
This Reaction uses a Standard Rate Law

Equation: OCOCOO => oxalate
Forward k: 1600000
This Reaction uses a Standard Rate Law

Equation: OCOCOO + H3O => OCOCOOH + H2O
Forward k: 10000000000
This Reaction uses a Standard Rate Law

Equation: OCOCOOH + H2O => OCOCOO + H3O
```

```
Forward k: 1000000
This Reaction uses a Nonstandard Rate Law

Equation: OCOCOOH => CO2 + formate
Forward k: 10000000000
This Reaction uses a Standard Rate Law

Equation: HOOC-COOH + H2O => HOOC-COO + H3O
Forward k: 1100000000
This Reaction uses a Nonstandard Rate Law

Equation: HOOC-COO + H3O => HCOO-COOH + H2O
Forward k: 20000000000
This Reaction uses a Standard Rate Law

Equation: HOOC-COO + H2O => oxalate + H3O
Forward k: 1000000
This Reaction uses a Nonstandard Rate Law

Equation: oxalate + H3O => HOOC-COO + H2O
Forward k: 20000000000
This Reaction uses a Standard Rate Law

Equation: HOOC-COOH + OHrad + H2O => COO + CO2 + H3O
Forward k: 1400000
This Reaction uses a Nonstandard Rate Law

Equation: HOOC-COOH + esolv + H2O => HOOC-C(OH)2rad + OH
Forward k: 25000000000
This Reaction uses a Nonstandard Rate Law

Equation: HOOC-COOH + Hrad => HOOC-C(OH)2rad
Forward k: 3300000
This Reaction uses a Standard Rate Law

Equation: HOOC-COO + OHrad => COO + CO2 + H2O
Forward k: 19000000
This Reaction uses a Standard Rate Law

Equation: HOOC-COO + esolv + H2O => HOOC-C(OH)2rad + 2 OH
Forward k: 3200000000
This Reaction uses a Nonstandard Rate Law

Equation: HOOC-COO + Hrad + H2O => HOOC-C(OH)2rad + OH
Forward k: 16000
This Reaction uses a Nonstandard Rate Law

Equation: oxalate + OHrad => COO + CO2 + OH
Forward k: 970000
This Reaction uses a Standard Rate Law

Equation: oxalate + esolv + H2O => HOOC-C(OH)2rad + 3 OH
Forward k: 4.8e7
This Reaction uses a Nonstandard Rate Law

Equation: oxalate + Hrad => HOOC-C(OH)2rad + 2 OH
Forward k: 10000
```

```
This Reaction uses a Standard Rate Law

Equation: oxalate + O2_minus => COO + CO2 + HO2 + OH
Forward k: 200
This Reaction uses a Standard Rate Law

Equation: 2 HOOC-C(OH)2rad => HOOC-COOH + HOOC-CHO
Forward k: 1000000000
This Reaction uses a Standard Rate Law

Equation: HOOC-C(OH)2rad + O2 =>  HOOC-COOH + HO2
Forward k: 100000
This Reaction uses a Standard Rate Law

Equation: CO2 + OHrad + H2O => CO3rad + H3O
Forward k: 1000000
This Reaction uses a Nonstandard Rate Law

Equation: HCO3 + COO => CO3rad + formate
Forward k: 20000
This Reaction uses a Standard Rate Law

Equation: HCO3 + OHrad => CO3rad + H2O
Forward k: 8500000
This Reaction uses a Standard Rate Law

Equation: CO3 + OHrad => CO3rad + OH
Forward k: 390000000
This Reaction uses a Standard Rate Law

Equation: CO3rad + O2_minus => CO3 + O2
Forward k: 650000000
This Reaction uses a Standard Rate Law

Equation: CO3rad + OHrad => CO2 + HO2_minus
Forward k: 3000000000
This Reaction uses a Standard Rate Law

Equation: CO3rad + COO => CO3 + CO2
Forward k: 50000000
This Reaction uses a Standard Rate Law

Equation: CO3rad + HO2_minus => HCO3 + O2_minus
Forward k: 30000000
This Reaction uses a Standard Rate Law

Equation: 2 CO3rad => 2 COO + O2
Forward k: 14000000
This Reaction uses a Standard Rate Law

Equation: CO3rad + formate => HCO3 + COO
Forward k: 150000
This Reaction uses a Standard Rate Law

Equation: CO3rad + H2O2 => HCO3 + HO2_minus
Forward k: 430000
This Reaction uses a Standard Rate Law
```

**Simulation 3 – Full System (Transient Illumination, Batch $H_2O$) – Calculation Setup**

Equation: reservoir => H2O + reservoir
Forward k: 0 (M $s^{-1}$)
This Reaction uses a Standard Rate Law

Equation: H2O + hv => Hrad + OHrad
Forward k: 0.045e-2
This Reaction uses a Nonstandard Rate Law

Equation: 6 H2O + 100 hv => 6 esolv
Forward k: 0.045e-2
This Reaction uses a Nonstandard Rate Law

→ All subsequent steps are identical to the Full System, steady-state simulations.

This Reaction Scheme Contains the Following Species/Property Definitions:

A (initial concentration  8300)
A-GLY_ads (initial concentration  0)
A-oxalate_ads (initial concentration  0)
Buffer (initial concentration  1)
BufferH1 (initial concentration  0)
BufferH2 (initial concentration  0)
Cat (initial concentration  2.5e-2)
Cat-DHA (initial concentration  0)
Cat-PFI (initial concentration  0)
CO2 (initial concentration  1e-4)
CO2g (initial concentration  1.52e-6)
CO3 (initial concentration  0)
CO3rad (initial concentration  0)
COO (initial concentration  0)
COOH (initial concentration  0)
DHA (initial concentration  0)
esolv (initial concentration  0)
Formate (initial concentration  1e-4)
GLY (initial concentration  0)
H2 (initial concentration  0)
H2CO3 (initial concentration  0)
H2O (initial concentration  55.4)
H2O2 (initial concentration  0)
H2O_source (initial concentration  55.4)
H3O (initial concentration  1e-9)
HCO3 (initial concentration  0.01)
HCO3rad (initial concentration  0)
HCOO-COOH (initial concentration  0)
HCOOH (initial concentration  0)
HO2 (initial concentration  0)
HO2_minus (initial concentration  0)
HO3 (initial concentration  0)
hole (initial concentration  0)
HOOC-C(OH)2rad (initial concentration  0)
HOOC-CHO (initial concentration  0)
HOOC-COO (initial concentration  0)
HOOC-COOH (initial concentration  0)
Hrad (initial concentration  0)

```
hv (initial concentration  10)
O2 (initial concentration  0)
O2_anode (initial concentration  0)
O2_minus (initial concentration  0)
O3_minus (initial concentration  0)
O_minus (initial concentration  0)
OCOCOO (initial concentration  0)
OCOCOOH (initial concentration  0)
OH (initial concentration  1e-5)
OHrad (initial concentration  0)
Oxalate (initial concentration  0)
Ozone (initial concentration  0)
side_product_0 (initial concentration  0)
side_product_1 (initial concentration  0)
side_product_2 (initial concentration  0)
```

## S.18 Calculation of Radiolysis Rates and Rate Constants from Nuclear Decay Source G Values

| Species | 10 MeV proton G values (per 100 eV) | Proton Energy (eV) | Beam Current (A) | Species Generation Rate (mol/sec) |
|---|---|---|---|---|
| $e^{-}_{(aq)}$ | 0.9 | 1.00E+07 | 5.00E-05 | 4.66E-05 |
| $H^{\cdot}$ | 0.57 | 1.00E+07 | 5.00E-05 | 2.95E-05 |
| $H_2$ | 0.64 | 1.00E+07 | 5.00E-05 | 3.32E-05 |
| $OH^{\cdot}$ | 1.18 | 1.00E+07 | 5.00E-05 | 6.11E-05 |
| $H_2O_2$ | 0.74 | 1.00E+07 | 5.00E-05 | 3.83E-05 |
| $HO_2$ | 0.03 | 1.00E+07 | 5.00E-05 | 1.55E-06 |
| $H^{+}$ | 1.1 | 1.00E+07 | 5.00E-05 | 5.70E-05 |
| OH- | 0.2 | 1.00E+07 | 5.00E-05 | 1.04E-05 |

| Species | 2 MeV proton G values | Proton Energy (eV) | Beam Current (A) | Species Generation Rate (mol/sec) |
|---|---|---|---|---|
| $e^{-}_{(aq)}$ | 0.3 | 2.00E+06 | 5.00E-05 | 3.11E-06 |
| $H^{\cdot}$ | 0.2 | 2.00E+06 | 5.00E-05 | 2.07E-06 |
| $H_2$ | 0.9 | 2.00E+06 | 5.00E-05 | 9.33E-06 |
| $OH^{\cdot}$ | 0.63 | 2.00E+06 | 5.00E-05 | 6.53E-06 |
| $H_2O_2$ | 0.76 | 2.00E+06 | 5.00E-05 | 7.88E-06 |
| $HO_2$ | 0.05 | 2.00E+06 | 5.00E-05 | 5.18E-07 |
| $H^{+}$ | 0.36 | 2.00E+06 | 5.00E-05 | 3.73E-06 |
| OH- | 0.06 | 2.00E+06 | 5.00E-05 | 6.22E-07 |

| Species | 10 MeV γ ray G values | Proton Energy (eV) | Beam Current (A) | Species Generation Rate (mol/sec) |
|---|---|---|---|---|
| $e^{-}_{(aq)}$ | 2.6 | 1.00E+07 | 5.00E-05 | 1.35E-04 |
| $H^{\cdot}$ | 0.66 | 1.00E+07 | 5.00E-05 | 3.42E-05 |
| $H_2$ | 0.45 | 1.00E+07 | 5.00E-05 | 2.33E-05 |
| $OH^{\cdot}$ | 2.7 | 1.00E+07 | 5.00E-05 | 1.40E-04 |
| $H_2O_2$ | 0.7 | 1.00E+07 | 5.00E-05 | 3.63E-05 |
| $HO_2$ | 0.02 | 1.00E+07 | 5.00E-05 | 1.04E-06 |
| $H^{+}$ | 3.1 | 1.00E+07 | 5.00E-05 | 1.61E-04 |
| OH- | 0.5 | 1.00E+07 | 5.00E-05 | 2.59E-05 |

**Table S3**

Rates of production for primary products of radioloysis of water from irradiation by 10 MeV protons, 2 MeV protons and 2 MeV gamma rays as described by Pastina et al[4]. G values describe the number of primary species products generated per particle emission. Simulations we run such that the relative rates of production for these radiolysis products were as shown here, with molar fluxes of each species in the simulation given by the quantity $r = k_{rad}[proton/gamma]$.

Example results for substitution of primary $OH^{\cdot}$, $H^{\cdot}$ and $e^{-}_{(aq)}$ production VUV illumination for these radiolytic processes are shown below. These substitutions, with the G values reported for these particular radiolytic sources fail to result in autocatalytic formate evolution. However, tuning radiolytic sources appropriately, such that their G values for $H^{\cdot}$, $OH^{\cdot}$, and $e^{-}_{(aq)}$ approach the quantum efficiencies of those species yields under VUV illumination, may possibly allow the use of radiolytic sources, such as depleted nuclear wastes, as energy sources for formate carboxylation in this autocatalytic cycle.

**S.18 Calculation of Radiolysis Rates and Rate Constants from Nuclear Decay Source G Values - Simulation of Autocatalysis Driven by Radiolysis from Nuclear Decays ($^{1}H^{+}$ irradiation).**

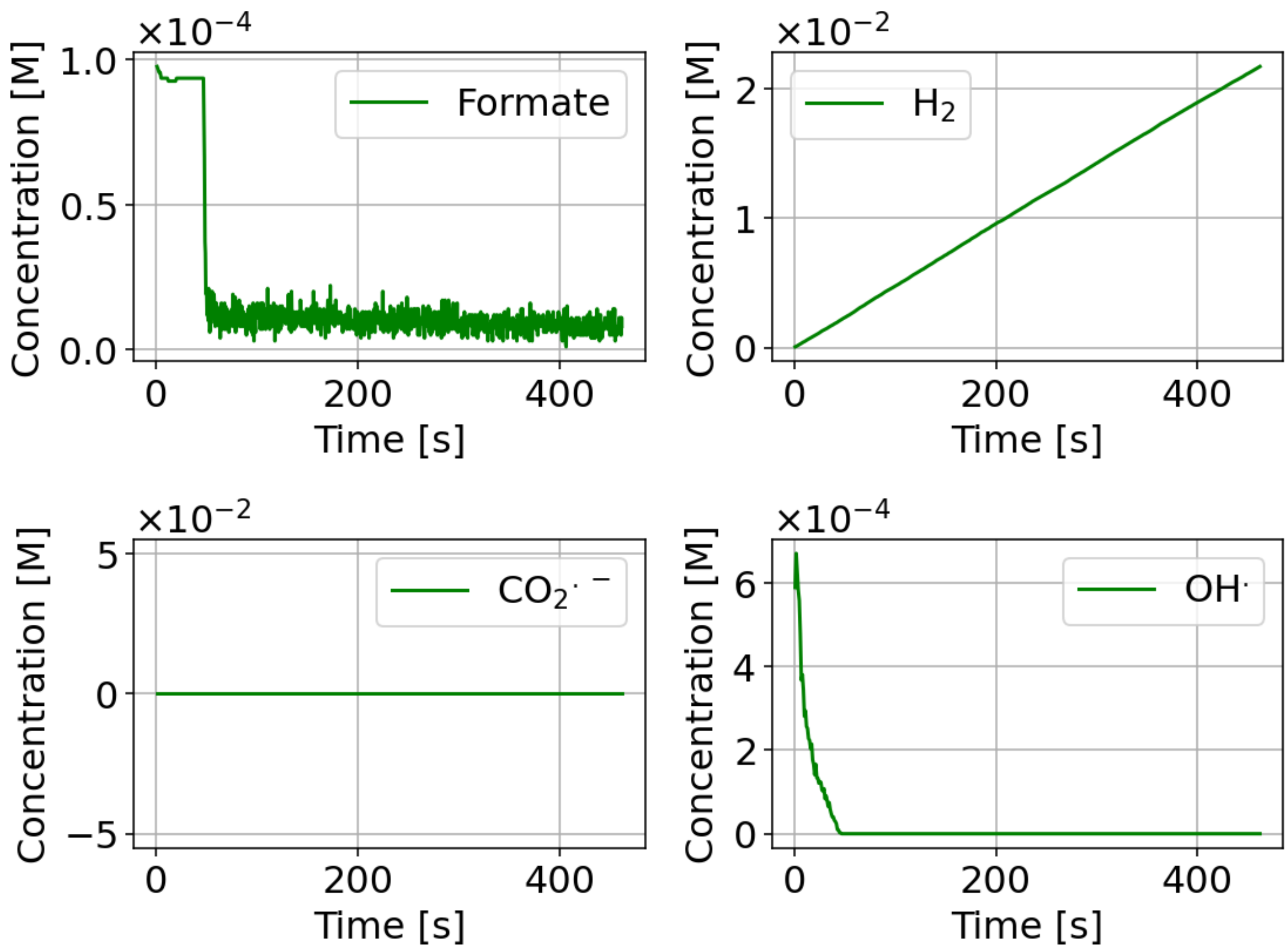

**Figure S31**

(*Top*) Formate autocatalysis simulated for radiolysis using 2 MeV proton irradation.

(*Bottom*) Formate autocatalysis simulated for radiolysis using 10 MeV proton irradation.

Calculations correspond to particle concentrations of 0.001 mol cm$^{-3}$ and a first-order rate constant ($k_{rad}$) for irradiation of 4.5x10$^{-4}$ s$^{-1}$.

**S.18 Calculation of Radiolysis Rates and Rate Constants from Nuclear Decay Source G Values - Simulation of Autocatalysis Driven by Radiolysis from Nuclear Decays ($^{1}H^{+}$ irradiation).**

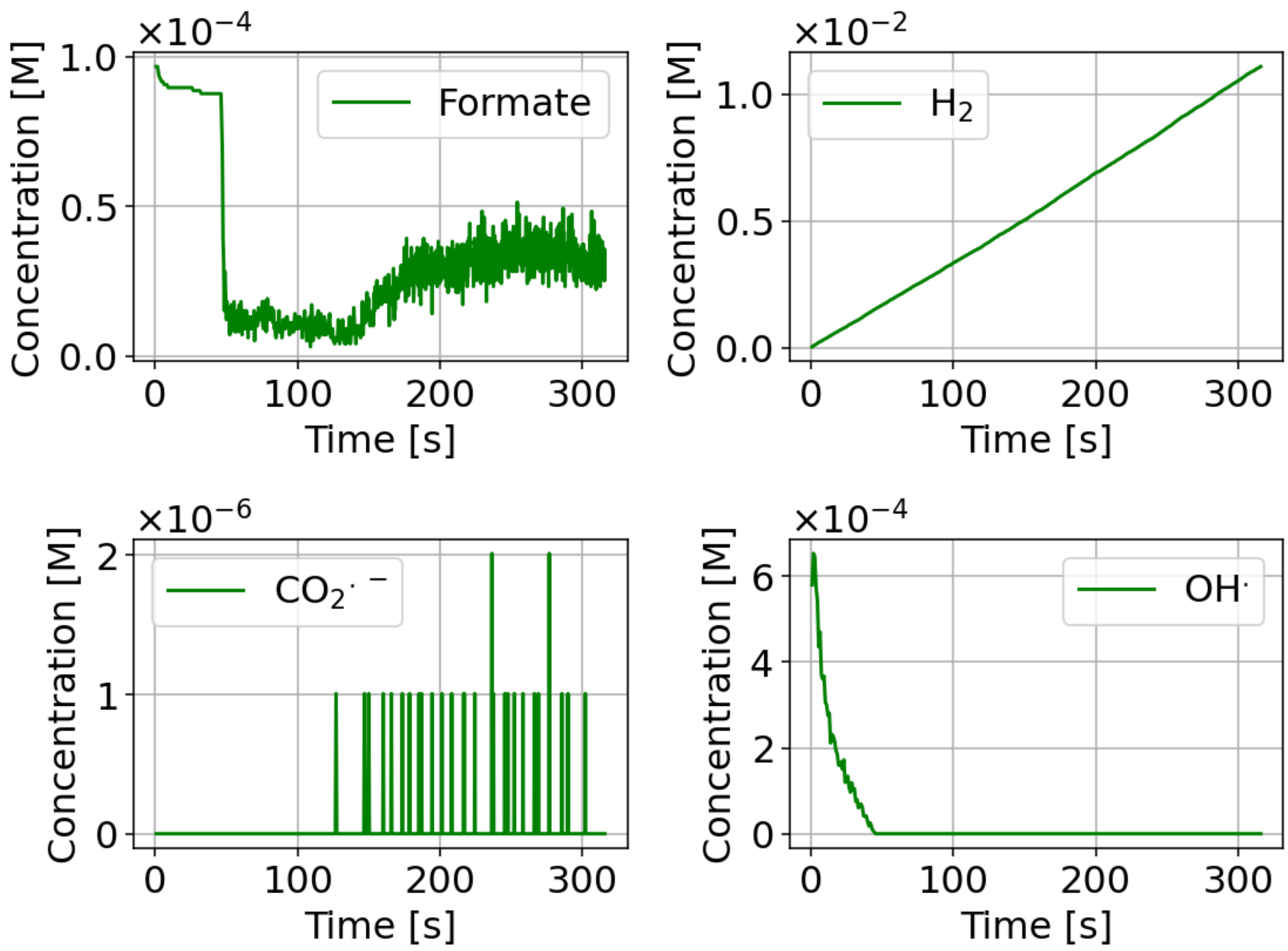

**Figure S32**

(*Top*) Formate autocatalysis simulated for radiolysis using 2 MeV proton irradation.

(*Bottom*) Formate autocatalysis simulated for radiolysis using 10 MeV proton irradation.

Calculations correspond to particle concentrations of 0.001 mol $cm^{-3}$ and a first-order rate constant ($k_{rad}$) for irradiation of $4.5x10^{-4}$ $s^{-1}$.

**S.18 Calculation of Radiolysis Rates and Rate Constants from Nuclear Decay Source G Values - Simulation of Autocatalysis Driven by Radiolysis from Nuclear Decays (γ rays).**

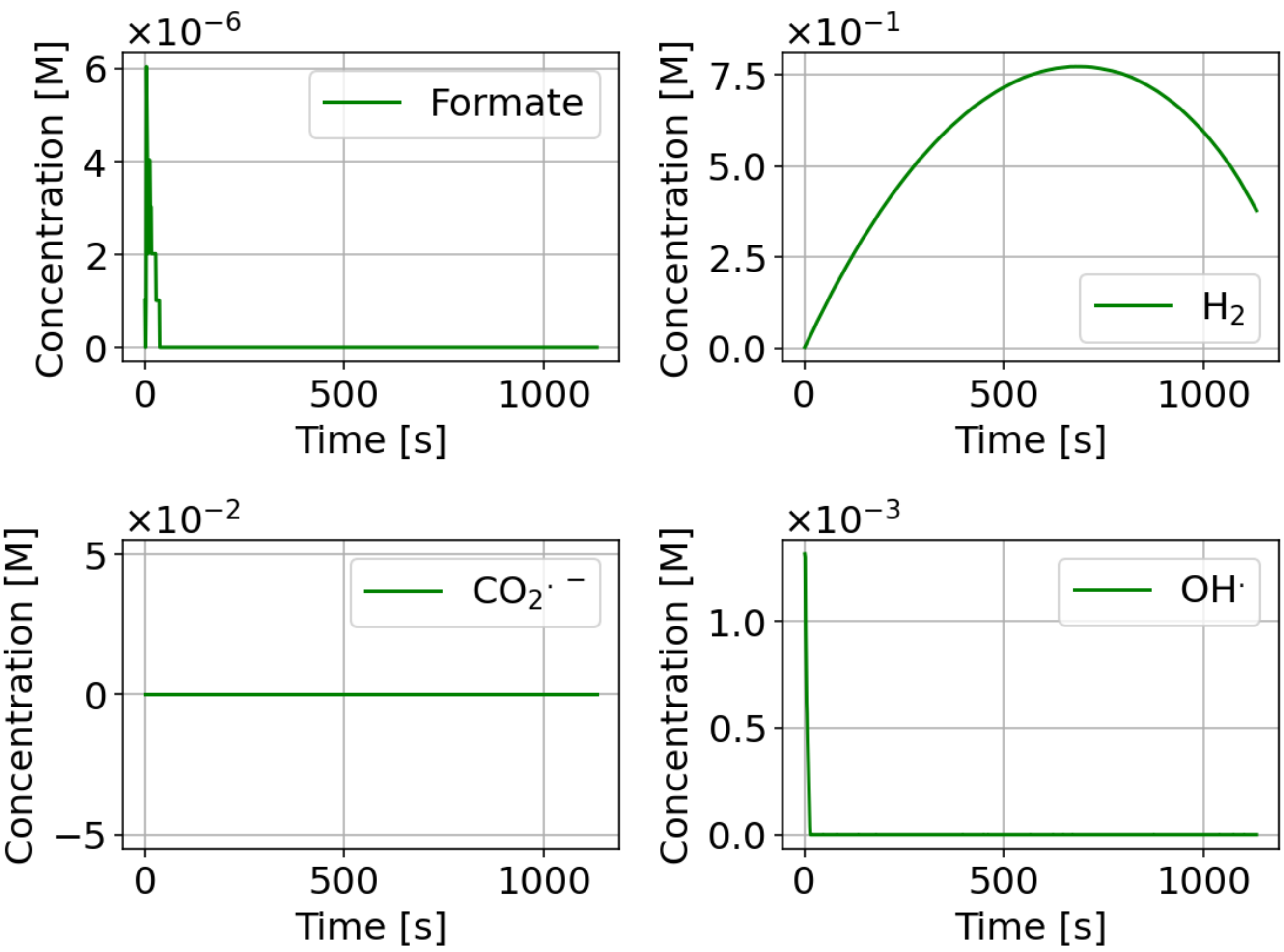

**Figure S33**

Formate autocatalysis simulated for radiolysis using 10 MeV γ rays.

Calculations correspond to particle concentrations of 0.001 mol $cm^{-3}$ and a first-order rate constant ($k_{rad}$) for irradiation of $4.5x10^{-4}$ $s^{-1}$.